\documentclass[pra,superscriptaddress,reprint]{revtex4-2}
\usepackage{graphicx}
\usepackage{bm}
\usepackage{ulem}
\usepackage{amsmath}
\usepackage{amssymb}
\usepackage{hhline}

\usepackage{color}

\newcommand{\bfq}{{\bf q}}
\newcommand{\bfE}{{\bf E}}
\newcommand{\bfJ}{{\bf J}}

\newcommand{\bfH}{{\bf H}}

\newcommand{\bfM}{{\bf M}}

\newcommand{\bfj}{{\bf j}}

\newcommand{\bra}{\langle}
\newcommand{\ket}{\rangle}

\begin{document}

\title{Fundamentals and advances in transverse thermoelectrics}

\author{Hiroto Adachi}
\email{hiroto.adachi@okayama-u.ac.jp}
\affiliation{Research Institute for Interdisciplinary Science, Okayama University, Okayama 700-8530, Japan}

\author{Fuyuki Ando}
\affiliation{Research Center for Magnetic and Spintronic Materials, National Institute for Materials Science, Tsukuba 305-0047, Japan}

\author{Takamasa Hirai}
\affiliation{Research Center for Magnetic and Spintronic Materials, National Institute for Materials Science, Tsukuba 305-0047, Japan}

\author{Rajkumar Modak}
\affiliation{Research Center for Magnetic and Spintronic Materials, National Institute for Materials Science, Tsukuba 305-0047, Japan}
\affiliation{Department of Advanced Materials Science, Graduate School of Frontier Sciences, The University of Tokyo, Kashiwa 277-8561, Japan}

\author{Matthew A. Grayson}
\affiliation{Department of Electrical and Computer Engineering, Northwestern University, Evanston, Illinois 60208, USA}

\author{Ken-ichi Uchida}
\email{UCHIDA.Kenichi@nims.go.jp}
\affiliation{Research Center for Magnetic and Spintronic Materials, National Institute for Materials Science, Tsukuba 305-0047, Japan}
\affiliation{Department of Advanced Materials Science, Graduate School of Frontier Sciences, The University of Tokyo, Kashiwa 277-8561, Japan}

\begin{abstract}
Transverse thermoelectric effects interconvert charge and heat currents in orthogonal directions due to the breaking of either time-reversal symmetry or structural symmetry, enabling simple and versatile thermal energy harvesting and solid-state cooling/heating within single materials. In comparison to the complex module structures required for the conventional Seebeck and Peltier effects, the transverse thermoelectric effects provide the complete device structures, potentially resolving the fundamental issue of multi-module degradation of thermoelectric conversion performance. This review article provides an overview of all currently known transverse thermoelectric conversion phenomena and principles, as well as their characteristics, and reclassifies them in a unified manner. The performance of the transverse thermoelectric generator, refrigerator, and active cooler is formulated, showing that thermal boundary conditions play an essential role to discuss their behaviors. Examples of recent application research and material development in transverse thermoelectrics are also introduced, followed by a discussion of future prospects.
\end{abstract}

\maketitle

%%%%%%%%%%%%%%%%%%%%%%%%
\section{Introduction}
%%%%%%%%%%%%%%%%%%%%%%%%

\begin{figure*}
  \begin{center}
    \includegraphics[width=13.2cm]{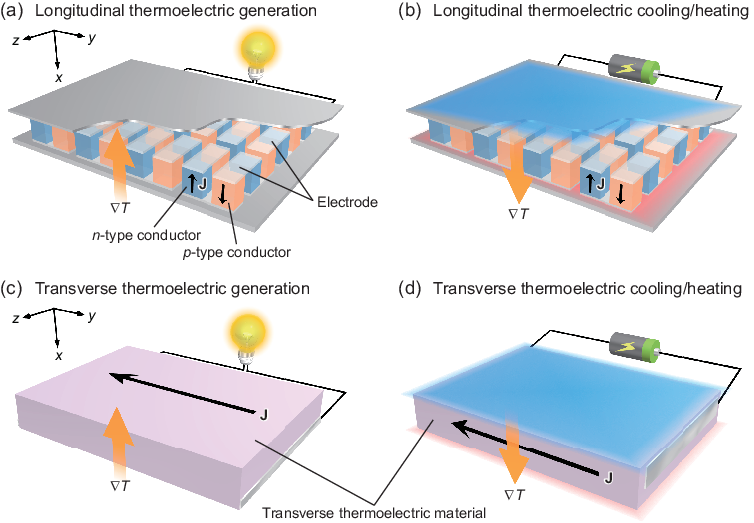}
  \end{center}
  \caption{Schematics of (a) longitudinal thermoelectric generation, (b) longitudinal thermoelectric cooling/heating, (c) transverse thermoelectric generation, and (d) transverse thermoelectric cooling/heating. $\bfJ$ ($\parallel y$) and $\nabla T$ ($\parallel x$) denotes the charge current and temperature gradient, respectively. The direction of $\bfJ$ in (a) and (c) corresponds to the direction of the electric field driven by the thermoelectric effects. In (b) and (d), the direction of $\bfJ$ is shown for the cooling operation. }
  \label{fig:long-vs-trans}
\end{figure*}

Thermoelectric conversion, which converts thermal and electrical energy in solids, is one of the promising technologies for realizing a sustainable society \cite{Yan22}. The most widely used thermoelectric generation principle is the Seebeck effect, discovered by T. J. Seebeck in 1821, which can be used to generate an electric voltage and current in the direction parallel to a temperature gradient $\nabla T$, {\it i.e.} the longitudinal direction. The ratio of the generated longitudinal electric field $E$ to the applied temperature gradient $\nabla T$ is called the Seebeck coefficient $S = E/\nabla T$, which represents the thermopower due to the Seebeck effect. As shown in Fig. \ref{fig:long-vs-trans}(a), a thermoelectric conversion module based on the longitudinal Seebeck effect has a structure in which a large number of pairs of $p$-type and $n$-type conductors are connected electrically in series and thermally in parallel. Since the Seebeck coefficient of $p$-type ($n$-type) conductors is positive (negative), the thermoelectric voltage of each conductor is added consecutively, and the output of the module is proportional to the number of the conductor pairs. Although each conductor only generates a thermoelectric voltage of the order of mV, by integrating them, a large, practical voltage can be obtained. In the reciprocal process, when a charge current $\bfJ$ is applied to such a module, the longitudinal heat current generated by the Peltier effect can be used to heat or cool the surface of the module depending on the $\bfJ$ direction [Fig. \ref{fig:long-vs-trans}(b)]. 

The thermoelectric conversion performance of materials is often evaluated using the dimensionless figure of merit $zT$ with $T$ being the absolute temperature. As discussed in Sec. \ref{Sec:Formulation-time} in more detail, $zT$ for the Seebeck effect is proportional to the electrical conductivity and the square of the Seebeck coefficient and inversely proportional to the thermal conductivity. For a long time, $zT$ exceeding 1 was an indicator to realize practical applications of the thermoelectric conversion. As a result of active materials research in the 21st century, various materials with $zT$ far exceeding 1 have been synthesized and discovered \cite{Yan22}. However, despite the revolutionary progress in materials research, the range of applications of thermoelectric power generation and cooling/heating technologies is still limited. Part of the reason for this situation is the complicated structure of the longitudinal thermoelectric modules based on the Seebeck and Peltier effects [Figs. \ref{fig:long-vs-trans}(a) and \ref{fig:long-vs-trans}(b)]. In this module structure, which has four junctions per pair of $p$-type and $n$-type conductors, the interfacial electrical and thermal resistances increase significantly; even if excellent thermoelectric characteristics are obtained at the material level, the energy conversion efficiency decreases when the multicomponent modules are constructed \cite{Yan22}. There are also problems with the thermal and mechanical durability of the modules and the manufacturing cost due to complexity. 

%%%%%%%%%%%%%%%%%%%%%%%%%%%%%%%%%%%%% 
%\begin{figure*}[tb]
%%%%%%%%%%%%%%%%%%%%%%%%%%%%%%%%%%%%

\begin{figure} 
  \begin{center}
    \includegraphics[width=6.5cm]{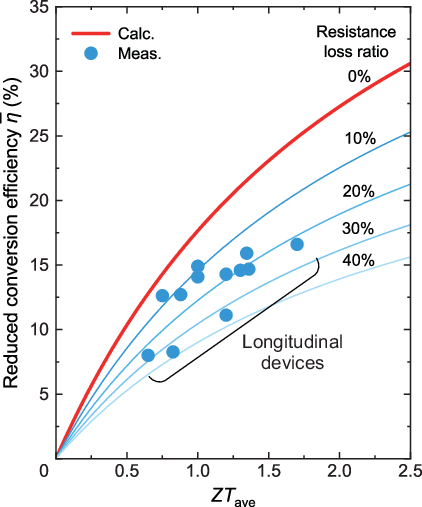}
  \end{center}
  \caption{Calculated and observed reduced conversion efficiency $\bar{\eta}$ as a function of the averaged figure of merit $ZT_{\rm ave}$. The blue curves are calculated with changing the resistance loss ratio from 10\% to 40\%. The blue data points are estimated from the experimental results in Refs. \cite{Hao16,Deng18,KELK,Skomedal16,Jia22,Liu21,Wang24,Jiang22,Ando23,Zong17,Yu20,Xing20}. }
  \label{fig:ZTvsEfficiency}
\end{figure}

To quantify this situation, Fig. \ref{fig:ZTvsEfficiency} compares the ideal reduced conversion efficiency $\bar{\eta}$ (red curve), {\it i.e.} the efficiency normalized by the Carnot efficiency, with the $\bar{\eta}$ values observed in actual longitudinal thermoelectric modules (blue plots) as a function of the device figure of merit $ZT_{\rm ave}$, where $ZT_{\rm ave}$ is defined as the averaged material $zT$ within the module measurement temperature range \cite{Snyder17} (note that lowercase $z$ and uppercase $Z$ represent the material and device figures of merit, respectively). The $\bar{\eta}$ values refer to various thermoelectric modules, including Bi$_2$Te$_3$-based \cite{Hao16,Deng18,KELK}, PbTe-based \cite{Jia22}, GeTe-based \cite{Jiang22,Ando23}, MgSi$_2$-based \cite{Skomedal16}, Mg$_3$Sb$_2$-based \cite{Liu21,Wang24}, Skutterudite-based \cite{Zong17}, and half-Heusler-alloy-based \cite{Yu20,Xing20} systems. In all cases, the observed efficiencies are much lower than the ideal values. Despite the development of thermoelectric modules with high $ZT_{\rm ave}$, the conversion efficiency has not increased drastically due to excess resistance, and the larger $ZT_{\rm ave}$, the greater the deviation from the ideal value. Following Refs. \cite{Ando23,Xing20,Rowe98,Shittu20}, the resistance loss ratio is defined to be the ratio of the total excess resistance, including junctions and electrodes, to the total resistance of thermoelectric materials themselves. For simplicity, we assume that the resistance loss ratio is the same for both electrical and thermal resistances, and plot the $ZT_{\rm ave}$ dependence of $\bar{\eta}$ as the resistance loss ratio varies from 10\% to 40\% (Fig. \ref{fig:ZTvsEfficiency}). These curves agree well with the experimental data, indicating that high-performance longitudinal thermoelectric modules exhibit a resistance loss of several tens of percent. 

%%%%%%%%%%%%%%%%%%%%%%%%%%%%%%%%%%%%% 
%\begin{figure}[bt]
%%%%%%%%%%%%%%%%%%%%%%%%%%%%%%%%%%%%

One way to solve these problems is to implement transverse thermoelectric conversion \cite{Uchida-Heremans22}. Using the transverse thermoelectric effects, it is possible to generate an electric voltage and current in a direction perpendicular to $\nabla T$, {\it i.e.} the transverse direction [Figs. \ref{fig:long-vs-trans}(c) and  \ref{fig:long-vs-trans}(d)]. The material figure of merit $z_{xy}T= S_{yx}^2 T/\rho_{yy} \kappa_{xx} $ for transverse thermoelectrics is defined in a similar way to that for longitudinal thermoelectrics, where the transverse thermopower $S_{yx}$ is defined as the generated electric field in the $y$ direction normalized by applied $\nabla T$ in the $x$ direction (see Secs. \ref{Sec:Formulation-time} and \ref{Sec:Formulation-structural} for details). Here, $\kappa_{xx}$ is the thermal conductivity for the $x$ direction and $\rho_{yy}$ is the electrical resistivity for the $y$ direction. The voltage and power induced by the transverse thermoelectric effects can be increased by increasing the length and area of the material in the direction perpendicular to $\nabla T$, respectively, and it is not necessary to form a large number of junction structures [Fig. \ref{fig:long-vs-trans}(c)]. For this reason, the transverse thermoelectric conversion is suitable for reusing thermal energy distributed over a wide area. Furthermore, transverse thermoelectric conversion modules do not require junctions, and there are no problems of interfacial electrical and thermal resistances and thermal degradation on the hot side [note that two electrical contacts for extracting output can be attached to the cold side, as depicted in Fig. \ref{fig:long-vs-trans}(c)]. Thus, the efficiency calculated from the material’s $z_{xy}T$ (red curve in Fig. \ref{fig:ZTvsEfficiency}) can be expected in experimental modules. The transverse thermoelectric conversion is also expected to improve the durability of the modules and reduce manufacturing costs. While the transverse thermoelectric conversion has many advantages, there are still various issues that remain to be addressed, such as the facts that the transverse thermopower has not yet reached a practical level and that the importance of thermal boundary conditions, which are essential for discussing the performance of the transverse thermoelectric devices, has not been fully recognized. Importantly, since various thermoelectric conversion principles are being studied independently in different fields, there is currently no unified paradigm for classifying these phenomena. 

In this article, we review fundamentals and advances of transverse thermoelectrics. This article is organized as follows. First, in Sec.~\ref{Sec:Classification}, we systematically organize the classification of the transverse thermoelectric conversion phenomena and summarize their mechanisms and characteristics. The transverse thermoelectric conversion phenomena require the symmetry breaking for charge and heat carriers. Depending on how this symmetry is broken, the transverse thermoelectric conversion phenomena can be broadly classified into two categories: those that occur due to time-reversal symmetry breaking and those due to structural symmetry breaking. Here, the structural symmetry breaking in this article means that the sample has a certain structural asymmetry with respect to an axis taken in the direction of a charge or heat current, at either microscale or macroscale. In other words, this terminology is defined as the breaking of sample's spatial inversion symmetry about the left and right sides of that axis. In Secs.~\ref{Sec:Formulation-time} and \ref{Sec:Formulation-structural}, we formulate the performance of a transverse thermoelectric generator, refrigerator, and active cooler for the cases with time-reversal symmetry breaking and structural symmetry breaking, respectively. Although a thermoelectric device also acts as a heat pump, this heating mode is not dealt with in this article due to the limited space. Unlike the longitudinal thermoelectric conversion based on the Seebeck and Peltier effects, we show that thermal boundary conditions in the electric field direction play an essential role in the performance of the transverse thermoelectric conversion. For the transverse thermoelectric generator, the expressions of the figure of merit and efficiency vary depending on whether the thermal boundary conditions are isothermal or adiabatic, but the important point is that they are interchangeable with each other; the transverse thermoelectric generation is not inherently more or less efficient than the longitudinal thermoelectric generation, contrary to the discussion in Ref. \cite{Mizuguchi-Nakatsuji19}. Importantly, depending on the thermal boundary conditions, a correction term due to the transverse temperature gradient and Seebeck effect occurs in the transverse thermopower. The experiments shown in Sec.~\ref{Sec:Measurement} demonstrate the importance of the thermal boundary condition and the contribution of the correction term. In Sec.~\ref{Sec:Application}, we summarize research activities for developing applications of the transverse thermoelectric conversion and introduce new concepts for further performance improvements. Section~\ref{Sec:Conclusion} is devoted to the conclusions and prospects. We hope that this article serves as a cornerstone for basic and applied research on transverse thermoelectrics and provides an opportunity for interdisciplinary fusion of research on various transverse thermoelectric conversion phenomena that have been studied in different communities.

%%%%%%%%%%%%%%%%%%%%%%%
\section{Classification of transverse thermoelectric conversion phenomena} \label{Sec:Classification}
%%%%%%%%%%%%%%%%%%%%%%%

\begin{figure*} 
  \begin{center}
    \includegraphics[width=14.5cm]{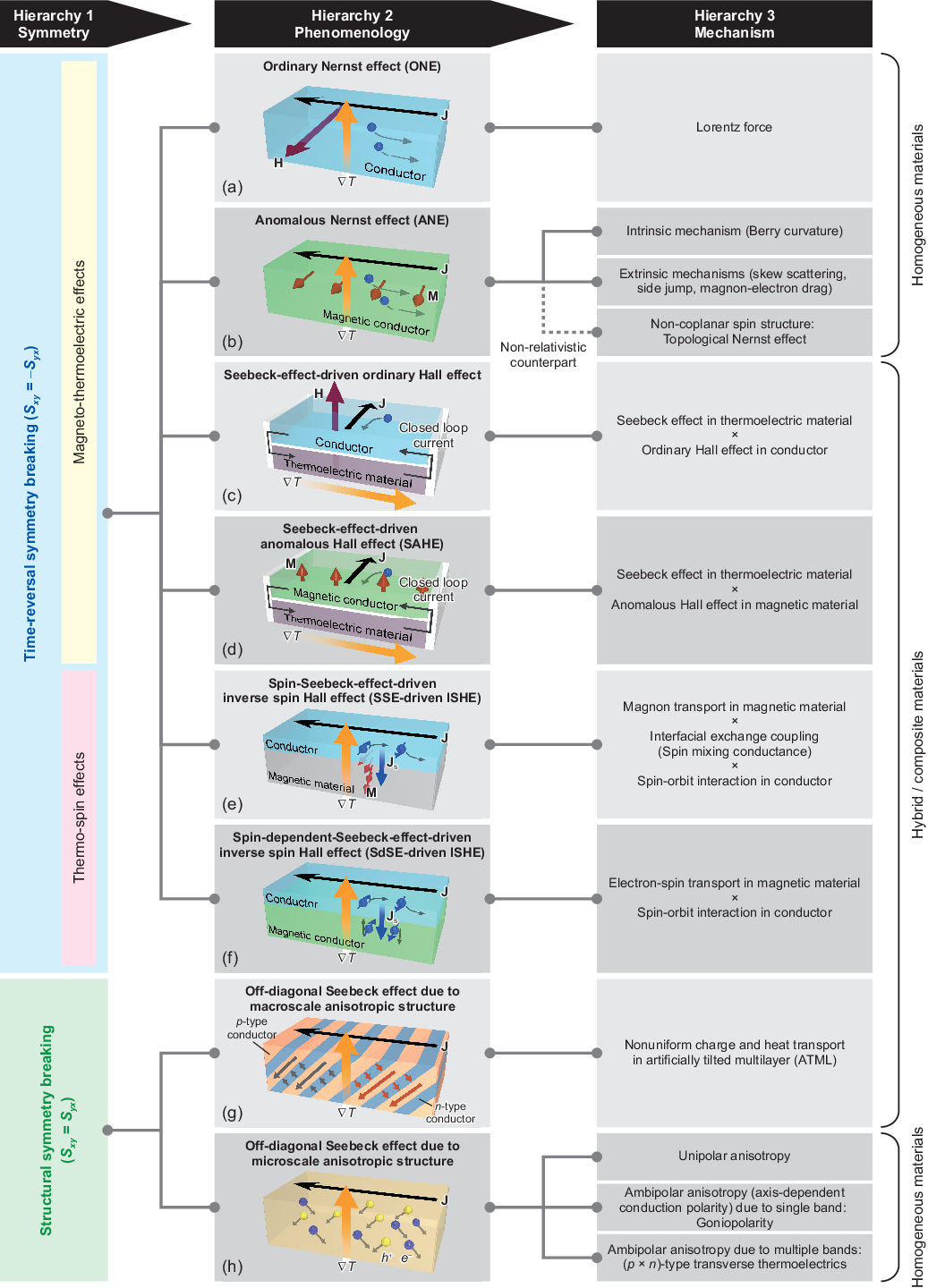}
  \end{center}
  \caption{Classification of the transverse thermoelectric effects and schematics of (a) the ordinary Nernst effect (ONE), (b) anomalous Nernst effect (ANE), (c) Seebeck-effect-driven ordinary Hall effect, (d) Seebeck-effect-driven anomalous Hall effect (SAHE), (e) spin-Seebeck-effect-driven (SSE-driven) inverse spin Hall effect (ISHE), (f) spin-dependent-Seebeck-effect-driven (SdSE-driven) ISHE, (g) off-diagonal Seebeck effect due to macroscale anisotropic structure, and (h) off-diagonal Seebeck effect due to microscale anisotropic structure. $\bfH$, $\bfM$, $\bfJ_{\rm s}$, $e^-$, and $h^+$ denote the magnetic field, magnetization, spatial direction of a spin current, conduction electrons, and holes, respectively. In (g), the gray (orange) arrows show the direction of charge (heat) currents in the constituent conductors of the artificially tilted multilayer (ATML). The thermopower tensor is antisymmetric $S_{xy} = - S_{yx}$ when a magnetic field or magnetism breaks time-reversal symmetry, or symmetric $S_{xy} = S_{yx}$ when a crystal lattice or layered macrostructure breaks structural symmetry.}
  \label{fig:classification}
\end{figure*}

The purpose of this section is to establish a unified classification for various transverse thermoelectric conversion phenomena. The transverse thermoelectric conversion phenomena caused by time-reversal symmetry breaking include the magneto-thermoelectric and thermo-spin effects, which have been studied mainly in the field of spin caloritronics \cite{Bauer12,Boona14,Uchida21}. Here, the magneto-thermoelectric effects refer to the thermoelectric effects depending on magnetic fields or magnetization, while the thermo-spin effects refer to the conversion between a heat current and a spin current, {\it i.e.} a flow of spin angular momentum. Some of these phenomena were discovered in the 21st century, and they are still in their infancy. Although the transverse thermoelectric conversion phenomena caused by structural symmetry breaking have been studied in the field of thermoelectrics for a long time \cite{Babin74,Goldsmid11}, they are often called by completely different terminologies, which may confuse non-specialists. For example, the transverse thermoelectric conversion in a material having charge carriers with opposite signs along different crystal orientations, {\it i.e.} axis-dependent conduction polarity (ADCP), is called goniopolarity \cite{He19,Scudder21,Scudder22} when it is caused by a single band, but called ($p \times n$)-type transverse thermoelectrics \cite{Zhou13,Tang15} when it is caused by multiple bands, despite the fact that the details of the band structure cannot be determined only by thermoelectric measurements. We argue that the leading classification of the transverse thermoelectric effects should be based on the symmetry and phenomenology, which can be determined from thermoelectric measurements alone. Therefore, we propose a classification of the transverse thermoelectric effects comprising three hierarchies: symmetry (hierarchy 1), phenomenology (hierarchy 2), and mechanism (hierarchy 3). Figure \ref{fig:classification} shows the currently known transverse thermoelectric generation phenomena organized based on this classification, which is a refinement of the classification in Ref. \cite{Uchida-Heremans22}. In the following subsections, we outline the mechanisms and characteristics of each phenomenon. Although we will mainly focus on the thermoelectric generation, the phenomena in Fig. \ref{fig:classification} have their Onsager reciprocals, enabling transverse thermoelectric cooling/heating (Table \ref{table:classification}). Recently, a higher-order transverse thermoelectric effect called the transverse Thomson effect \cite{Takahagi25} has been observed experimentally, but it is not discussed in this article. 

%%%%%%%%%%%%%%%%%%%%%%%%%%%%%%%%%%%%% 
%\begin{figure*}[tb]
%%%%%%%%%%%%%%%%%%%%%%%%%%%%%%%%%%%%

%%%%%%%%%%%%%%%%%%%%%%%%%%%%%%%%%%%%
\begin{table*}
  \caption{List of the transverse thermoelectric generation phenomena and their Onsager reciprocals. \label{table:classification}}
%\begin{indented}
 %\item[]\begin{tabular}{@{}ll}
  %\br
  \begin{tabular}{ll} \hline
Heat-to-charge current conversion & Charge-to-heat current conversion \\ \hline\hline
%\br
 Ordinary Nernst effect & Ordinary Ettingshausen effect \\
 Anomalous Nernst effect & Anomalous Ettingshausen effect \\
 Seebeck-effect-driven ordinary Hall effect & Ordinary-Hall-effect-driven Peltier effect \\
 Seebeck-effect-driven anomalous Hall effect & Anomalous-Hall-effect-driven Peltier effect \\
 Spin-Seebeck-effect-driven inverse spin Hall effect & Spin-Hall-effect-driven spin Peltier effect \\
 Spin-dependent-Seebeck-effect-driven inverse spin Hall effect & Spin-Hall-effect-driven spin-dependent Peltier effect \\
 Off-diagonal Seebeck effect & Off-diagonal Peltier effect \\ \hline
%\br
\end{tabular}
%\end{indented}
\end{table*}
%%%%%%%%%%%%%%%%%%%%%%%%%%%%%%%%%%%%

%%%%%%%%%%%%%%%%%%%%%%%
\subsection{Transverse thermoelectric effects with time-reversal symmetry breaking}
%%%%%%%%%%%%%%%%%%%%%%%

The majority of the transverse thermoelectric effects require time-reversal symmetry breaking. In other words, the transverse thermoelectric effects listed in this subsection occur in conductors under magnetic fields or in magnetic materials with magnetization. Here, ``magnetization'' in this article includes not only spontanious magnetization of ferromagnets and ferrimagnets but also weak magnetization of canted antiferromagnets. 

\subsubsection{Ordinary Nernst effect. }

The oldest known transverse thermoelectric conversion phenomenon is the Nernst effect, discovered in 1886 by A. V. Ettingshausen and W. Nernst \cite{Ettingshausen-Nernst86}. They found that when $\nabla T$ and the magnetic field $\bfH$ are applied to a conductor in directions orthogonal to each other, a thermoelectric voltage is generated in the direction of the cross product of $\nabla T$ and $\bfH$. The transverse thermopower due to the Nernst effect in normal metals is proportional to the magnitude of the magnetic field $H$, while that in semimetals often exhibits a nonlinear dependence on $H$ due to the concerted contributions of electron and hole transport. This phenomenon is often called the Nernst-Ettingshausen effect but in this article, we simply call it the Nernst effect and its reciprocal the Ettingshausen effect. Nowadays, this phenomenon is widely called the ordinary Nernst effect (ONE) to distinguish it from the anomalous Nernst effect (ANE), described later. 

ONE is caused by the Lorentz force acting on conduction electrons and/or holes in a conductor under a magnetic field [Fig. \ref{fig:classification}(a)]. The ONE-induced transverse thermopower in typical metals is very small, but it is known that several Dirac/Weyl semimetals and low-dimensional materials exhibit high transverse thermoelectric conversion performance due to ONE \cite{Jandl-Birkholz94,Pan22-ONE,Li22,Yang23,Pasquale24}. For example, the figure of merit of BiSb alloys reaches $z_{xy}T>0.3$ at 100–200 K when an external magnetic field of $\sim$1 T is applied \cite{Jandl-Birkholz94}. However, the drawback of ONE is that an external magnetic field must be applied to generate the transverse thermopower. To overcome this drawback, it has been demonstrated that the thermoelectric power generation based on ONE can be achieved without applying an external magnetic field by embedding permanent magnets in a module, where the remanent magnetization of the permanent magnets applies stray magnetic fields to the adjacent conductors \cite{Murata24}. However, there are still issues to be addressed, such as the fact that the fill factor of ONE materials that are responsible for the power generation decreases by the amount of the embedded permanent magnets. 

\subsubsection{Anomalous Nernst effect. }

ANE is a phenomenon in which a thermoelectric voltage is generated in the direction of the cross product of $\nabla T$ and the spontaneous magnetization $\bfM$ in a magnetic material [Fig. \ref{fig:classification}(b)]. The transverse thermopower $S_{\rm T}$ in a magnetic material under a magnetic field is phenomenologically described as the summation of the contributions proportional to $H$, {\it i.e.} ONE, and magnetization $M$, {\it i.e.} ANE \cite{Uchida-Zhou-Sakuraba21}: 
\begin{equation} \label{eq:ANE-definition}
S_{\rm T} = Q_{H} \mu_0 H + Q_{M} \mu_0 M, 
\end{equation}
where $\mu_0$ is the vacuum permeability and $Q_H$ ($Q_M$) is the proportionality factor of the $H$-dependent ($M$-dependent) term. Equation (\ref{eq:ANE-definition}) shows that the anomalous Nernst coefficient $S_{\rm ANE}$ ($= Q_{M} \mu_{0}M_{\rm s}$ with $M_{\rm s}$ being the saturation magnetization) can be extracted by extrapolating the $H$ dependence of $S_{\rm T}$ from the high field region, in which $M$ is saturated, to zero field. Thus, if $\bfM$ is aligned in one direction, ANE can generate a thermoelectric voltage and power even without applying an external magnetic field. Note that the magnitude of the transverse thermopower due to ANE is determined by $S_{\rm ANE}$, not $Q_{M}$; $Q_{M}$ is just a phenomenological parameter defined for convenience to separate ANE from ONE. Based on Eq. (\ref{eq:ANE-definition}), $S_{\rm ANE}$ is often compared in terms of $M$ to discuss the scaling behavior. However, since ANE and related transport properties are estimated for uniformly magnetized materials, they should be compared in terms of $M_{\rm s}$, not $M$. In fact, $S_{\rm ANE}$ is not correlated with $M_{\rm s}$ in many magnetic materials; the scaling behavior does not hold even for simple ferromagnetic metals, such as Fe, Ni, and Co \cite{Uchida-Zhou-Sakuraba21}. 

The mechanism of ANE differs from that of ONE. In general, ANE originates from an intrinsic mechanism, {\it i.e.} a fictitious magnetic field in a momentum space derived from the Berry curvature in electronic band structures, and/or an extrinsic mechanism, {\it i.e.} skew scattering, side jump, and magnon-electron drag \cite{Pu08,He19-Joule,Toyama2024}. Recent experiments and calculations show that ANE can be enhanced through the intrinsic mechanism by the topology of the electronic structures, and it becomes one of the hot topics in condensed matter physics. Although $S_{\rm ANE}$ is in the order of $0.1~\mu {\rm V/K}$ in conventional ferromagnetic metals ({\it e.g.} Fe, Ni, and Co), Co-based Heusler compounds ({\it i.e.} Co$_2$MnGa and Co$_2$MnAl$_{1-x}$Si$_x$) exhibit a transverse thermopower of $S_{\rm ANE} \sim 6~\mu {\rm V/K}$ due to their topological electronic structure \cite{Sakai18,Reichlova18,Guin19,Sakuraba20,Sumida20,Oyanagi25}. In addition to the Co-based Heusler compounds, large $S_{\rm ANE}$ has been observed in various systems including the binary Fe-based alloys \cite{Nakayama19,Zhou-Sakuraba20,Sakai20,Chen22,Fujiwara23}, Mn-based pnictide ({\it i.e.} YbMnBi$_2$) \cite{Pan22-ANE}, Co-based shandite ({\it i.e.} Co$_3$Sn$_2$S$_2$) \cite{Guin19-AM,Ding19,Noguchi24}, U-based compound ({\it i.e.} UCo$_{0.8}$Ru$_{0.2}$Al) \cite{Asaba21}, magnetic multilayer films \cite{Uchida15,Seki21}, amorphous metals with nanoscale precipitates \cite{Gautam24,Park25,Park-ArXiv1,Park-ArXiv2}, and rare-earth permanent magnets ({\it i.e.} SmCo$_5$-type magnets) \cite{Miura19,Miura20}. 

To realize practical applications of ANE, it is necessary to find and develop magnetic materials with larger $S_{\rm ANE}$. Guidelines for designing $S_{\rm ANE}$ can be obtained by separating it into two components \cite{Sakuraba20,Lee04}:
\begin{equation} \label{eq:ANE-components}
S_{\rm ANE} = \rho_{xx} \alpha_{xy} - \rho_{\rm AHE} \alpha_{xx},
\end{equation} 
where $\rho_{xx}$, $\rho_{\rm AHE}$, and $\alpha_{xx}$ ($\alpha_{xy}$) are the longitudinal electrical resistivity, anomalous Hall resistivity, and diagonal (off-diagonal) component of the thermoelectric conductivity tensor, respectively. Here, the thermoelectric conductivity, also called the Peltier conductivity, is the tensor that relates an induced charge current density to an applied temperature gradient, the unit of which is A/Km. The first term on the right-hand side of Eq. (\ref{eq:ANE-components}) shows the direct transverse thermoelectric conversion due to $\alpha_{xy}$; a recent trend in improving $S_{\rm ANE}$ is to find materials with large $\alpha_{xy}$ caused by the Berry curvature of the electronic bands near the Fermi level. The second term on the right-hand side of Eq. (\ref{eq:ANE-components}) appears due to the anomalous Hall effect (AHE) \cite{Nagaosa10} induced by the longitudinal carrier flow through the Seebeck effect. However, even though the output of ANE has improved in recent years, $S_{\rm ANE}$ is still more than an order of magnitude smaller than the Seebeck coefficients of thermoelectric materials in practical use. Therefore, further performance improvements through materials research are necessary to realize thermoelectric applications of ANE. 

Here, we should mention the non-relativistic counterpart of ANE: the topological Nernst effect [Fig. \ref{fig:classification}(b)]. In the breaking of the spin-rotation invariance, the Nernst effect occurs in the absence of the spin-orbit interaction due to non-coplanar spin structures, {\it e.g.} in skyrmions and non-coplanar antiferromagnets \cite{Smejkal20,Hirschberger20,Kolincio21,Khanh25}. Although only a limited number of materials exhibit the topological Nernst effect, this phenomenon also functions as the transverse thermoelectric conversion.

\subsubsection{Seebeck-effect-driven ordinary and anomalous Hall effects. }

As part of efforts to further improve magneto-thermoelectric conversion performance, a transverse thermoelectric conversion mechanism appearing in hybrid materials composed of magnetic and thermoelectric materials was proposed and demonstrated in 2021 \cite{Zhou21}. This mechanism, originally named the Seebeck-driven transverse thermoelectric generation, operates when a closed circuit is constructed by connecting the ends of the magnetic and thermoelectric materials and is driven by the concerted action of the Seebeck effect in the thermoelectric material and AHE in the magnetic material. In other words, this mechanism corresponds to artificially modulating the second term on the right-hand side of Eq. (\ref{eq:ANE-components}) by constructing hybrid materials. However, the name ``Seebeck-driven transverse thermoelectric generation'' could be confused with the transverse thermoelectric conversion due to the off-diagonal Seebeck effect described below. In this article, we thus call this mechanism the Seebeck-effect-driven AHE (SAHE). 

SAHE has a high degree of freedom in material design. The transverse thermopower in the hybrid structure depicted in Fig. \ref{fig:classification}(d) is expressed as
\begin{equation} \label{eq:SANE}
S_{\rm T} = S_{\rm ANE} - \frac{\rho_{\rm AHE}}{\rho_{\rm TE}/r + \rho_{\rm M}}(S_{\rm TE} - S_{\rm M}),
\end{equation} 
where $\rho_{\rm M(TE)}$ and $S_{\rm M(TE)}$ are the longitudinal resistivity and Seebeck coefficient of the magnetic (thermoelectric) material, respectively, and $r$ is the size ratio determined by the dimensions of the magnetic and thermoelectric materials \cite{Zhou21}. Here, we consider a situation in which magnetic and thermoelectric materials are separated by an insulating layer to avoid the shunting effect. When the in-plane areas of the magnetic and thermoelectric materials are the same, $r$ is determined by the thickness ratio of the two: $r = t_{\rm TE} / t_{\rm M}$ with $t_{\rm M(TE)}$ being the thickness of the magnetic (thermoelectric) material. The second term on the right-hand side of Eq. (\ref{eq:SANE}) represents the contribution of SAHE. By optimizing the combination and size ratio of the magnetic and thermoelectric materials, it is possible to obtain a much larger transverse thermopower than $S_{\rm ANE}$ alone. When a thin film of a magnetic material is combined with a slab of a thermoelectric material, $r$ becomes very large, and $S_{\rm T}$ approaching $100~\mu {\rm V/K}$ has been observed \cite{Zhou21}. When a slab of a magnetic material is combined with a slab of a thermoelectric material, $S_{\rm T}$ is limited to around 10-$20~\mu {\rm V/K}$ due to smaller $r$ but such all-bulk hybrid structure is more suitable for increasing output power \cite{Zhou23}. Recently, direct-contact SAHE has also been demonstrated, in which a magnetic material is directly deposited on the surface of a thermoelectric material, rather than forming a closed circuit \cite{Zhou24}. Direct-contact SAHE functions with a simpler device structure, but its transverse thermopower decreases due to the shunting effect compared to the value for the structure depicted in Fig. \ref{fig:classification}(d). Since there is a vast amount of research data accumulated over many years on AHE and the Seebeck effect, there is still plenty of room for improving the performance of SAHE. However, its versatility is lower than that of ANE; whereas ANE works in both in-plane and perpendicularly magnetized configurations, SAHE works only in the perpendicularly magnetized configuration. Furthermore, even if $S_{\rm T}$ is enhanced, $z_{xy}T$ and power generation efficiency of SAHE are not always higher than those of ANE \cite{Yamamoto21}. 

It is worth mentioning that, if a magnetic material showing AHE is replaced with a nonmagnetic material showing the ordinary Hall effect, the Seebeck-effect-driven ordinary Hall effect should occur [Fig. \ref{fig:classification}(c)]. In principle, the transverse thermoelectric conversion due to the Seebeck-effect-driven ordinary Hall effect exists in any hybrid material comprising two or more materials under a magnetic field. However, at present, there are no reports where its contribution is explicitly observed. 

\subsubsection{Spin Seebeck and spin-dependent Seebeck effects. }

As is clear from the phenomena discussed above, magnetic materials play an important role in the development of the transverse thermoelectric conversion but magnetization has been treated as a macroscopic parameter that determines the symmetry of the phenomena. Since the beginning of the 21st century, there has been a lot of activity in the field of spintronics, which aims to utilize the spin degree of freedom, an origin of magnetism, from a microscopic perspective. In this field, various transport phenomena related to a spin current have been discovered and elucidated. As part of this activity, thermoelectrics and thermal transport have been integrated into spintronics, giving rise to the field of spin caloritronics \cite{Bauer12,Boona14,Uchida21}. Spin caloritronics has rapidly progressed since the discovery of the spin Seebeck effect (SSE), in which a spin current is generated from a heat current in a magnetic material \cite{Uchida08,Uchida10-Nmat,Jaworski10,Uchida10-APL}. 

SSE enables spin-current-driven transverse thermoelectric generation in combination with the spin-to-charge current conversion phenomenon called the inverse spin Hall effect (ISHE) \cite{Azevedo05,Saitoh06,Valenzuela06} [Fig. \ref{fig:classification}(e)]. When $\nabla T$ is applied perpendicular to a bilayer structure consisting of a metal film on a magnetic material having in-plane $\bfM$, a spin current with the spatial direction $\bfJ_{\rm s}$ is generated near the metal/magnetic-material interface due to SSE. This spin current is then converted into a charge current in a direction perpendicular to both $\nabla T$ and $\bfM$ through the spin-orbit interaction, {\it i.e.} ISHE, in the metal film. In spin caloritronics, SSE in the configuration shown in Fig. \ref{fig:classification}(e) is called ``longitudinal'' SSE because $\nabla T$ and $\bfJ_{\rm s}$ are parallel to each other. As a result of converting the spin current into the charge current in the orthogonal direction by ISHE, SSE can be the mechanism of the ``transverse'' thermoelectric conversion. We also note that, instead of ISHE, the inverse Rashba-Edelstein effect at the junction interface of different materials can be used to induce the spin-current-driven transverse thermoelectric conversion \cite{Rojas-Sanchez13,Yagmur16}. 

The energy carrier of SSE is the collective motion of thermally excited localized magnetic moments, {\it i.e.} spin waves or magnons \cite{Uchida10-Nmat,Uchida10-APL,Xiao10,Adachi11,Adachi13,Rezende14,Kikkawa15,Jin15}. Thus, it can operate even if the magnetic layer is an insulator, enabling the thermoelectric conversion using an insulator that is not be possible with other phenomena \cite{Kirihara12,Uchida16}. At present, the thermopower generated by SSE is on the same order as the anomalous Nernst coefficient, which is insufficient for thermoelectric applications. However, physics and material science researches are underway to improve the efficiency of the heat-to-spin current conversion in magnetic materials and spin-to-charge current conversion in metals, as well as to scale up materials from simple bilayer films to multilayers \cite{Ramos15} and bulk composites \cite{Boona16}. 

The spin currents are carried by elementary particles or quasiparticles having spin angular momentum. Historically, conduction-electron spin currents and magnon spin currents have been actively studied in spintronics and spin caloritronics. As mentioned above, the carriers of the spin currents generated by SSE are magnons. In magnetic conductors, the spin currents can be generated also via conduction-electron transport under $\nabla T$, which is called the spin-dependent Seebeck effect (SdSE). This was directly observed by Slachter et al. in 2010 using a non-local method \cite{Slachter10}. As the name suggests, this phenomenon originates in the spin dependence of the Seebeck coefficient in a magnetic conductor. It is also possible to convert the spin current generated by SdSE into a transverse voltage using ISHE \cite{Iguchi18}; SdSE can also be the driving principle for the transverse thermoelectric conversion [Fig. \ref{fig:classification}(f)]. However, it is difficult to distinguish the transverse thermoelectric power generated by this mechanism from the magnon-driven SSE and the interface effect on ANE.

%%%%%%%%%%%%%%%%%%%%%%%
\subsection{Transverse thermoelectric effects with structural symmetry breaking}
%%%%%%%%%%%%%%%%%%%%%%%

In the transverse thermoelectric conversion, there are also phenomena that do not depend on magnetic fields or magnetization. Such phenomena are driven by the anisotropy of the structure and/or electronic transport properties. As mentioned above, various names are used for the transverse thermoelectric conversion phenomena due to structural symmetry breaking. In this article, we classify them into the phenomena due to macroscale and microscale anisotropy structures.

\subsubsection{Off-diagonal Seebeck effect due to macroscale anisotropic structure. }

In an artificial multilayer structure constructed by alternately stacking two conductors and cutting the stack at a certain angle, anisotropic transport properties occurs even if the constituent conductors exhibit isotropic electron/hole transport. When the off-diagonal components of the thermopower tensor become finite due to such anisotropy, the artificially tilted multilayer (ATML) functions as a transverse thermoelectric conversion element \cite{Babin74,Goldsmid11,Zahner98,Kyarad06,Kanno09,Kanno12,Takahashi13,Sakai14,Mu19,Li20,Zhou22,Yue22,Uchida24,Hirai24,Ando25,Lee25} [Fig. \ref{fig:classification}(g)]. Since this transverse thermoelectric conversion originates from the Seebeck coefficient, it is called the off-diagonal Seebeck effect, which is clearly different from the Nernst effects that appear under time-reversal symmetry breaking. As described later, the phenomenological formulation differs in part between the transverse thermoelectric conversion principles in systems with time-reversal symmetry breaking and structural symmetry breaking. 

The transverse thermoelectric conversion based on the off-diagonal Seebeck effect has been studied for a long time \cite{Babin74,Goldsmid11}, and it is possible to design a transverse thermopower and figure of merit by selecting appropriate constituent materials and optimizing their tilt angle and thickness ratio. While the figure of merit for the off-diagonal Seebeck effect in ATML is improved by a high transverse thermopower, high effective electrical conductivity, and low effective thermal conductivity for the hybrid material, the characteristics required of each constituent material differ from those of conventional thermoelectric materials. In addition to the large difference in the Seebeck coefficient between the two materials, the contrast in the electrical and thermal conductivities between them is important, which induces nonuniform charge and heat current distributions [Fig. \ref{fig:classification}(g)]. In other words, even if the difference in the Seebeck coefficient is large, if the electrical and thermal conductivities of the two materials are the same, the transverse thermoelectric conversion does not occur. The transverse thermopower takes its maximum value when the tilt angle is 45$^\circ$, but the other parameters exhibit different behaviors. The smaller the tilt angle, the higher the effective electrical conductivity in the direction of the charge current and the lower the thermal conductivity in the direction of the heat current in ATML. Based on these situations, the maximum value of the figure of merit for the off-diagonal Seebeck effect in ATML is obtained at an angle smaller than 45$^\circ$ \cite{Hirai24,Ando25}. However, if the tilt angle is too small, it is difficult to process and synthesize ATML in practice and its thermoelectric performance deteriorates due to the influence of boundary conditions \cite{Kanno09,Kanno12}. Thus, the contrast in the electrical and thermal conductivities should not be too large. Many studies on ATMLs have been conducted using sintered bulk materials and $z_{xy}T$ values exceeding 0.2 have been achieved around room temperature \cite{Ando25}. The transverse thermoelectric conversion performance of ATML can be calculated analytically, but in many systems, the experimentally obtained $z_{xy}T$ is much smaller than the ideal value due to the interfacial electrical and thermal resistances between the two materials. Despite this situation, in SmCo$_5$/Bi$_{0.2}$Sb$_{1.8}$Te$_3$ (BST) ATML developed by Ando et al., extremely small interfacial electrical and thermal resistances were achieved, and $z_{xy}T$ close to the ideal value was obtained experimentally, highlighting the importance of interface engineering in ATML \cite{Ando25}. The transverse thermoelectric conversion performance of SmCo$_5$/BST ATML has been further improved by utilizing the anisotropic electrical and thermal conductivities of the SmCo$_5$ magnets and hybridizing their ANE; $z_{xy}T$ now reaches 0.3 around room temperature \cite{Lee25}.

\subsubsection{Off-diagonal Seebeck effect due to microscale anisotropic structure. }

Even without constructing ATMLs, the transverse thermoelectric conversion can be achieved in single crystals showing anisotropic electronic conduction properties. In the single crystals of several metals such as Sb, Gd, Ho, Er, and Bi, the Seebeck coefficient is of the same sign but anisotropic \cite{Chandrasekhar59,Sill65,Saunders65}. Such materials having ``unipolar'' anisotropy make it possible to construct a transverse thermoelectric conversion element simply by cutting them so that the off-diagonal components of the thermopower tensor are finite. It is difficult to construct high-output power generation or cooling/heating devices from such single-crystal metals because of their small transverse thermopower caused by the unipolar anisotropy in the Seebeck coefficient. In fact, due to their unipolar nature, the temperature gradient can never be purely orthogonal to the electric field \cite{Shao19}. Nevertheless, the off-diagonal Seebeck effect in single-crystalline metals is being considered for use in heat flux sensors because of their low thermal resistance \cite{McAfee23}. 

It is also possible for single crystals to exhibit ADCP, though early experimental reports \cite{Chung03,Gu05,Ong10,Cohn12} did not recognize their utility as transverse thermoelectrics until it was pointed out \cite{Zhou13}. In recent years, even more single-crystalline materials with ADCP have been discovered \cite{He19,Scudder21,Scudder22,Ochs24,Goto24,Manako24,Ohsumi24}. Unlike the aforementioned unipolar anisotropy, such materials exhibit ``ambipolar anisotropy’’, in which the sign of the Seebeck coefficient reverses depending on the crystal orientation. Therefore, by cutting the material along the direction between the axis showing the large positive Seebeck coefficient and the axis showing the large negative Seebeck coefficient and by applying a temperature gradient to allow the conduction electrons and holes to diffuse in the transverse direction, excellent transverse thermoelectric performance can be obtained through the off-diagonal Seebeck effect [see Fig. \ref{fig:classification}(h)]. For example, it has been reported that $z_{xy}T$ of single-crystalline Re$_4$Si$_7$ reaches 0.7 at 980 K \cite{Scudder21}. This value is outstanding compared to other principles, but there are issues such as the limited choice of materials, the need for large single crystals, and the fact that the large figure of merit is obtained only at high temperatures. 

The transverse thermoelectric conversion arising from ADCP is referred to by various names. When ambipolar anisotropy arises from a single metallic band with an open Fermi surface and alternating positive and negative Gaussian curvature, such a metallic band is called a goniopolar band, and such materials are referred to as goniopolar materials \cite{He19}. When ambipolar anisotropy arises in semiconductors or semimetals whose separate $n$-type and $p$-type bands results in $n$-type conduction dominating one axis and $p$-type conduction dominating orthogonally, such a materials are called ($p \times n$)-type transverse thermoelectrics \cite{Zhou13,Tang15} (see Hierarchy 3 in Fig. \ref{fig:classification}). However, as mentioned at the beginning of Sec. \ref{Sec:Classification}, the details of the band structures cannot be determined from thermoelectric measurements alone, and the terminology of the thermoelectric effects should be determined based on their symmetry and phenomenology. Thus, we have merged the effects mentioned in this subsection into “off-diagonal Seebeck effect due to microscale anisotropic structure” (see Hierarchy 2 in Fig. \ref{fig:classification}).

%%%%%%%%%%%%%%%%%%%%%%%%%%%%%%%%%%%%%%%%%%%%%%%%%%%%%%%%%%%%%%%%%%%%%%%%%%%%%%%%%%%%%%%%%%%
\section{Formulation of transverse thermoelectric conversion with time-reversal symmetry breaking} \label{Sec:Formulation-time}
%%%%%%%%%%%%%%%%%%%%%%%%%%%%%%%%%%%%%%%%%%%%%%%%%%%%%%%%%%%%%%%%%%%%%%%%%%%%%%%%%%%%%%%%%%%

In this section, we present the formulation of the efficiency of both longitudinal and transverse thermoelectric devices. This subject has a long history as seen in many textbooks~\cite{Ioffe57,Heikes-Ure61,Harman-Honig67,Goldsmid10} and publications~\cite{Uchida16,Harman58,Sherman60,Wright62}. However, less known is the fact that an apparent expression for the efficiency varies with the choice of independent variables appearing in the transport equations, as well as with the symmetry of the transport coefficient matrices. For the longitudinal device, choosing the charge current density and the temperature gradient $(\bfj, {\bf \nabla} T)$ as independent variables leads to the conventional expression for the efficiency~\cite{Ioffe57, Heikes-Ure61}. An alternative choice of the electric field and the temperature gradient $(\bfE, {\bf \nabla} T)$ as independent variables yields another expression for the efficiency~\cite{Mahan98}. Of course, these two efficiency expressions are physically the same and equivalent, but expressed by seemingly different parameters. 

On the other hand, in the case of the transverse device, the thermal boundary condition in the transverse direction, or more precisely in the direction of the electric field, inevitably specifies the allowed choice of independent variables. For example, under the adiabatic condition, the charge and heat current densities $(\bfj, \bfq)$ are the appropriate independent variables, while under the isothermal condition the charge current density and temperature gradient $(\bfj, {\bf \nabla}T)$ are the appropriate independent variables. As expected, these two thermal boundary conditions yield two different expressions for the efficiency~\cite{Horst63,Delves64}. In the presence of the Nernst effect but the absence of the Seebeck effect, these two efficiency expressions are the same and equivalent, similar to the case of the longitudinal device. However, in the simultaneous presence of the Nernst and Seebeck effects, these two expressions are no longer the same quantity, representing two physically different situations. Moreover, inclusion of the thermal Hall effect (THE) or the off-diagonal component of the thermal conductivity tensor further modifies the efficiency expression (note that THE is often called the Righi-Leduc effect \cite{Onose08}). 

Below, following the line of argument of Ref.~\cite{Osterle-Angrist63}, we present as simple as possible formulation of the performance of longitudinal and transverse thermoelectric devices. In Sec.~\ref{Sec:3.1}, we present two formulations for the performance of longitudinal thermoelectric generators that are expressed by two different figures of merit. Next, in Sec.~\ref{Sec:3.2}, we discuss the performance formulation for transverse thermoelectric generators, and show how the efficiency expression differs depending on the isothermal or adiabatic boundary condition in the transverse direction. Then, in Sec.~\ref{Sec:3.3}, we discuss the performance of transverse refrigerators and active coolers~\cite{Adams19}. In Sec.~\ref{Sec:Formulation-structural}, we discuss the case for the off-diagonal Seebeck effect~\cite{He19,Scudder22,Tang15}. 

In this section we discuss transverse devices with time-reversal symmetry breaking, while in the next section we discuss transverse devices with structural symmetry breaking. An important distinction between time-reversal symmetry breaking and structural symmetry breaking is that the off-diagonal components of the thermopower, electronic thermal conductivity, and electrical resistivity tensors [see Eqs.~(\ref{eq:alpha_mat01})--(\ref{eq:rho_mat01}) below] are antisymmetric ($S_{xy} = -S_{yx}$, $\kappa_{xy}= - \kappa_{yx}$, $\rho_{xy}= - \rho_{yx}$) for the former and symmetric ($S_{xy} = S_{yx}$, $\kappa_{xy}=\kappa_{yx}$, $\rho_{xy}=\rho_{yx}$) for the latter \cite{Tang16}. It is a useful distinction for clarifying how fundamentally different these phenomena are from a tensor representational standpoint.

%%%%%%%%%%%%%%%%%%%%%%%%%%%%%%%%%%%%% 
%\begin{figure}[tb]
\begin{figure} 
  \begin{center}
    \includegraphics[width=6.2cm]{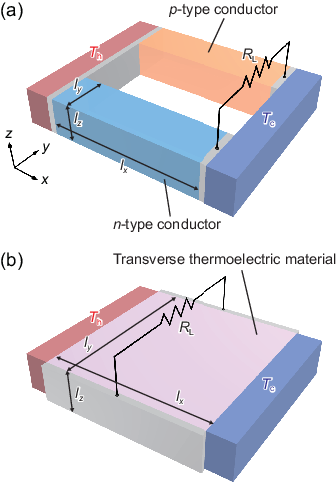}
  \end{center}
  \caption{Schematics of the longitudinal thermoelectric device [(a)] considered in Sec.~\ref{Sec:3.1} and transverse thermoelectric device [(b)] considered in Sec.~\ref{Sec:3.2}. In (a), $p$-type and $n$-type thermoelectric materials are thermally connected in parallel and electrically connected in series, with both cross sections being the same as $A=l_y l_z$ and the device is sandwiched between a hot reservoir $T_{\rm h}$ and a cold reservoir $T_{\rm c}$. In (b), a transverse thermoelectric material is sandwiched between $T_{\rm h}$ and $T_{\rm c}$. $l_x$, $l_y$, and $l_z$ denote the length of the material along the $x$, $y$, and $z$ directions, respectively. 
  }
  \label{fig:TEdev01}
\end{figure}
%%%%%%%%%%%%%%%%%%%%%%%%%%%%%%%%%%%%

%%%%%%%%%%%%%%%%%%%%%%%%%%%%%%%%%%%%
\subsection{{Revising performance of longitudinal thermoelectric generator } \label{Sec:3.1}}
%%%%%%%%%%%%%%%%%%%%%%%%%%%%%%%%%%%%

Before formulating the efficiency of a transverse thermoelectric generator, we revisit the efficiency of a longitudinal thermoelectric generator. The system is shown in Fig.~\ref{fig:TEdev01}(a), where $p$-type and $n$-type thermoelectric materials are thermally connected in parallel and electrically connected in series. This device operates under a temperature bias between a hot reservoir $T_{\rm h}$ and a cold reservoir $T_{\rm c}$.

First, we present the conventional formulation of the efficiency~\cite{Ioffe57,Heikes-Ure61} using the charge current density and the temperature gradient $(\bfj, {\bf \nabla} T)$ as independent variables. Next, we present an alternative formulation of the efficiency using the electric field and the temperature gradient $(\bfE, {\bf \nabla} T)$ as independent variables~\cite{Mahan98}.

%%%%%%%%%%%%%%%%%%%%%%%%%%%%%%%%
\subsubsection{Conventional expression. } 
%%%%%%%%%%%%%%%%%%%%%%%%%%%%%%%%
We begin with the following transport equations for the heat current density $\bfq$ and electric field $\bfE$~\cite{Landau-electrodyn}: 
%%%
\begin{eqnarray}
  \begin{pmatrix} q_x \\ q_y \end{pmatrix} &=& T {S}_\nu
  \begin{pmatrix} j_x \\ j_y \end{pmatrix}
  - {\kappa}_\nu 
  \begin{pmatrix} \nabla_x T \\ \nabla_y T \end{pmatrix},  \qquad \label{eq:q_conv01} \\
  \begin{pmatrix} E_x \\ E_y \end{pmatrix} &=& 
  {\rho}_\nu \begin{pmatrix} j_x \\ j_y \end{pmatrix} 
  + {S}_\nu 
  \begin{pmatrix} \nabla_x T \\ \nabla_y T \end{pmatrix} , \label{eq:E_conv01}
\end{eqnarray}
%%% 
where $S_\nu$, $\kappa_\nu$, and $\rho_\nu$ are respectively the Seebeck coefficient, thermal conductivity, and electrical resistivity of each thermoelectric element with the carrier index $\nu= {p, n}$. Although we assume an isotropic material in the above transport equations, this is merely for the sake of simplicity. In the case of an anisotropic material, we can start from the transport equations represented with tensor coefficients (see, {\it e.g.} Eq.~(26.12) of Ref.~\cite{Landau-electrodyn}). Then, after multiplying the $x$ component of Eq.~(\ref{eq:q_conv01}) with the cross section of each element $A= l_y l_z$, as well as integrating the $x$ component of Eq.~(\ref{eq:E_conv01}) along the $x$ axis, we obtain 
%%%
\begin{eqnarray}
  Q_x &=& T S J_x + K \Delta T, \label{eq:q_conv02} \\ 
  \Delta V &=&  - R J_x + S \Delta T, \label{eq:E_conv02}
\end{eqnarray}
%%%
where $Q_x= A q_x $, $\Delta V = - E_x l_x$, $J_x= A j_x$, $\Delta T = T_{\rm h}- T_{\rm c}$, $S= S_{p} - S_{n}$, $R = (\rho_{p} + \rho_{n}) l_x/A$, and $K= (\kappa_{p} + \kappa_{n}) A/l_x$. The above linear-response relation can be expressed by using a matrix as 
%%%
\begin{equation}
  \begin{pmatrix} Q_x \\ \Delta V \end{pmatrix}
  = \widehat{\cal A} \begin{pmatrix} J_x \\ \Delta T \end{pmatrix}, \label{eq:linear-res01}
\end{equation}
%%%
where the matrix $\widehat{\cal A}$ is defined by 
%%%
\[
      \widehat{\cal A}= \begin{pmatrix} T S & K \\ -R & S \end{pmatrix}.
\]
%%%
The determinant of $\widehat{\cal A}$ is given by $\det \widehat{\cal A} = \kappa \rho (1+ z T)$, where $\kappa= \kappa_{p}+ \kappa_{n}$, $\rho= \rho_{p}+ \rho_{n}$, and the quantity 
%%%
\begin{equation}
  z T = \frac{T S^2 }{\kappa \rho} 
  \label{eq:zT_conv01} 
\end{equation}
%%%
is the figure of merit of the longitudinal device. Note that, because of the conditions $R >0 $ and $K>0$ which are set by the second law of thermodynamics~\cite{Landau-electrodyn}, the sign of $\det \widehat{\cal A}$ is always positive. This means that the linear-response relation [Eq.~(\ref{eq:linear-res01})] is physically stable, such that there is no constraint on the value of $z T$ except that it should have positive sign. Therefore, the figure of merit for the longitudinal device is unbounded, namely, 
%%%
\begin{equation}
  0 \le z T < \infty.
  \label{eq:FOM_zT_bound01}
\end{equation}
%%%

Since the linear-response relations have been set up, we are now ready to calculate the efficiency of the Seebeck generator [Fig.~\ref{fig:TEdev01}(a)]. In the following, in line with the argument of Ref.~\cite{Osterle-Angrist63}, we consider only a term linear in $\Delta T$. This means that the Joule heating $R J_x^2$, which otherwise enters the efficiency evaluation through the thermal power input, can be safely neglected in our calculation. The Carnot efficiency is given by
%%%
\begin{equation}
  \eta_{\rm C}= \frac{\Delta T}{T}
\end{equation}
%%%
where $T= (T_{\rm h}+ T_{\rm c})/2$, and the reduced efficiency $\overline{\eta}= \eta/\eta_{\rm C}$ relative to the Carnot efficiency is defined by 
%%%
\begin{equation}
  \overline{\eta} = \frac{T}{\Delta T}\frac{{\cal P}_{\rm out}}{{\cal Q}_{\rm h}}, \label{eq:eta_def01}
\end{equation}
%%%
where ${\cal Q}_{\rm h}$ is the thermal power input from the hot reservoir and ${\cal P}_{\rm out}= R_{\rm L} J_x^2$ is the electrical power output dissipated by the load resistance $R_{\rm L}$. 

In order to evaluate the right-hand side of Eq.~(\ref{eq:eta_def01}), we need to know the charge current $J_x$ determined by
%%%
\begin{equation}
  J_x= \frac{V_{\rm emf}}{R_{\rm tot}}, \label{eq:Jx01}
\end{equation}
%%%
where $V_{\rm emf}$ is the electromotive force and $R_{\rm tot}= R + R_{\rm L}$ is the total resistance of the circuit. Using $V_{\rm emf}= S \Delta T$ and ${\cal Q}_{\rm h}= Q_x$, the efficiency can be expressed as a function of $u= R_{\rm L}/R $ as $\overline{\eta} = {u} \{ {{(1+u)^2}/{z T}+ 1+u} \}^{-1} $, where $z T$ is defined in Eq.~(\ref{eq:zT_conv01}). The maximum reduced efficiency $\overline{\eta}$ is obtained for the stationary value $u^*= \sqrt{1+zT}$, which yields 
%%%
\begin{equation}
  \overline{\eta} = \frac{\sqrt{1+ zT}-1}{\sqrt{1+zT}+1}.
  \label{eq:eta_conv01}
\end{equation}
%%%

%%%%%%%%%%%%%%%%%%%%%%%%%%%%%
\subsubsection{Alternative expression. }
%%%%%%%%%%%%%%%%%%%%%%%%%%%%
In the preceding subsection, we have formulated the efficiency of the longitudinal thermoelectric device by choosing $(\bfj, {\bf \nabla} T)$ as independent variables. Below, we discuss an alternative formulation of the efficiency by choosing $(\bfE, {\bf \nabla} T)$ as independent variables.  

We first invert Eqs.~(\ref{eq:q_conv02}) and (\ref{eq:E_conv02}) as follows: 
%%%
\begin{eqnarray}
  Q_x &=& - \frac{T S }{R} \Delta V + K' \Delta T, \label{eq:q_alt01} \\ 
  J_x &=&  - \frac{1}{R}\Delta V + \frac{S}{R} \Delta T, \label{eq:E_alt01}
\end{eqnarray}
%%%
where $K' = K + T S^2 /R$. This defines the thermal conductivity at zero electric field, 
%%%
\begin{equation}
  \kappa' = \kappa+ \frac{T S^2 }{\rho}. 
  \label{eq:k_alt01}
\end{equation}
%%%
The above linear-response relation can be represented as 
%%%
\begin{equation}
  \begin{pmatrix} Q_x \\ J_x \end{pmatrix}
  = \widehat{\cal B} \begin{pmatrix} \Delta V \\ \Delta T \end{pmatrix}, \label{eq:linear-res02}
\end{equation}
%%%
where the matrix $\widehat{\cal B}$ is defined by
%%%
\[ 
  \widehat{\cal B}= \begin{pmatrix} -T S /R & K' \\ -1/R  & S/R \end{pmatrix}.
\]
%%%
The determinant of $\widehat{\cal B}$ is calculated to be $\det \widehat{\cal B} = ({R}/{K'}) (1- z' T) $, where
%%%
\begin{equation}
  z'T = \frac{T S^2 }{\kappa' \rho}
  \label{eq:zT_alt01}
\end{equation}
%%%
is a new figure of merit defined by using $\kappa'$~\cite{Mahan98}. Note that, for Eq.~(\ref{eq:linear-res02}) to be physically stable, $\det \widehat{\cal B}$ should be positive, from which we obtain a condition $1- z' T >0$. This means that the figure of merit is bounded as 
%%%
\begin{equation}
  0 \le z' T < 1.
  \label{eq:FOM_zT'_bound01}
\end{equation}
%%%

We next calculate the reduced efficiency for the present choice of independent variables. Using ${\cal Q}_{\rm h}= Q_x$, $V_{\rm emf}= S \Delta T$, and $J_x= V_{\rm emf}/R_{\rm tot}$ with $R_{\rm tot}= R_{\rm L} + R $, the reduced efficiency [Eq.~(\ref{eq:eta_def01})] is calculated to be $\overline{\eta} = {u} \{ {(1+u)^2}/{z'T}- (1+u) \}^{-1} $, where $u=R_{\rm L}/R$. The maximum reduced efficiency $\overline{\eta}$ is obtained for the stationary value $u^*= \sqrt{1-z'T}$, which yields 
%%%
\begin{equation}
  \overline{\eta} = \frac{1- \sqrt{1- z'T}}{1+ \sqrt{1- z'T}}.
  \label{eq:eta_alt01}
\end{equation}
%%%

%%%%%%%%%%%%%%%%%%%%%%%%%%%%%%%%%%%%%%%%%%%%%%%%%
\subsubsection{Summary of longitudinal device. }
%%%%%%%%%%%%%%%%%%%%%%%%%%%%%%%%%%%%%%%%%%%%%%%%%

To summarize this subsection, we have formulated the efficiency of a longitudinal thermoelectric generator based on the Seebeck effect in two different ways. In the former we have calculated the maximum efficiency [Eq.~(\ref{eq:eta_conv01})] by choosing $(\bfj, {\bf \nabla} T)$ as independent variables, whereas in the latter we have calculated the maximum efficiency Eq.~(\ref{eq:eta_alt01}) by choosing $(\bfE, {\bf \nabla} T)$ as independent variables~\cite{Mahan98}. We note that, in deriving the two different efficiency expressions, we assume no physical difference in the condition of the thermoelectric device; the difference in the efficiency expressions stems from the difference in the choice of the independent variables. In other words, the two seemingly different efficiency expressions for the longitudinal device correspond to two different representations of the same quantity in two different parameters $zT$ and $z'T$. Indeed, from Eqs.~(\ref{eq:zT_conv01}), (\ref{eq:k_alt01}), and (\ref{eq:zT_alt01}), we find that $zT$ and $z'T$ are related by 
%%%
\begin{equation}
  z'T = \frac{zT}{1+ zT}.
  \label{eq:zT_conv-zT_alt01}
\end{equation}
%%%
Then, substituting Eq.~(\ref{eq:zT_conv-zT_alt01}) into the efficiency expression of Eq.~(\ref{eq:eta_alt01}), we come to the efficiency expression of Eq.~(\ref{eq:eta_conv01}). In this sense, we note that these two expressions are equivalent, and achieving $z'T=1$ limit is as difficult as achieving $zT= \infty$ limit.

%%%%%%%%%%%%%%%%%%%%%%%%%%%%%%%%%%%%%%%%%%%%%
\subsection{Performance of transverse thermoelectric generator \label{Sec:3.2} }
%%%%%%%%%%%%%%%%%%%%%%%%%%%%%%%%%%%%%%%%%%%%%
In this subsection, we formulate the efficiency of a transverse thermoelectric generator as shown in Fig.~\ref{fig:TEdev01}(b), where a transverse thermoelectric material showing ONE/ANE is sandwiched between $T_{\rm h}$ and $T_{\rm c}$. Below, it will be shown that a thermal boundary condition in the direction of the electric field [along the $y$ axis in Fig.~\ref{fig:TEdev01}(b)] plays a crucial role in the formulation. 

In the literature~\cite{Uchida16,Harman-Honig67,Wright62,Horst63,Delves64,Osterle-Angrist63}, the efficiency of the transverse device has been discussed by assuming that only the Nernst/Ettingshausen effects are present in the device, while the Seebeck/Peltier effects are absent. In this subsection, we discuss a more realistic situation where both the Nernst/Ettingshausen and Seebeck/Peltier effects are present. Below, we show that the efficiency expression is substantially modified by the simultaneous presence of the Nernst/Ettingshausen and Seebeck/Peltier effects. Moreover, we discuss the influence of THE, which also affects the efficiency expression. 

We use the following transport equations~\cite{Landau-electrodyn}:
%%%
\begin{eqnarray}
  \begin{pmatrix} q_x \\ q_y \end{pmatrix} &=& T \widehat{S} 
  \begin{pmatrix} j_x \\ j_y \end{pmatrix}
  - \widehat{\kappa}
  \begin{pmatrix} \nabla_x T \\ \nabla_y T \end{pmatrix},  \qquad \label{eq:qT_adia01} \\ 
  \begin{pmatrix} E_x \\ E_y \end{pmatrix} &=&
  \widehat{\rho} \begin{pmatrix} j_x \\ j_y \end{pmatrix} 
  + \widehat{S} 
  \begin{pmatrix} \nabla_x T \\ \nabla_y T \end{pmatrix} , \label{eq:ET_adia01}
\end{eqnarray}
%%%
where the transport coefficient matrices are given by 
%%%
\begin{equation}
  \widehat{S} =
  \begin{pmatrix} S_{xx} & S_{xy} \\ S_{yx} & S_{yy} \end{pmatrix},
  \label{eq:alpha_mat01}
\end{equation}
\begin{equation}
  \widehat{\kappa} =
    \begin{pmatrix} \kappa_{xx} & \kappa_{xy} \\ \kappa_{yx} & \kappa_{yy} \end{pmatrix},
    \label{eq:kappa_mat01}
\end{equation}
\begin{equation}
  \widehat{\rho} =
  \begin{pmatrix} \rho_{xx} & \rho_{xy} \\ \rho_{yx} & \rho_{yy} \end{pmatrix}. 
    \label{eq:rho_mat01}
\end{equation}
%%%
Note that the resistivity in the above equation is measured under a zero temperature gradient, which coincides with the conventional definition of electrical resistance~\cite{Callen-textbook}. Under the open-circuit condition along the $x$ axis ($j_x= 0$), equations necessary for the efficiency calculation are given by 
%%%
\begin{eqnarray}
  q_x &=& -\kappa_{xx} \nabla_x T - \kappa_{xy} \nabla_y T + T S_{xy}  j_y, \label{eq:qx_T_adia02} \\ 
  q_y &=& T S_{yy} j_y - \kappa_{yy} \nabla_y T - \kappa_{yx} \nabla_x T, \label{eq:qy_T_adia02} \\ 
  E_y &=&  \rho_{yy} j_y + S_{yy} \nabla_y T+ S_{yx} \nabla_x T. \label{eq:Ey_T_adia02}  
\end{eqnarray}
%%%

%%%%%%%%%%%%%%%%%%%%%%%%%%%%%%%%%%
\subsubsection{Adiabatic condition in transverse direction. } 
%%%%%%%%%%%%%%%%%%%%%%%%%%%%%%%%%%

We first discuss the adiabatic boundary condition in the direction of the electric field [along the $y$ axis in Fig.~\ref{fig:TEdev01}(b)], since this is the ``natural'' case for the transverse device according to Ref.~\cite{Osterle-Angrist63}. In this subsections, we are concerned with the case where the transport coefficient matrices in Eqs.~(\ref{eq:alpha_mat01})--(\ref{eq:rho_mat01}) have antisymmetric off-diagonal components, {\it i.e.} $S_{yx}= - S_{xy}$, $\kappa_{yx}= - \kappa_{xy}$, and $\rho_{yx}= - \rho_{xy}$. This symmetry applies to ONE and ANE devices. 

Now, using the adiabatic condition along the $y$ axis ($q_y=0$), the above equations can be solved for $\nabla_x T$ and $E_y$, yielding 
%%%
\begin{eqnarray}
  \nabla_x T &=& \frac{T {S}^{\rm (a)}_{xy} }{{\kappa}^{\rm (a)}_{xx}} j_y
  - \frac{1}{{\kappa}^{\rm (a)}_{xx}} q_x, \label{eq:gradTx_T_adia03} \\
  E_y &=&
  \rho_{yy}^{\rm (a)} j_y
  - \frac{{S}^{\rm (a)}_{yx}}{{\kappa}^{\rm (a)}_{xx}} q_x. \label{eq:Ey_T_adia03} 
\end{eqnarray}
%%%
In these equations, we defined the following new coefficients renormalized by the Seebeck effect and THE: 
%%%
\begin{eqnarray*}
  {S}^{\rm (a)}_{xy} &=& S_{xy}- S_{yy} \Theta_x , \\
  {S}^{\rm (a)}_{yx} &=&  S_{yx}- S_{yy} \Theta_y ,  \\ 
  {\kappa}^{\rm (a)}_{xx} &=& \kappa_{xx} (1- \Theta_x \Theta_y r_{yx}), \\
  {\kappa}^{\rm (a)}_{yy} &=& \kappa_{yy} (1- \Theta_{x}\Theta_{y} r_{yx}), \\
  {S}^{\rm (a)}_{yy} &=& S_{yy} \sqrt{1- \Theta_x \Theta_y r_{yx}}, 
\end{eqnarray*}
%%%
where $\Theta_x= \kappa_{xy}/\kappa_{yy}$, $\Theta_y= \kappa_{yx}/\kappa_{yy}$, and $r_{yx}= \kappa_{yy}/\kappa_{xx}$. Note that for an isotropic material ($\kappa_{xx}= \kappa_{yy}$), we have $\Theta \equiv \Theta_x = - \Theta_y$ due to the symmetry $\kappa_{yx}= -\kappa_{xy}$. From this and $S_{yx}= -S_{xy}$, we obtain 
%%%
\begin{eqnarray}
  {S}^{\rm (a)}_{xy} &=& - {S}^{\rm (a)}_{yx} = S_{xy}- S_{yy} \Theta, \label{eq:thermopower-iso-vs-ad} \\
  {\kappa}^{\rm (a)}_{xx} &=& {\kappa}^{\rm (a)}_{yy} = \kappa_{xx} (1+ \Theta^2), \\
  {S}^{\rm (a)}_{yy} &=& S_{yy} \sqrt{1+ \Theta^2}. 
\end{eqnarray}
%%%
Note also that, due to the variable change from $\nabla_x T$ to $q_x$, the adiabatic resistivity 
%%%
\begin{eqnarray}
  {\rho}_{yy}^{\rm (a)} &=& \rho_{yy} +
  \frac{T {S}_{yy}^{\rm (a) \, 2} }{{\kappa}^{\rm (a)}_{yy}}
  + \frac{T {S}^{\rm (a)}_{xy} {S}^{\rm (a)}_{yx}}{{\kappa}^{\rm (a)}_{xx}} \nonumber \\
  &=&
  \rho_{yy} +
  \frac{T {S}_{yy}^{\rm (a) \, 2} }{{\kappa}^{\rm (a)}_{yy}}
  - 
  \frac{T {S}_{yx}^{\rm (a) \, 2} }{{\kappa}^{\rm (a)}_{xx}}   
\label{eq:rho_T_adia01}
\end{eqnarray}
%%%
appears in Eq.~(\ref{eq:Ey_T_adia03}), where we used the symmetry $S_{yx}= - S_{xy}$ and $\kappa_{yx}= - \kappa_{xy}$ to move to the second line. Now, multiplying Eq.~(\ref{eq:gradTx_T_adia03}) with $A= l_y l_z$ and integrating Eq.~(\ref{eq:Ey_T_adia03}) along the $y$ axis, we obtain 
%%%
\begin{eqnarray}
  \Delta T &=& - \frac{{S}^{\rm (a)}_{xy}T }{{K}^{\rm (a)} (l_x/l_y)} J_y + \frac{1}{{K}^{\rm (a)}} Q_x , \label{eq:DeltT01}\\
  \Delta V &=& - R^{\rm (a)} J_y + \frac{{S}^{\rm (a)}_{yx}}{{K}^{\rm (a)}(l_x/l_y)} Q_x ,
  \label{eq:DeltV01}
\end{eqnarray}
%%%
where ${K}^{\rm (a)} = {\kappa}^{\rm (a)}_{xx} A/l_x $, ${R}^{\rm (a)}= {\rho}_{yy}^{\rm (a)} l_y/(l_x l_z)$.

The linear-response relation described above can be expressed as 
%%%
\begin{equation}
  \begin{pmatrix} \Delta T \\ \Delta V \end{pmatrix}
  = \widehat{\cal C} \begin{pmatrix} J_y \\ Q_x \end{pmatrix}, \label{eq:linear-T_adia01}
\end{equation}
%%%
where the matrix $\widehat{\cal C}$ is defined by
%%%
\begin{equation}
  \widehat{\cal C}=
  \begin{pmatrix} \frac{ T S^{\rm (a)}_{yx} }{{K}^{\rm (a)}(l_x/l_y)} & \frac{1}{{K}^{\rm (a)}} \\ - {R}^{\rm (a)} & \frac{{S}^{\rm (a)}_{yx}}{{K}^{\rm (a)}(l_x/l_y)} \end{pmatrix}.
  \label{eq:matC01}
\end{equation}
%%%
Note that in moving from Eqs.~(\ref{eq:DeltT01}) and (\ref{eq:DeltV01}) to Eqs.~(\ref{eq:linear-T_adia01}) and (\ref{eq:matC01}), we used the symmetry $S_{yx}= -S_{xy}$ and $\kappa_{yx}= -\kappa_{xy}$. The determinant of $\widehat{\cal C}$ is calculated as $\det \widehat{\cal C} = ({{R}^{\rm (a)}}/{{K}^{\rm (a)}}) (1+ z_{xy}^{\rm (a)} T) $, where the quantity 
%%%
\begin{equation}
  z_{xy}^{\rm (a)} T = 
 \frac{T {S}_{yx}^{\rm (a) \, 2}}{{\kappa}^{\rm (a)}_{xx} {\rho}_{yy}^{\rm (a)}} 
  \label{eq:ZT_Tadia01}
\end{equation}
%%%
is the adiabatic figure of merit for the transverse device. Note that $\det \widehat{\cal C}$ is positive for positive values of $\rho^{\rm (a)}$. This means that, under the condition $\rho^{\rm (a)}>0$, the linear-response relation [Eq.~(\ref{eq:linear-T_adia01})] is physically stable, such that there is no constraint on $z_{xy}^{\rm (a)} T$ except that it should have positive sign. Therefore, $z_{xy}^{\rm (a)} T$ is unbounded, namely,
%%%
\begin{equation}
  0 \le z_{xy}^{\rm (a)} T < \infty.
  \label{eq:FOM_T_adia_bound01}
\end{equation}
%%%

The remaining steps to calculate the reduced efficiency [Eq.~(\ref{eq:eta_def01})] of the transverse device in the adiabatic limit are basically the same as in the previous section. The thermal power input from the hot reservoir is given by ${\cal Q}_{\rm h}= Q_x$ and the current flowing through the circuit is given by $J_y= V_{\rm emf}/R_{\rm tot}$, where $R_{\rm tot}= R_{\rm L}+ {R}^{\rm (a)}$ with $R_{\rm L}$ being the load resistance. Using $V_{\rm emf}= (- {S}^{\rm (a)}_{xy}(l_y/l_x)/ {K}^{\rm (a)}) {\cal Q}_{\rm h}$, the reduced efficiency is calculated as $\overline{\eta} = {u} \{ {(1+u)^2}/{z_{xy}^{\rm (a)} T}+ (1+u) \}^{-1} $, where $u= R_{\rm L}/R_{\rm tot}$. Then, the maximum reduced efficiency $\overline{\eta}$ is obtained for the stationary value $u^*= \sqrt{1+ z_{xy}^{\rm (a)} T}$, which yields 
%%%
\begin{equation}
  \overline{\eta}= \frac{\sqrt{1+ z_{xy}^{\rm (a)} T}-1}{\sqrt{1+ z_{xy}^{\rm (a)}T}+1}.
  \label{eq:eta_T_adia01} 
\end{equation}
%%%

%%%%%%%%%%%%%%%%%%%%%%%%%%%%
\subsubsection{Isothermal condition in transverse direction. }
%%%%%%%%%%%%%%%%%%%%%%%%%%%%
Next, we discuss the isothermal boundary condition in the direction of the electric field. As in the previous subsection, we are concerned with the case where the transport coefficient matrices in Eqs.~(\ref{eq:alpha_mat01})--(\ref{eq:rho_mat01}) have antisymmetric off-diagonal components, {\it i.e.} $S_{yx}= - S_{xy}$, $\kappa_{yx}= - \kappa_{xy}$, and $\rho_{yx}= - \rho_{xy}$, whose symmetry applies to the ONE/ANE devices. Assuming the vanishing temperature gradient along the $y$ axis in Eqs.~(\ref{eq:qx_T_adia02}), (\ref{eq:qy_T_adia02}), and (\ref{eq:Ey_T_adia02}), we obtain 
%%%
\begin{eqnarray}
  q_x &=& -\kappa_{xx} \nabla_x T + T S_{xy}  j_y, \label{eq:qx_T_iso01} \\ 
  E_y &=&  \rho_{yy} j_y + S_{yx} \nabla_x T. \label{eq:Ey_T_iso01} 
\end{eqnarray}
%%%
Note that, in contrast to the adiabatic case, there is no contribution from THE. Now, multiplying Eq.~(\ref{eq:qx_T_iso01}) with $A= l_y l_z$ and integrating Eq.~(\ref{eq:Ey_T_iso01}) along the $y$ axis, we obtain 
%%%
\begin{eqnarray}
  Q_x &=& T S_{xy}  (l_y/l_x) J_y + K \Delta T, \label{eq:qx_T_iso02} \\
  \Delta V &=& -R J_y + S_{yx} (l_y/l_x) \Delta T, \label{eq:DV_T_iso02} 
\end{eqnarray}
%%%
where ${K} = {\kappa}_{xx} A/l_x $ and ${R}= {\rho}_{yy} l_y/(l_x l_z)$. The above linear-response relation can be represented as 
%%%
\begin{eqnarray}
  \begin{pmatrix} Q_x \\ \Delta V \end{pmatrix}
  &=& \widehat{\cal D} \begin{pmatrix} J_y \\ \Delta T \end{pmatrix}, \label{eq:linear-T_iso01}
\end{eqnarray}
%%%
where the matrix $\widehat{\cal D}$ is defined by
%%%
\begin{equation}
  \widehat{\cal D}= 
\begin{pmatrix} -T S_{yx} (l_y/l_x) & K \\ -R  &  S_{yx} (l_y/l_x) \end{pmatrix}.
\label{eq:matD01}
\end{equation}
%%%
Note that in moving from Eqs.~(\ref{eq:qx_T_iso02}) and (\ref{eq:DV_T_iso02}) to Eqs.~(\ref{eq:linear-T_iso01}) and (\ref{eq:matD01}), we used the symmetry $S_{yx}= - S_{xy}$. The determinant of $\widehat{\cal D}$ is calculated to be  
%%%
\[ 
  \det \widehat{\cal D} = K R (1- z_{xy}^{\rm (i)} T), 
\]
%%%
where
%%%
\begin{equation}
  z_{xy}^{\rm (i)}T = 
  \frac{T S_{yx}^2 }{\rho_{yy} \kappa_{xx}} 
\label{eq:ZT_T_iso01}
\end{equation}
%%%
is the isothermal figure of merit for the transverse device. Note that, for Eq.~(\ref{eq:linear-T_iso01}) to be physically stable, $\det \widehat{\cal D}$ should be positive. Then, we have a condition $1- z_{xy}^{\rm (i)}  T >0$, which means that $z_{xy}^{\rm (i)} T$ is bounded, namely, 
%%%
\begin{equation}
  0 \le z_{xy}^{\rm (i)}  T < 1.
  \label{eq:FOM_T_iso_bound01}  
\end{equation}
%%%

Finally, we calculate the reduced efficiency $\overline{\eta}$ [Eq.~(\ref{eq:eta_def01})]. Using $V_{\rm emf}= S_{xy} (l_y/l_x) \Delta T$ and $R_{\rm tot}= R+ R_{\rm L}$ in Eq.~(\ref{eq:Jx01}), $\overline{\eta}$ can be calculated as $\overline{\eta} = {u} \{(1+u)^2/{z_{xy}^{\rm (i)} T}-(1+u) \}^{-1} $, 
%%%
where $u= R_{\rm L}/R$. The maximum efficiency is obtained for the stationary value $u^* = \sqrt{1- z_{xy}^{\rm (i)} T}$, which yields 
%%%
\begin{equation}
  \overline{\eta}= \frac{1- \sqrt{1- z_{xy}^{\rm (i)} T}}{1+ \sqrt{1- z_{xy}^{\rm (i)} T}}.
  \label{eq:eta_T_iso01}
\end{equation}
%%%

%%%%%%%%%%%%%%%%%%%%%%%%%%%%%%%%%%%%%%%%%%%%%%%
\subsubsection{Figure of merit defined by Delves.  \label{sec:3.2.3}}
%%%%%%%%%%%%%%%%%%%%%%%%%%%%%%%%%%%%%%%%%%%%%%
In this part, we comment on the figure of merit defined by Delves~\cite{Delves64}. Here, Delves referred to those as ``A authors'' who start from our Eqs.~(\ref{eq:qx_T_iso01}) and (\ref{eq:Ey_T_iso01}), and denoted the corresponding figure of merit as ${z_E}^* T$ with asterisk. When we compare our Eqs.~(\ref{eq:qx_T_iso01}) and (\ref{eq:Ey_T_iso01}) with Eqs.~(4) and (5) of Ref.~\cite{Delves64}, we find that Delves's $\kappa^*_{11}$ corresponds to our $\kappa_{xx}$ in Eq.~(\ref{eq:qx_T_iso01}). Note that $\kappa^*_{11}$ coincides with the usual definition of the thermal conductivity under zero charge current. 

Delves~\cite{Delves64} also referred to those as ``B authors'' who start by inverting Eqs.~(\ref{eq:qx_T_adia02}) and (\ref{eq:Ey_T_adia02}) in the following form: 
%%%
\[ 
   q_x = -\kappa''_{xx} \nabla_x T - \frac{T S_{xy}}{\rho} E_y,
\]
\[ 
      j_y =  \frac{1}{\rho_{yy}} E_y - \frac{S_{yx}}{\rho_{yy}} \nabla_x T,   
\]
%%%
where
%%%
\begin{equation}
  \kappa''_{xx} = 
    \kappa_{xx} -  
    \frac{T S_{yx}^2}{\rho_{yy}}
  \label{eq:kappa'xx01}
\end{equation}
%%%
is the thermal conductivity under zero electric field, which is denoted by Delves as $\kappa_{11}$ without asterisk~\cite{Delves64}. Note that THE is discarded here, and the symmetry $S_{yx}= - S_{xy}$ is used. Multiplying these two equations with their cross sections $l_y l_z$ and $l_x l_z$, we obtain 
%%%
\begin{eqnarray}
  Q_x &=& -\frac{ T S_{xy} (l_y/l_x)}{R} \Delta V + K'' \Delta T, \label{eq:Qx_Dlv01} \\
  J_y &=& -\frac{1}{R} \Delta V + \frac{S_{yx} (l_y/l_x)}{R} \Delta T, \label{eq:Jy_Dlv01} 
\end{eqnarray}
%%%
where $K''= \kappa''_{xx} l_y l_z/l_x$ and $R= \rho_{yy} l_x l_z /l_y$. Then, using $V_{\rm emf}= -S_{xy} (l_y/l_x) \Delta T$ and $J_y= V_{\rm emf}/R_{\rm tot}$, where $R_{\rm tot}= R_{\rm L}+ {R}^{\rm (a)}$, the reduced efficiency [Eq.~(\ref{eq:eta_def01})] is calculated to be $\overline{\eta} = {u}\{ {(1+u)^2}/{z'_{xy} T}+ 1+u \}^{-1} $, where $u= R/R_{\rm L}$, and the corresponding figure of merit is given by 
%%%
\begin{equation}
  z''_{xy} T = 
    \frac{S^2_{yx}}{\rho_{yy} \kappa''_{xx}}, 
  \label{eq:z''_xyT01}
\end{equation}
%%%
and from Eqs.~(\ref{eq:ZT_T_iso01}) and (\ref{eq:kappa'xx01}), this quantity satisfies 
%%%
\begin{equation}
  z''_{xy} T = \frac{z_{xy}^{\rm (i)} T }{1-  z_{xy}^{\rm (i)} T}. 
  \label{eq:zT_Dlv01}  
\end{equation}
%%%
Note that $z''_{xy} T$ is denoted by Delves as $z_E T$~\cite{Delves64}. The maximum efficiency is obtained for $s^*= \sqrt{1+ z''_{xy} T}$, which yields 
%%%
\begin{equation}
  \overline{\eta} = \frac{\sqrt{1+ z''_{xy} T}-1}{\sqrt{1+ z''_{xy} T}+1}.
\end{equation}
%%%

From the above argument, we find the connection among ${z_E}^* T$ and $z_E T$ defined by Delves~\cite{Delves64}, and $z_{xy}^{\rm (i)} T$ [Eq.~(\ref{eq:ZT_T_iso01})] and $z''_{xy} T$ [Eq.~(\ref{eq:z''_xyT01})] defined in this work. In short, the relations are given by 
%%%
\begin{eqnarray}
  {z_E}^* T &=& z_{xy}^{\rm (i)}T, \\
  z_E T &=& z''_{xy} T.
\end{eqnarray}
%%%
To check this correspondence, we substitute the above relations into Eq.~(\ref{eq:zT_Dlv01}), then we find 
%%%
\begin{equation}
  {z_E}T = \frac{{z_E}^* T }{1- {z_E}^* T}, \quad 
  {z_E}^*T = \frac{{z_E} T }{1+ {z_E} T}, 
  \label{eq:zT_Dlv02}  
\end{equation}
%%%
which is exactly the same as Eq.~(4) [the second one among two Eq.~(4)'s] of Ref.~\cite{Delves64}. Therefore, ${z_E}^* T$ defined by Delves~\cite{Delves64} coincides with the isothermal figure of merit. Moreover, in the limit of neglecting the Seebeck effect and THE, $z_E T$ coincides with the adiabatic figure of merit, $z^{\rm (a)}_{xy}$ [Eq.~(\ref{eq:ZT_Tadia01})]. In this sense, two formulations of the figure of merit in Refs. \cite{Horst63} and \cite{Delves64} are the same and equivalent. Therefore there is no conflict between them, and the corresponding description about their conflict~\cite{Uchida-Heremans22,Scudder22} should be corrected accordingly.

%%%%%%%%%%%%%%%%%%%%%%%%%%%%%%%%%%%%%%%%%%%%%%%%
\subsubsection{Summary of transverse device. }
%%%%%%%%%%%%%%%%%%%%%%%%%%%%%%%%%%%%%%%%%%%%%%%

%%%%%%%%%%%%%%%%%%%%%%%%%%%%%%%%%%%%% 
%\begin{figure}[tb]
\begin{figure}
  \begin{center}
    \includegraphics[width=8.0cm]{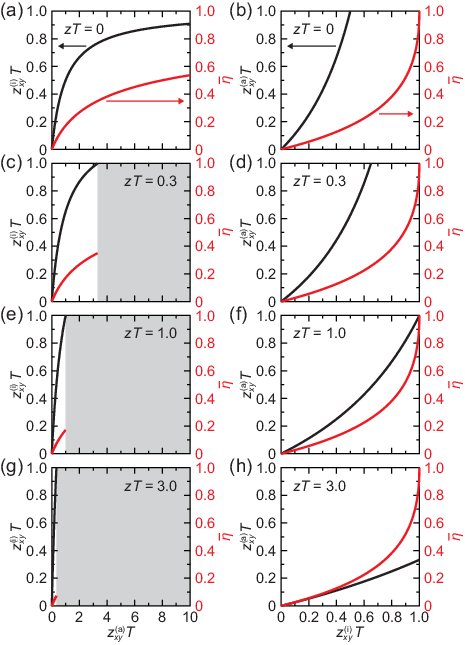}
  \end{center}
  \caption{Relations between $z_{xy}^{\rm (i)}$, $z_{xy}^{\rm (a)}$, and $\overline{\eta}$ for various values of $zT$. $z_{xy}^{\rm (i)}$ ($z_{xy}^{\rm (a)}$) and $zT$ denote the transverse figure of merit in the isothermal (adiabatic) condition and the longitudinal figure of merit for the Seebeck effect, respectively. }
  \label{fig:zT-efficiency-calc}
\end{figure}
%%%%%%%%%%%%%%%%%%%%%%%%%%%%%%%%%%%%

To summarize this subsection, we have formulated the maximum efficiency of a transverse thermoelectric generator for two thermal boundary conditions in the transverse direction: the adiabatic boundary condition [Eq.~(\ref{eq:eta_T_adia01})], and isothermal boundary condition [Eq.~(\ref{eq:eta_T_iso01})]. We note here that, for the Seebeck-effect-driven anomalous/ordinary Hall effects and SSE/SdSE-driven ISHEs, the efficiency expression is different from what we have discussed in the previous subsections, because these effects involve composite/hybrid materials \cite{Yamamoto21,Uchida16}. For example, the efficiency for the SSE device depends on the thickness ratio of the two layers of a magnetic insulator and attached metal. 

Let us discuss the result obtained in Sec. \ref{Sec:3.2}. In the absence of the Seebeck effect and THE ($S_{yy}=0$ and $\Theta=0$) which is customary assumed in the literature~\cite{Uchida16,Harman-Honig67,Wright62,Horst63,Delves64,Osterle-Angrist63}, from Eqs.~(\ref{eq:rho_T_adia01}), (\ref{eq:ZT_Tadia01}), and (\ref{eq:ZT_T_iso01}), we obtain a simple relation similar to Eq.~(\ref{eq:zT_conv-zT_alt01}): 
%%%
\begin{equation}
  z_{xy}^{\rm (i)} T = \frac{z_{xy}^{\rm (a)}T}{1+ z_{xy}^{\rm (a)}T}. 
  \label{eq:FOM_T_adia-iso01}
\end{equation}
%%%
Substituting Eq.~(\ref{eq:FOM_T_adia-iso01}) into Eq.~(\ref{eq:eta_T_adia01}), we obtain Eq.~(\ref{eq:eta_T_iso01}). That is, these two expressions for the efficiency coincide with each other. If we invert Eq.~(\ref{eq:FOM_T_adia-iso01}), we have 
%%%
\begin{equation}
  z_{xy}^{\rm (a)} T = \frac{z_{xy}^{\rm (i)}T}{1- z_{xy}^{\rm (i)}T},  
  \label{eq:FOM_T_adia-iso01b}
\end{equation}
%%%
from which we see that achieving $z_{xy}^{\rm (i)}T = 1$ limit is equivalent to achieving $z_{xy}^{\rm (a)}T = \infty$ limit. The equivalence of the energy conversion efficiency in the adiabatic and isothermal conditions can be visually confirmed in Figs. \ref{fig:zT-efficiency-calc}(a) and \ref{fig:zT-efficiency-calc}(b), where $zT$ for the Seebeck effect is set to be zero because of $S_{yy}=0$. 

Next, we consider the case in the presence of the Seebeck effect while in the absence of THE, {\it i.e.} $S_{yy} \neq 0$ and $\Theta=0$. In this case, from Eqs.~(\ref{eq:rho_T_adia01}), (\ref{eq:ZT_Tadia01}), and (\ref{eq:ZT_T_iso01}), we have a relation
%%%
\begin{equation}
  z_{xy}^{\rm (a)} T = \frac{z_{xy}^{\rm (i)}T}{1- z_{xy}^{\rm (i)} T + zT },
  \label{eq:FOM_T_adia-iso02}
\end{equation}
%%%
where, in comparison to Eq.~(\ref{eq:FOM_T_adia-iso01b}), the denominator has an additional contribution coming from the Seebeck effect, $zT= S_{yy}^2T/\kappa_{yy} \rho_{yy}$. 
This equation is solved for $z_{xy}^{\rm (i)}T$, giving 
%%%
\begin{equation}
  z_{xy}^{\rm (i)} T = \frac{z_{xy}^{\rm (a)}T (1+ zT)}{1+ z_{xy}^{\rm (a)}T}.
  \label{eq:FOM_T_adia-iso03}
\end{equation}
%%%
Substituting the above equation into Eq.~(\ref{eq:FOM_T_iso_bound01}), we obtain a constraint more severe than Eq.~(\ref{eq:FOM_T_adia_bound01}),  
%%%
\begin{equation}
0 \le  z_{xy}^{\rm (a)} T < \frac{1}{zT}, 
  \label{FOM_T_adia_bound02}
\end{equation}
%%%
which means that the smaller the Seebeck effect, the larger the upper limit of the adiabatic figure of merit for the transverse device. This behavior can be visually found in Figs. \ref{fig:zT-efficiency-calc}(c), \ref{fig:zT-efficiency-calc}(e), and \ref{fig:zT-efficiency-calc}(g). In contrast, as shown in Figs. \ref{fig:zT-efficiency-calc}(d), \ref{fig:zT-efficiency-calc}(f), and \ref{fig:zT-efficiency-calc}(h), the relation between $z_{xy}^{\rm (i)} T$ and $\overline{\eta}$ is independent of $zT$ because the Seebeck effect does not occur in the $y$ direction under the isothermal condition. 

Finally, in the generic situation of the simultaneous presence of the Seebeck effect and THE, {\it i.e.} $S_{yy} \neq 0$ and $\Theta \neq 0$, a similar analysis is much more involved and the results are not summarized in the form presented above.

%%%%%%%%%%%%%%%%%%%%%%%%%%%%%%%%%%%%%%%%%%%%%%%%%%%%%%%%%%%%
\subsection{Performance of transverse thermoelectric refrigerator \label{Sec:3.3}}
%%%%%%%%%%%%%%%%%%%%%%%%%%%%%%%%%%%%%%%%%%%%%%%%%%%%%%%%%%%%

In this subsection, we consider the performance of a transverse thermoelectric refrigerator/heat pump. Here, as mentioned in Introduction, we focus only on the cooling mode and discuss the maximum attainable temperature difference of the device as a measure of its performance. Note that the thermoelectric refrigerator discussed here is different from compressor-based refrigerators and magnetocaloric refrigerators \cite{Franco18}. In the previous two sections, we discarded the Joule heating effect as we used an approximation that is valid up to linear order in the temperature bias. By contrast, in order to investigate the maximum temperature difference, we need to take the Joule heating into account, and determine the temperature distribution in the refrigerator. 

In discussing the maximum attainable temperature difference of a refrigerator, we need to evaluate the rate of the heat removal ${\cal Q}_{\rm c}$ from the cold reservoir. Then, when we use Fig.~\ref{fig:TEdev01}(b) for the present problem, first of all we need to replace the load resistance with a current source for applying a charge current density ${\bf j}$ to the device. Besides, it is convenient to interchange the role of the hot and cold reservoirs as $T_{\rm h} \leftrightarrow T_{\rm c}$~\cite{Harman-Honig67} merely due to the fact that this choice simplifies the calculation by allowing us to calculate ${\cal Q}_{\rm c}$ at the origin $x=0$. Therefore, in this subsection, we assume that the left-hand side (right-hand side) of the device in Fig.~\ref{fig:TEdev01}(b) is in contact with $T_{\rm c}$ ($T_{\rm h}$). 

We first need to obtain the differential equation that determines temperature distribution in the device. The rate of heat evolution is given by $-{\bm \nabla} \cdot {\bf q}_E$, where ${\bf q}_E= {\bf q}+ \phi {\bf j} $ is the energy current density with $\phi$ being the electrochemical potential. Then, assuming the steady-state condition, the temperature distribution in the device is determined by~\cite{Domenicali54} 
%%%
\begin{equation}
  - {\bm \nabla} \cdot {\bf q}_E = -{\bm \nabla} \cdot {\bf q}+ {\bf E \cdot j} = 0. \label{eq:Domenicali01}
\end{equation}
%%%
Below, since we assume the boundary condition $j_x = 0$, the Joule heating term is given by ${\bf E \cdot j}= E_y j_y$ in Eq.~(\ref{eq:Domenicali01}).

%%%%%%%%%%%%%%%
\subsubsection{Isothermal condition in transverse direction. \label{Subsec:3.3.1}}
%%%%%%%%%%%%%%%

We first discuss the isothermal boundary condition in the direction of the electric field [along the $y$ axis in Fig.~\ref{fig:TEdev01}(b)], because the result for this case is known in the literature~\cite{Harman-Honig62b,Kooi63}. Applying the isothermal condition $\nabla_y T= 0$ to Eqs.~(\ref{eq:qx_T_adia02})--(\ref{eq:Ey_T_adia02}), we have 
%%%
\begin{eqnarray}
  q_x &=& -\kappa_{xx} \nabla_x T + T S_{xy}  j_y, \label{eq:qx_T_iso01b} \\
  q_y &=& -\kappa_{yx} \nabla_x T + T S_{yy}  j_y, \label{eq:qy_T_iso01b} \\
  E_y &=&  \rho_{yy} j_y + S_{yx} \nabla_x T. \label{eq:Ey_T_iso01b} 
\end{eqnarray}
%%%
Then, substituting these equations into Eq.~(\ref{eq:Domenicali01}), we obtain
%%%
\begin{equation}
  \kappa_{xx} \nabla_x ^2 T + T S_{yx} \nabla_x j_y + 2 S_{yx} j_y \nabla_x T + \rho_{yy} j_y^2 = 0, \label{eq:Domenicali_T01}
\end{equation}
%%%
where we used the symmetry $S_{xy}= - S_{yx}$, as well as the equations $\nabla_y \nabla_x T = \nabla_x \nabla_y T = 0$ and ${\bm \nabla}\cdot {\bf j}= \nabla_y j_y= 0$ under $j_x= 0$.

In order to solve Eq.~(\ref{eq:Domenicali_T01}), we follow an approximate method introduced in Refs.~\cite{Harman-Honig67,Harman-Honig62b} as this method is known to reproduce the exact solution obtained in Ref.~\cite{Kooi63}. We first replace $j_y$ with its spatial average $\bra j_y \ket_{\rm s}$, substitute it into Eq.~(\ref{eq:Domenicali_T01}), and obtain 
%%%
\begin{equation}
  \nabla_x^2 T + a \nabla_x T + b  = 0, \label{eq:Domenicali_T02}
\end{equation}
%%%
where $a= 2 S_{yx} \bra j_y \ket_{\rm s}/\kappa_{xx}$ and $b= \rho_{yy} \bra j_y \ket_{\rm s}^2/\kappa_{xx} $. We next seek for a solution of the above equation in the form of power series expansion as 
%%%
\begin{equation}
  T(x)= T_{\rm c} + c_1 x + \frac{c_2}{2} x^2, \label{eq:Tdist01}
\end{equation}
%%%
where $c_1$ and $c_2$ are coefficients to be determined. Then, substituting Eq.~(\ref{eq:Tdist01}) into Eq.~(\ref{eq:Domenicali_T02}) and setting $x=0$, we obtain $a c_1 + c_2 + b = 0$. Besides, using the boundary condition $T(x=l_x)= T_{\rm h}$, we have $c_1 l_x + \frac{c_2}{2} l_x^2 = \Delta T $, where $\Delta T = T_{\rm h}- T_{\rm c}$ as before. From these two equations, by discarding terms higher in $a^2, ab$, the two coefficients $c_1$ and $c_2$ are approximately obtained as $c_1 = {\Delta T}/{l_x} + ({a}/{2}) \Delta T + ({b}/{2}) l_x$ and $c_2 = -a {\Delta T}/{l_x} - b$, and the temperature distribution is given by
%%%
\begin{equation}
  T(x) = \left[ T_{\rm c}+ \left( \frac{x}{l_x} \right) \Delta T \right]
  - \frac{1}{2} \left[ b+ \frac{a \Delta T}{l_x}\right] x(x-l_x).
  \label{eq:Tdist02}
\end{equation}
%%%

Given the temperature distribution above, we are now in a position to calculate the rate of the heat removal from the cold reservoir ${\cal Q}_{\rm c}= A q_x(x=0)$, where $A= l_y l_z$. Substituting Eq.~(\ref{eq:Tdist02}) into Eq.~(\ref{eq:qx_T_iso01b}) and multiplying $A$, we obtain ${\cal Q}_{\rm c} = - K \Delta T - ({R}/{2}) J_y^2 - T_{\rm h} S_{yx} (l_y/l_x) J_y $, where $J_y= l_x l_z \bra j_y \ket_{\rm s}$, and $K$ and $R$ are defined below Eq.~(\ref{eq:DV_T_iso02}). We complete the square with respect to $J_y$ in the above equation as ${\cal Q}_{\rm c} = - K \Delta T- ({R}/{2}) \left( J_y - J_y^*  \right)^2 + ({T_{\rm h}^2 S^2_{yx}}/{2 R})  (l_y/l_x)^2 $, where $J_y^*= -{T_{\rm h} S_{yx} (l_y/l_x)}/{R}$. Then, we see that the maximum value ${\cal Q}_{\rm c} = ({T_{\rm h}^2 S^2_{yx} (l_y/l_x)^2})/({2 R}) - K \Delta T $ is obtained for $J_y= J_y^*$. Finally, the maximum temperature difference $\Delta T_{\rm max}$ is obtained for zero heat load ${\cal Q}_{\rm c}$ at the cold reservoir, yielding 
%%%
\begin{equation}
  \Delta T_{\rm max} = \frac{1}{2} z^{\rm (i)}_{xy} T^2_{\rm h},
  \label{eq:DeltTmax01}
\end{equation}
%%%
where $z^{\rm (i)}_{xy}$ is defined by Eq.~(\ref{eq:ZT_T_iso01}). Note that, as emphasized in Ref.~\cite{Harman-Honig67}, Eq.~(\ref{eq:DeltTmax01}) is the same as the result obtained by the exact treatment~\cite{Kooi63}. Note also that the isothermal figure of merit $z^{\rm (i)}_{xy}$ is bounded as expressed by Eq.~(\ref{eq:FOM_T_iso_bound01}).

%%%%%%%%%%%%%%%
\subsubsection{Adiabatic condition in transverse direction. 
\label{Subsec:3.3.2}}
%%%%%%%%%%%%%%%
We next consider the adiabatic condition in the direction of the electric field [along the $y$ axis in Fig.~\ref{fig:TEdev01}(b)]. Taking account of the adiabatic condition $q_y=0$, we start from Eqs.~(\ref{eq:gradTx_T_adia03}) and (\ref{eq:Ey_T_adia03}), and rewrite them as follows: 
%%%
\begin{eqnarray}
  q_x &=& - {\kappa}^{\rm (a)}_{xx} \nabla_x T + T {S}^{\rm (a)}_{xy} j_y,
  \label{eq:qx_T_adia03b} \\
  E_y &=&
  \rho_{yy}^{\rm (b)} j_y
  + {S}^{\rm (a)}_{yx} {\nabla_{x}} T,  \label{eq:Ey_T_adia03b} 
\end{eqnarray}
%%%
where a new quantity $\rho^{\rm (b)}_{yy}$ is defined by
%%%
\begin{equation}
  \rho^{\rm (b)}_{yy} = \rho_{yy} + \frac{T {S}^{\rm (a) \, 2}_{yy}}{{\kappa}^{\rm (a)}_{yy}}, \label{eq:rho_b01}
\end{equation}
%%%
and ${\kappa}^{\rm (a)}_{xx}$,  ${\kappa}^{\rm (a)}_{yy}$, ${S}^{\rm (a)}_{yy}$, ${S}^{\rm (a)}_{xy}$, and ${S}^{\rm (a)}_{yx}$ are defined below Eq.~(\ref{eq:Ey_T_adia03}). Note that $\rho^{\rm (b)}_{yy}$ in Eq.~(\ref{eq:Ey_T_adia03b}) is different from the adiabatic resistivity defined by Eq.~(\ref{eq:rho_T_adia01}), and its appearance in the present situation is already commented on in Table 7.7.1 of Ref.~\cite{Harman-Honig67}. Note also that to obtain the above equation, we used the symmetry ${S}^{\rm (a)}_{xy}= -{S}^{\rm (a)}_{yx}$. Then, after substituting Eqs.~(\ref{eq:qx_T_adia03b}) and (\ref{eq:Ey_T_adia03b}) into Eq.~(\ref{eq:Domenicali01}), we obtain the same equation as Eq.~(\ref{eq:Domenicali_T02}) but with the coefficients $a$ and $b$ being replaced by
%%%
\begin{eqnarray*}
  a &=& \frac{2 {S}^{\rm (a)}_{yx} \bra j_y \ket_{\rm s}}{{\kappa}^{\rm (a)}_{xx}}, \\
  b &=& \frac{\rho^{\rm (b)}_{yy} \bra j_y \ket^2_{\rm s}}{{\kappa}^{\rm (a)}_{xx}}. 
\end{eqnarray*}
%%%
Since the temperature distribution is obtained in the same way, we now proceed exactly in the same manner as in the preceding subsection, and obtain the rate of the heat removal from the cold reservoir as ${\cal Q}_{\rm c} = - {K}^{\rm (a)} \Delta T - ({R^{\rm (b)}}/{2}) J_y^2 - T_{\rm h} {S}^{\rm (a)}_{yx} (l_y/l_x) J_y $, where ${K}^{\rm (a)}= {\kappa}^{\rm (a)}_{xx} (l_y l_z/l_x)$, $R^{\rm (b)}= \rho^{\rm (b)}_{yy} l_y/(l_x l_z)$, and $J_y= l_x l_z \bra j_y \ket_{\rm s}$. This is transformed as ${\cal Q}_{\rm c} = - {K}^{\rm (a)} \Delta T
- ({R^{\rm (b)}}/{2}) \left( J_y - J_y^* \right)^2 + ({T_{\rm h}^2 {S}^{\rm (a) \, 2}_{yx}}/{2 R^{\rm (b)}} ) (l_y/l_x)^2 $, where $J_y^*= - {T_{\rm h} {S}^{\rm (a)}_{yx}(l_y/l_x) }/{R^{\rm (b)}}$. From this result we see that ${\cal Q}_{\rm c}$ is maximized for $J_y = J_y^*$, yielding the following maximum temperature difference
%%%
\begin{equation}
  \Delta T_{\rm max} = \frac{1}{2} z^{\rm (b)}_{xy} T^2_{\rm h},
  \label{eq:DeltTmax02}
\end{equation}
%%%
where
%%%
\begin{equation}
  z_{xy}^{\rm (b)} T = 
     \frac{T {S}^{\rm (a) \, 2}_{yx} }{{\kappa}^{\rm (a)}_{xx} {\rho}_{yy}^{\rm (b)}}
\label{eq:ZT_b01} 
\end{equation}
%%%
is the corresponding figure of merit for the transverse refrigerator in the adiabatic operation. Note the difference between $z_{xy}^{\rm (b)}$ [Eq.~(\ref{eq:ZT_b01})] and $z_{xy}^{\rm (a)}$ [Eq.~(\ref{eq:ZT_Tadia01})], where $z_{xy}^{\rm (b)}$ is bounded. Indeed, in the absence of THE, we have
%%%
\begin{equation}
  0 \le z_{xy}^{\rm (b)} < \frac{1}{1+ zT} 
\end{equation}
%%%
due to Eq.~(\ref{eq:FOM_T_iso_bound01}). 

Finally, we note the practical aspect of the transverse thermoelectric cooler. In this subsection, we have assumed a rectangular-shaped device [Fig. \ref{fig:TEdev01}(b)] to formulate the maximum temperature difference and figure of merit for the transverse thermoelectric refrigerator. However, the actual cooling performance depends on a geometric shape of the refrigerator due to the competition between the transverse thermoelectric conversion and Joule heating, as well as heat release conditions at the hot side. In a similar manner to the conventional Peltier refrigerator that achieves a large temperature difference with multistage cascading \cite{Goldsmid17}, the temperature difference in the transverse thermoelectric refrigerator can be increased by constructing a so-called infinite-stage device \cite{Harman63,Polash21}. Also in this device, the absence of interfacial thermal resistance at stacking boundaries is a significant advantage of the transverse thermoelectric conversion.

%%%%%%%%%%%%%%%%%%%%%%%%%%%%%%%%%%%%%%%%%%%%%%%%%%%%%%%%%%%%
\subsection{Performance of transverse active cooler
\label{Sec:3.4}}
%%%%%%%%%%%%%%%%%%%%%%%%%%%%%%%%%%%%%%%%%%%%%%%%%%%%%%%%%%%%

Before ending this section, we would like to discuss the active cooling~\cite{Adams19}. As discussed there, the active cooling system is designed to dissipate heat from the hot part of an object, so that the heat drained from the hot part is maximized while $\Delta T$ is minimized. The key is that, since the heat is drained from the hot part, the Fourier heat conduction and Peltier heat act in a constructive way. Then, the quantity that characterizes the performance is the effective thermal conductivity $\kappa^{\rm eff}$ that drains the heat from the hot part. While this idea is developed for the longitudinal device, below we apply the same argument to the transverse device with time-reversal symmetry breaking.  

In Sec.~\ref{Sec:3.3} when discussing the transverse refrigerator, we interchanged the role of hot and cold reservoirs as $T_{\rm h} \leftrightarrow T_{\rm c}$. Because the following calculation applies to the active cooling device, here we need to undo this interchange and assume that the left-hand side (right-hand side) of the device is in contact with $T_{\rm h}$ ($T_{\rm c}$) as in Fig.~\ref{fig:TEdev01}(b). 

First, we consider the case under the isothermal condition in the transverse direction. Starting from Eq.~(\ref{eq:Domenicali_T01}) and performing the same calculation as in Sec.~\ref{Subsec:3.3.1}, we obtain the temperature distribution, 
%%%
\begin{equation}
  T(x) = \left[ T_{\rm h}- \left( \frac{x}{l_x} \right) \Delta T \right]
  - \frac{1}{2} \left[ b- \frac{a \Delta T}{l_x}\right] x(x-l_x), 
  \label{eq:Tdist03}
\end{equation}
%%%
where $a= 2 S_{yx} \bra j_y \ket_{\rm s}/\kappa_{xx}$ and $b= \rho_{yy} \bra j_y \ket_{\rm s}^2/\kappa_{xx} $. Note that this expression is also obtained by interchanging $T_{\rm h}$ and $T_{\rm c}$ in Eq.~(\ref{eq:Tdist02}). Then, substituting this temperature distribution into Eq.~(\ref{eq:qx_T_iso01b}), we obtain the rate of heat removal from the hot reservoir ${\cal Q}_{\rm h} = l_y l_z q_x(x=0)$ as ${\cal Q}_{\rm h} = K \Delta T - ({R}/{2}) J_y^2 -T_{\rm c} S_{yx} (l_y/l_x) J_y $, where $J_y= l_x l_z \bra j_y \ket_{\rm s}$, and $K$ and $R$ are defined below Eq.~(\ref{eq:DV_T_iso02}). Completing the square with respect to $J_y$ in the above equation, we obtain ${\cal Q}_{\rm h} = K \Delta T - ({R}/{2}) \big( J_y - J_y^* \big)^2 + ({T_{\rm c}^2 S_{yx}^2 }/{2R}) (l_y/l_x)^2 $, where $J_y^* = -T_{\rm c} S_{yx} (l_y/l_x)^2/2R $. Then, by setting $ J_y = J_y^*$, the maximum heat removal from the hot reservoir ${\cal Q}_{\rm h}^*$ is obtained as ${\cal Q}_{\rm h}^* = K \Delta T + ({T_{\rm c}^2 S_{yx}^2}/{2R}) (l_y/l_x)^2$. Now we introduce the heat current density $q_{\rm h}^*= {\cal Q}^*_{\rm h}/(l_y l_z)$, and rewrite the above equation as $q_{\rm h}^* = \kappa_{xx} \left( ({\Delta T}/{l_x}) \right) + ({T_{\rm c}^2 S_{yx}^2})/({2 \rho_{yy} l_x})$. Then, defining $\widetilde{\nabla}_x T=  \Delta T/l_x$ which is positive by definition, the above equation can be represented as 
%%%
\begin{equation}
q_{\rm h}^* = \kappa^{\rm eff}_{xx} \widetilde{\nabla}_x T, 
\label{eq:kappa_eff01}
\end{equation}
%%%
where the effective thermal conductivity is defined by 
%%%
\begin{equation}
\kappa^{\rm eff}_{xx} = \kappa_{xx}
+ \frac{{\rm PF}^{\rm (i)} T_{\rm c}^2}{2 \Delta T}, 
\label{eq:kappa-PF01}
\end{equation}
%%%
and the isothermal power factor for the transverse device is given by 
%%%
\begin{equation}
 {\rm PF}^{\rm (i)} = \frac{S_{yx}^2}{\rho_{yy}}. 
\end{equation}
%%%

Next, we consider the case under the adiabatic condition in the transverse direction. In this case, repeating the same discussion as in Sec.~\ref{Subsec:3.3.2}, we only need to make the following substitutions to the isothermal result:
%%%
\[ \rho_{yy} \to \rho^{\rm (b)}_{yy},  \]
\[ \kappa_{xx} \to \kappa^{\rm (a)}_{xx}, \]
\[ S_{yx} \to S_{yx}^{\rm (a)}, \]
%%%
where $\rho_{yy}^{\rm (b)}$ is defined by Eq.~(\ref{eq:rho_b01}) and other quantities are defined below Eq.~(\ref{eq:Ey_T_adia03}). Note that $\rho_{yy}^{\rm (b)}$ explicitly depends on temperature. Since this is the temperature felt by the driving charge current parallel to the $y$ axis, we approximate $T= (T_{\rm h}+ T_{\rm c})/2$. Then, repeating the same calculation as above, we obtain Eq.~(\ref{eq:kappa_eff01}), in which the effective thermal conductivity is given by 
%%%
\begin{equation}
\kappa^{\rm eff}_{xx} = \kappa_{xx}
+ \frac{{\rm PF}^{\rm (a)} T_{\rm c}^2}{2 \Delta T}, 
\label{eq:kappa-PF02}
\end{equation}
%%%
and the adiabatic power factor for the transverse device is given by 
%%%
\begin{equation}
 {\rm PF}^{\rm (a)} = \frac{S_{yx}^{\rm (a) \,2}}{\rho_{yy}^{\rm (b)}}. 
\label{eq:PF02}
\end{equation}
%%%

%%%%%%%%%%%%%%%%%%%%%%%%%%%%%%%%%%%%%%%%%%%%%%%%%%%%%%%%%%%%%%%%%%%%%%
\section{Formulation of transverse thermoelectric conversion with structural symmetry breaking  \label{Sec:Formulation-structural} }
%%%%%%%%%%%%%%%%%%%%%%%%%%%%%%%%%%%%%%%%%%%%%%%%%%%%%%%%%%%%%%%%%%%%%%

So far, we have considered situations where the transport coefficient matrices in Eqs.~(\ref{eq:alpha_mat01})--(\ref{eq:rho_mat01}) have antisymmetric off-diagonal components. For the ONE/ANE devices, this condition is satisfied. However, the off-diagonal Seebeck effect is of geometric origin. Therefore, the off-diagonal components of the transport coefficient matrices are symmetric, satisfying $S_{yx}= S_{xy}$, $\kappa_{yx}= \kappa_{xy}$, and $\rho_{yx}= \rho_{xy}$. In this section, we discuss the efficiency of such thermoelectric generators in the first two subsections, and then formulate the performance of such thermoelectric refrigerators and active coolers in the last two subsections. This treatment goes into greater depth than prior review articles which consider a more introductory perspective to transverse thermoelectrics based on structural symmetry breaking \cite{Grayson18}. 

%%%%%%%%%%%%%%%%%%%%%%%%%%%%%%%%%%
\subsection{Transverse thermoelectric generator in adiabatic condition}
%%%%%%%%%%%%%%%%%%%%%%%%%%%%%%%%%%

Under the adiabatic boundary condition for the present system, instead of Eq.~(\ref{eq:rho_T_adia01}), we have 
%%%
\begin{equation}
  {\rho}_{yy}^{\rm (a)} = 
  \rho_{yy} +
  \frac{T {S}^{\rm (a) \, 2}_{yy}}{{\kappa}^{\rm (a)}_{yy}} + 
  \frac{T {S}^{\rm (a) \, 2}_{yx}}{{\kappa}^{\rm (a)}_{xx}}, 
  \label{eq:rho_T_adia02}
\end{equation}
%%%
where we used the symmetry $S_{yx}= S_{xy}$ and $\kappa_{yx}= \kappa_{xy}$. Then, instead of Eqs.~(\ref{eq:linear-T_adia01}) and (\ref{eq:matC01}), we have 
%%%
\begin{equation}
  \widehat{\cal C'}=
  \begin{pmatrix} \frac{- {S}^{\rm (a)}_{yx} T}{{K}^{\rm (a)}(l_x/l_y)} & \frac{1}{{K}^{\rm (a)}} \\ - {R}^{\rm (a)} & \frac{+{S}^{\rm (a)}_{yx}}{{K}^{\rm (a)}(l_x/l_y)} \end{pmatrix},
  \label{eq:matC02}
\end{equation}
%%%
where the determinant is also modified as $\det \widehat{\cal C'} = ({{R}^{\rm (a)}}/{{K}^{\rm (a)}}) (1- z_{xy}^{\rm (a)} T)$. Note that the adiabatic figure of merit $z_{xy}^{\rm (a)}$ in the above equation is defined by Eq.~(\ref{eq:ZT_Tadia01}), but the adiabatic resistivity therein is redefined by Eq.~(\ref{eq:rho_T_adia02}). Consistent with this change, instead of Eq.~(\ref{eq:FOM_T_adia_bound01}), the adiabatic figure of merit for the present system is bounded, {\it i.e.} 
%%%
\begin{equation}
  0 \le z^{\rm (a)}_{xy} T < 1. 
  \label{eq:FOM_S_adia_bound01}
\end{equation}
%%%
Then, instead of Eq.~(\ref{eq:eta_T_adia01}), the maximum efficiency of the present system is given by 
%%%
\begin{equation}
  \overline{\eta} = \frac{1- \sqrt{1- z_{xy}^{\rm (a)} T}}{1+ \sqrt{1- z_{xy}^{\rm (a)}T}}
  \label{eq:eta_T_adia02} 
\end{equation}
%%%
under the adiabatic boundary condition. 

%%%%%%%%%%%%%%%%%%%%%%%%%%%%%%%%%
\subsection{Transverse thermoelectric generator in isothermal condition}
%%%%%%%%%%%%%%%%%%%%%%%%%%%%%%%%%
Under the isothermal boundary condition for the present system, instead of Eq.~(\ref{eq:matD01}), we have 
%%%
\begin{equation}
  \widehat{\cal D'}= 
    \begin{pmatrix} S_{yx} T (l_y/l_x) & K \\ -R  & S_{yx} (l_y/l_x) \end{pmatrix},
    \label{eq:matD02}
\end{equation}
%%% 
where we used $S_{yx}= S_{xy}$. Then, the determinant also changes as $\det \widehat{\cal D'} = K R (1+ z_{xy}^{\rm (i)} T)$, which means that, instead of Eq.~(\ref{eq:FOM_T_iso_bound01}), the isothermal figure of merit for the present system is unbounded, {\it i.e.} 
%%%
\begin{equation}
  0 \le z^{\rm (i)}_{xy} T < \infty. 
  \label{eq:FOM_S_iso_unbound01}
\end{equation}
%%%
In line with this change, the maximum efficiency of the present system is given by 
%%%
\begin{equation}
  \overline{\eta} = \frac{\sqrt{1+ z_{xy}^{\rm (i)} T}-1}{\sqrt{1+ z_{xy}^{\rm (i)}T}+1}
  \label{eq:eta_T_iso02} 
\end{equation}
%%%
under the isothermal boundary condition. 

%%%%%%%%%%%%%%%%%
\subsection{Transverse thermoelectric refrigerator}
%%%%%%%%%%%%%%%%%

Here, we briefly comment on the performance of an off-diagonal Peltier refrigerator. Let us first discuss the isothermal boundary condition in the direction of charge current [along the $y$ axis in Fig.~\ref{fig:TEdev01}(b)]. In deriving Eq.~(\ref{eq:Domenicali_T02}) for the transverse thermoelectric refrigerators, we calculate the expression of the coefficient $a$ by using the symmetry $S_{xy}= - S_{yx}$ as 
%%%
\begin{eqnarray*}
  a &=& - \frac{S_{xy} j_y }{\kappa_{xx}} + \frac{S_{yx} j_y }{\kappa_{xx}} \nonumber \\
  &=&
  \frac{2 S_{yx} j_y }{\kappa_{xx}}, 
\end{eqnarray*}
%%%
then we perform replacement $j_y \to \bra j_y \ket_{\rm s}$. By contrast, in the present case, we have the symmetry $S_{xy}= S_{yx}$ such that the coefficient $a$ vanishes, {\it i.e.} $a = 0 $ for the off-diagonal Peltier device under the isothermal condition, while the other coefficient becomes $b= {\rho_{yy} j_y^2}/{\kappa_{xx}}$. Then, performing the same calculation as in Sec.~\ref{Sec:3.3}, the maximum attainable temperature difference is obtained as
%%%
\begin{equation}
  \Delta T_{\rm max} = \frac{1}{2} z^{\rm (i)}_{xy} T^2_{\rm c}, 
  \label{eq:DeltTmax03}
\end{equation}
%%%
where the appearance of the cold side temperature $T_{\rm c}$ is similar to the result for longitudinal Peltier devices~\cite{Heikes-Ure61}. Note the difference from Eq.~(\ref{eq:DeltTmax01}). Note also that $z^{\rm (i)}_{xy}$ for the off-diagonal Peltier device [Eq.~(\ref{eq:DeltTmax03})] is unbounded as shown in Eq.~(\ref{eq:FOM_S_iso_unbound01}). 

We next discuss the off-diagonal Peltier refrigerator under the adiabatic condition. Repeating the same calculation as in Sec.~\ref{Sec:3.3}, we find $a = 0$ and $b= {\rho^{\rm (b)}_{yy} j_y^2}/{\kappa^{\rm (a)}_{xx}}$, where $\rho^{\rm (b)}_{yy}$ is defined by Eq.~(\ref{eq:rho_b01}). Then, the maximum temperature difference is calculated to be
%%%
\begin{equation}
  \Delta T_{\rm max} = \frac{1}{2} z^{\rm (b)}_{xy} T^2_{\rm c}, 
  \label{eq:DeltTmax04}
\end{equation}
%%%
where $z_{xy}^{\rm (b)}$ is defined by Eq.~(\ref{eq:ZT_b01}).

%%%%%%%%%%%%%%%%%
\subsection{Transverse active cooler}
%%%%%%%%%%%%%%%%%

Finally, we discuss the performance of the active cooler based on the off-diagonal Peltier effect. Let us first consider the case of isothermal condition in the transverse direction. In this case, as in the previous section, we need to determine the temperature distribution by solving Eq.~(\ref{eq:Domenicali_T02}). Then, we obtain the temperature distribution expressed by Eq.~(\ref{eq:Tdist03}), where $a=0$ and $b= \rho_{yy}j_y^2/\kappa_{xx}$ should be noted. Then, substituting this temperature distribution into Eq.~(\ref{eq:qx_T_iso01b}), we obtain the rate of heat removal from the hot reservoir ${\cal Q}_{\rm h} = l_y l_z q_x(x=0)$ as ${\cal Q}_{\rm h} = K \Delta T - ({R}/{2}) J_y^2 + T_{\rm h} S_{yx} (l_y/l_x) J_y$, where $J_y= l_x l_z j_y $. Completing the square with respect to $J_y$ in the above equation, we obtain ${\cal Q}_{\rm h} = K \Delta T - ({R}/{2}) \big( J_y - J_y^* \big)^2  + ({T_{\rm h}^2 S_{yx}^2 }/{2R} ) (l_y/l_x)^2 $, where $J_y^* = -{T_{\rm h} S_{yx}}(l_y/l_x)/2R $. Then, by setting $ J_y = J_y^*$, the maximum heat removal from the hot reservoir ${\cal Q}_{\rm h}^*$ is obtained as ${\cal Q}_{\rm h}^* = K \Delta T + ({T_{\rm h}^2 S_{yx}^2}/{2R}) (l_y/l_x)^2$. Now we define $q_{\rm h}^*= {\cal Q}^*_{\rm h}/(l_y l_z)$, and rewrite the above equation as $q_{\rm h}^* = \kappa_{xx} \left( {\Delta T}/{l_x} \right) + ({T_{\rm h}^2 S_{yx}^2}/{2 \rho_{yy} l_x}) $. Then, with the definition of $\widetilde{\nabla}_x T=  \Delta T/l_x$, the above equation can be transformed to Eq.~(\ref{eq:kappa_eff01}), in which the effective thermal conductivity for the present case is given by 
%%%
\begin{equation}
\kappa^{\rm eff}_{xx} = \kappa_{xx}
+ \frac{{\rm PF}^{\rm (i)} T_{\rm h}^2}{2 \Delta T}, 
\label{eq:kappa-PF03}
\end{equation}
%%%
and the isothermal power factor for the transverse device is given by 
%%%
\begin{equation}
 {\rm PF}^{\rm (i)} = \frac{S_{yx}^2}{\rho_{yy}}. 
\end{equation}
%%%

We also discuss the case under the adiabatic condition in the transverse direction.  As in Sec.~\ref{Sec:3.4}, we only need to make the following substitutions to the isothermal case:
%%%
\[ \rho_{yy} \to \rho^{\rm (b)}_{yy},  \]
\[ \kappa_{xx} \to \kappa^{\rm (a)}_{xx}, \]
\[ S_{yx} \to S_{yx}^{\rm (a)}, \]
%%%
where $\rho_{yy}^{\rm (b)}$ is defined by Eq.~(\ref{eq:rho_b01}) and other quantities are defined below Eq.~(\ref{eq:Ey_T_adia03}). Note that $\rho_{yy}^{\rm (b)}$ explicitly depends on temperature, which is approximated $T= (T_{\rm h}+ T_{\rm c})/2$ as before. Then, repeating the same calculation as in the isothermal case, we obtain Eq.~(\ref{eq:kappa_eff01}), where the effective thermal conductivity is given by 
%%%
\begin{equation}
\kappa^{\rm eff}_{xx} = \kappa_{xx}
+ \frac{{\rm PF}^{\rm (a)} T_{\rm h}^2}{2 \Delta T}, 
\label{eq:kappa-PF04}
\end{equation}
%%%
and the adiabatic power factor for the transverse device is given by 
%%%
\begin{equation}
 {\rm PF}^{\rm (a)} = \frac{S_{yx}^{\rm (a) \,2}}{\rho_{yy}^{\rm (b)}}. 
 \label{eq:PF04}
\end{equation}
%%%

%%%%%%%%%%%%%%%%%%%%%%%%%%%%%%%%%%%%%%%%%
\section{Measurement of transverse thermopower} \label{Sec:Measurement}
%%%%%%%%%%%%%%%%%%%%%%%%%%%%%%%%%%%%%%%%%

The transverse thermoelectric effects can be investigated by measuring an electric field or voltage in a sample in the transverse direction ($y$ direction) while applying a uniform temperature gradient in the longitudinal direction ($x$ direction). In the measurement of the transverse thermoelectric effects due to structural symmetry breaking, which is independent of a magnetic field $H$ and/or magnetization $M$, the positions of electrodes for measuring the voltage must be carefully set up because asymmetry in the electrode positions cause the longitudinal thermopower to be superimposed on the transverse thermopower. It is easy to separate the transverse thermoelectric effects due to time-reversal symmetry breaking, which show an odd dependence on $H$ and/or $M$ in the $z$ direction, from the longitudinal thermopower. Here, the pure transverse thermopower can be extracted by measuring the $H$ dependence of the transverse voltage $V$ and calculating $(V(+H)-V(-H))/2$ with $V(+H)$ and $V(-H)$ respectively being $V$ obtained when positive and negative magnetic fields are applied. In the case of the longitudinal thermoelectric effects, where the temperature gradient and electric field are parallel, the voltage per temperature difference and the electric field per temperature gradient are the same value. In contrast, in the case of the transverse thermoelectric effects, these are different values depending on the aspect ratio of the sample. Therefore, to discuss the performance of the transverse thermoelectric effects, the electric field per temperature gradient, {\it i.e.} the transverse thermopower, must be compared. 

The following focuses on the measurement methods for the heat-to-charge current conversion phenomena (mainly, ANE). On the other hand, methods for clarifying the charge-to-heat current conversion phenomena have also been established \cite{Uchida-Iguchi21}. To measure the phenomena that output heat currents, including the anisotropic magneto-Peltier, anomalous Ettingshausen, and spin Peltier effects, various temperature detection techniques such as the direct contact with thermocouple wires \cite{Itoh17}, integration of on-chip thermocouples \cite{Flipse14}, lock-in thermography method \cite{Breitenstein10,Wid16,Daimon16,Seki18,Uchida18}, and lock-in thermoreflectance method \cite{Yamazaki20} are employed. In all of these methods, lock-in detection to separate thermoelectric responses from Joule heating and magnetic field dependence measurements to distinguish different magneto-thermoelectric and thermo-spin effects are indispensable. 

%%%%%%%%%%%%%%%%%%%%%%%%%%%%%%%%%%%%
\subsection{Measurement configuration for Nernst effect in thin films}
%%%%%%%%%%%%%%%%%%%%%%%%%%%%%%%%%%%%

%%%%%%%%%%%%%%%%%%%%%%%%%%%%%%%%%%%%% 
%\begin{figure*}[tb]
\begin{figure*}
  \begin{center}
    \includegraphics[width=12.7cm]{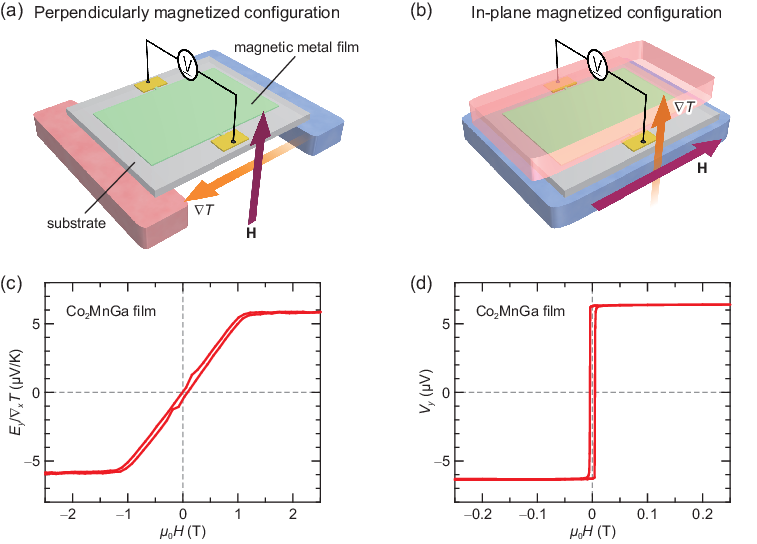}
  \end{center}
  \caption{(a),(b) Schematics of the perpendicularly magnetized (PM) and in-plane magnetized (IM) configurations for the measurements of ANE. (c) Magnetic field $H$ dependence of the transverse thermopower $E_y / \nabla_x T$ for the Co$_2$MnGa film at room temperature, measured in the PM configuration. $E_y$, $\nabla_x T$, and $\mu_0$ denote the transverse electric field, longitudinal temperature gradient, and vacuum permeability, respectively. (d) $H$ dependence of the transverse voltage $V_y$ for the Co$_2$MnGa film, measured in the IM configuration. This result was obtained when a heater power of 160 mW was applied to the top heat bath. The exact composition of this film is Co$_{53.0}$Mn$_{23.8}$Ga$_{23.2}$. The data in (c) and (d) are reconstructed from Fig. 3 of Ref. \cite{Uchida-Zhou-Sakuraba21}. }
  \label{fig:configuration}
\end{figure*}
%%%%%%%%%%%%%%%%%%%%%%%%%%%%%%%%%%%%

In the fields of spin caloritronics and topological materials science, ANE has been widely measured not only in bulk materials but also in thin film devices. In contrast to simple measurements for bulk materials, there are two experimental configurations for thin films because of their huge difference between the in-plane and out-of-plane dimensions, and their characteristics need to be understood \cite{Uchida-Zhou-Sakuraba21,Uchida15}. One configuration is the perpendicularly magnetized (PM) configuration, where ${\bf H}$ ($\nabla T$) is applied along the out-of-plane (in-plane) direction [Fig. \ref{fig:configuration}(a)]. The PM configuration is widely used for the quantitative estimation of $S_{\rm ANE}$ since the magnitude of $\nabla T$ can be precisely measured by various experimental techniques including thermometers attached on hot and cold sides of a substrate or heat baths \cite{Uchida15}, on-chip thermometers grown on the substrate \cite{Reichlova18}, and an infrared camera \cite{Sakuraba20,Sumida20,Nakayama19,Zhou-Sakuraba20,Seki21}. Thus, by estimating the ANE-induced electric field through the extraction of the zero-field intercept of the $H$ dependence of the transverse electric field $E_y$ and by normalizing it by $\nabla T$, one can determine $S_{\rm ANE}$. However, the PM configuration often requires the application of large $H$ to align the magnetization of films along the out-of-plane direction to overcome the strong demagnetization field. The other configuration is the in-plane magnetized (IM) configuration, where ${\bf H}$ ($\nabla T$)  is applied along the in-plane (out-of-plane) direction [Fig. \ref{fig:configuration}(b)]. The IM configuration is suitable for thermal energy harvesting and heat flux sensing, discussed later, as ANE-based thermoelectric generation works simply by forming films onto heat sources. However, the IM configuration is not suitable for estimating $S_{\rm ANE}$ quantitatively because the temperature difference between the top and bottom of thin films is hard to be quantified (note that it is possible to quantify $S_{\rm ANE}$ even in the IM configuration by measuring the out-of-plane thermal conductivity of thin films using the time-domain thermoreflectance method \cite{Yamazaki24}). Comparing the transverse thermopower in the PM and IM configurations is also used to separate the contribution of ANE from that of SSE-driven ISHE, which appears only in the IM configuration due to the symmetry of ISHE \cite{Kikkawa13,Kikkawa13-PRB}. Note that the situations discussed in this subsection are applicable to ONE in thin films. 

Figures \ref{fig:configuration}(c) and \ref{fig:configuration}(d) show an example of the experimental results of ANE measured in the PM and IM configurations, respectively, for a 50-nm-thick Co$_2$MnGa thin film, epitaxially grown on a single-crystalline MgO (001) substrate \cite{Uchida-Zhou-Sakuraba21}. In both the configurations, the observed transverse thermoelectric signals showed an odd dependence on $H$ and its magnitude saturated in the high-$H$ region, where the magnetization of the Co$_2$MnGa film saturated. This is a typical behavior of ANE. In the IM configuration, a large ANE-induced thermopower was observed to appear even at zero field because the strong demagnetization field in the out-of-plane direction stabilized the remanent magnetization in the in-plane direction [Fig. \ref{fig:configuration}(d)], a situation preferable for practical applications. In contrast, a tiny ANE-induced thermopower was observed at zero field in the PM configuration [Fig. \ref{fig:configuration}(c)]. To generate the ANE-induced thermopower in the absence of an external magnetic field in the PM configuration, it is necessary to use a magnetic thin film having large perpendicular magnetic anisotropy \cite{Mizuguchi-Nakatsuji19,Noguchi24}. 

%%%%%%%%%%%%%%%%%%%%%%%%%%%%%%%%%%%%
\subsection{Effect of thermal boundary conditions}
%%%%%%%%%%%%%%%%%%%%%%%%%%%%%%%%%%%%

As is clear from the formulation in Secs. \ref{Sec:Formulation-time} and \ref{Sec:Formulation-structural}, it is important to take the thermal boundary conditions in the electric field direction into account in the measurements of the transverse thermoelectric effects. This is because under adiabatic conditions, the transverse thermopower is modulated by the concerted action of the Seebeck effect and THE (off-diagonal thermal conductivity) for the transverse thermoelectric effects due to time-reversal symmetry breaking (structural symmetry breaking). It has been experimentally shown that this correction term can improve the performance of the transverse thermoelectric conversion when the sign of $S_{xy}$ and the correction term is the same \cite{Scudder22,Ando25-PRAppl}. 

%%%%%%%%%%%%%%%%%%%%%%%%%%%%%%%%%%%%% 
%\begin{figure*}[tb]
\begin{figure*}
  \begin{center}
    \includegraphics[width=16.4cm]{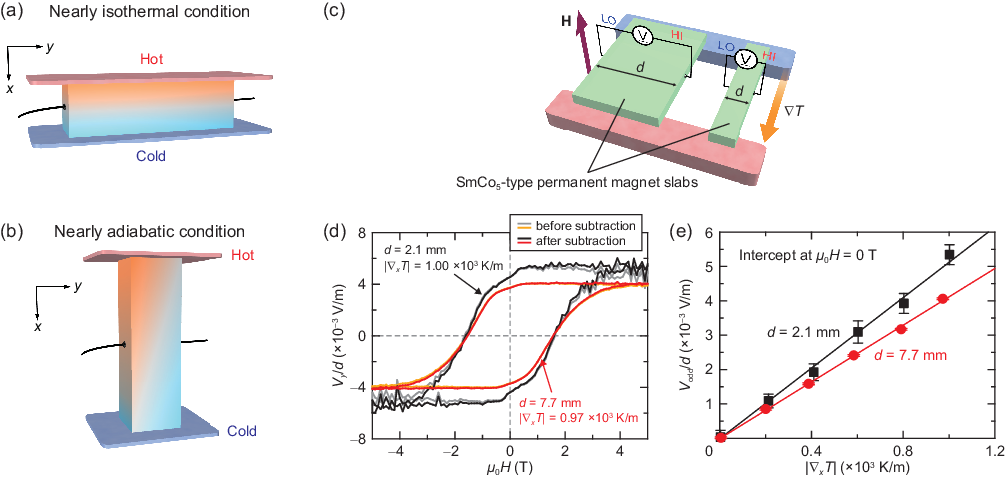}
  \end{center}
  \caption{(a),(b) Schematics of the measurement setups for the transverse thermoelectric generation in the nearly isothermal and adiabatic conditions. (c) Schematic of the measurement setup for the transverse thermopower in SmCo$_5$-type permanent magnet slabs with different values of the width $d$. (d) $H$ dependence of the transverse electric field $V_y/d$ for the SmCo$_5$ slabs with $d = 7.7~{\rm mm}$ (red and orange) and 2.1 mm (black and gray) at $|\nabla_x T| \sim 1 \times 10^3~{\rm K/m}$. The $H$-independent offset of the transverse electric field is subtracted from the raw data. The red and black (orange and gray) curves show the $H$ dependence of $V_y/d$ after (before) subtracting the $H$-linear component estimated by linear fitting of the signals in the high $H$ region. (e) $|\nabla_x T|$ dependence of the $M$-odd-dependent component of the transverse electric field $V_{\rm odd}/d$, which is estimated by extrapolating the $V_y/d$ signal in the high $H$ region to zero field.}
  \label{fig:bulkANE}
\end{figure*}
%%%%%%%%%%%%%%%%%%%%%%%%%%%%%%%%%%%%

In the following, we show the results of ANE measurements for SmCo$_5$-type bulk permanent magnets to confirm that the effect of the thermal boundary conditions is too large to be ignored even in ferromagnetic metals, of which THE is typically small. In SmCo$_5$, $S_{xy}$ and the correction term should have the same sign \cite{Miura19}. It is difficult to achieve an ideal adiabatic or isothermal condition in an experiment, but which condition is closer depends on the aspect ratio of the material \cite{Scudder22,Takahagi25}. When the length of the sample along the temperature gradient is much shorter than the width along the electric field, the thermal boundary condition along the electric field direction can be regarded as almost isothermal [Fig. \ref{fig:bulkANE}(a)]. When the length of the sample along the temperature gradient is long, it is closer to adiabatic [Fig. \ref{fig:bulkANE}(b)]. In most experiments, the sample dimension along the transverse electric field direction ($y$ direction) is smaller than that along the longitudinal temperature gradient $\nabla_x T$ direction ($x$ direction), resembling the adiabatic thermal boundary condition shown in Fig. \ref{fig:bulkANE}(b). In this configuration, a transverse temperature gradient along the $y$ direction $\nabla_y T$ is finite due to the contribution from THE. Consequently, according to Eq. (\ref{eq:thermopower-iso-vs-ad}), the observed transverse thermopower $E_y^* / (-\nabla_x T)$ is the combination of the isothermal transverse thermopower $E_y / (-\nabla_x T)|_{\nabla_y T =0}$ and the longitudinal thermopower $S_{yy}$ multiplied by the ratio of $\nabla_y T$ to $\nabla_x T$:
%%%
\begin{equation} \label{eq: thermopower-measurement1}
\frac{E_y^*}{(-\nabla_x T)} = \frac{E_y}{(-\nabla_x T)|_{\nabla_y T =0}} + S_{yy} \left( \frac{\nabla_y T}{\nabla_x T} \right).
\end{equation}
%%%
Therefore, unless the sample dimension along the transverse electric field is long enough to realize $\nabla_y T \sim 0$, the observed thermopower is either underestimated or overestimated, depending on the sign of the second term in the right-hand side of Eq. (\ref{eq: thermopower-measurement1}); the choice of sample dimensions plays a crucial role in transverse thermopower measurements. To accurately determine $S_{xy}$ from direct thermopower measurements, the sample dimension along the $y$ direction should be larger than that along the $x$ direction.

We now experimentally demonstrate the influence of sample dimensions on the transverse thermopower using a bulk polycrystalline SmCo$_5$ slab. The SmCo$_5$ block, commercially available from Magfine Corporation, Japan, was cut into two rectangular slabs with dimensions along the $x$, $y$, and $z$ directions respectively being 10.0, 7.7, and 0.8 mm and 10.0, 2.1, and 0.8 mm using a diamond wire saw, where only the length along the $y$ direction, $d$, is significantly different. The slabs have an easy axis of magnetization along $z$ direction. The measurement configuration is illustrated in Fig. \ref{fig:bulkANE}(c). The two SmCo$_5$ slabs were bridged between two heat baths separated by 8.0 mm and firmly attached to the heat baths. To apply a temperature difference between the heat baths, chip heaters were attached to the hot side. The sample holder was fixed at the center of a superconducting magnet, where a magnetic field ranging from $\mu_0 H = 5~{\rm T}$ to $-5~{\rm T}$ was applied along the $z$ direction. The $H$ dependence of the transverse voltage $V_y$ along the $y$ direction was recorded by attaching Cu wires at the ends of the slabs at various values of the temperature difference. To accurately measure the temperature gradient on the SmCo$_5$ slabs, the top surface of the slabs was coated with insulating black ink with an emissivity of $>0.94$ and steady-state temperature images were captured using an infrared camera.

Figure \ref{fig:bulkANE}(d) shows the $H$ dependence of the transverse electric field $V_y/d$ for the SmCo$_5$ slabs, recorded under a longitudinal temperature gradient of $|\nabla_x T| \sim 1 \times 10^3~{\rm K/m}$. In both the slabs, the transverse electric field changed its sign with large hysteresis with respect to the $H$ reversal. The $H$ dependence of $V_y/d$ corresponds to the magnetization curve of the SmCo$_5$ permanent magnet, indicating the appearance of ANE. To determine the magnitude of the $M$-dependent transverse thermopower for the SmCo$_5$ slabs, the $M$-odd-dependent component of $V_y/d$ under the saturation state, $V_{\rm odd}/d$, was extracted for each $|\nabla_x T|$ value by extrapolating the signal in the high $H$ region to zero field. Here, the finite $H$-linear component was observed only for the slab with $d = 2.1~{\rm mm}$, suggesting the presence of the Seebeck voltage between the SmCo$_5$/Cu-wire contacts induced by the ordinary ($H$-dependent) THE in SmCo$_5$ [Fig. \ref{fig:bulkANE}(d)]. Although the SmCo$_5$ slabs were cut from the same ingot, they exhibited different $V_{\rm odd}/d$ values [Fig. \ref{fig:bulkANE}(e)]. The $M$-dependent transverse thermopower for the SmCo$_5$ slab with $d = 2.1~{\rm mm}$ ($d = 7.7~{\rm mm}$) was found to be $5.1~\mu {\rm V/K}$ ($4.1~\mu {\rm V/K}$); the difference in the transverse thermopower is much larger than measurement errors. Recall the fact that the slab with $d = 2.1~{\rm mm}$ ($d = 7.7~{\rm mm}$) is closer to the adiabatic (isothermal) condition. Therefore, the transverse thermopower under the adiabatic condition is expected to be overestimated compared to that under the isothermal condition because of the same sign of $S_{xy}$ and the correction term induced by the Seebeck effect and anomalous ($M$-dependent) THE. This expectation is consistent with the observed difference in the $M$-dependent transverse thermopower. Here, we note that the correction term discussed theoretically in Sec. \ref{Sec:Formulation-time} is determined by the absolute Seebeck coefficient of a magnetic material ({\it i.e.} SmCo$_5$), while the observed difference in the transverse thermopower in Fig. \ref{fig:bulkANE} depends on the relative Seebeck coefficient between SmCo$_5$ and Cu. However, since the Seebeck coefficient of Cu ($<2~{\rm \mu V/K}$) is much smaller than that of SmCo$_5$ ($-19~{\rm \mu V/K}$ \cite{Miura19}) at room temperature, the magnitude and sign of the correction term in the present setup is determined mainly by the transport properties of SmCo$_5$. The experimental results shown here offer guidelines for determining the appropriate sample aspect ratio and measurement setup for transverse thermopower measurements, minimizing parasitic signals from the Seebeck effect and THE. We also note that considering typical sample dimensions, the estimation of $S_{\rm ANE}$ from the measurement of the anomalous Ettingshausen effect and the Onsager reciprocal relation reduces the influence of the correction term \cite{Takahagi25}.

A similar effect occurs not only in the Nernst effects but also in the transverse thermoelectric conversion with structural symmetry breaking. In the latter case, THE does not occur due to the absence of a magnetic field or magnetization, but the transverse thermopower can be modulated by the off-diagonal thermal conductivity derived from anisotropic transport properties. To accurately determine each transport coefficient, it is necessary to quantify or suppress the contribution coming from the correction term. However, since the performance of the transverse thermoelectric conversion can be improved by superimposing the correction-term contribution, its utilization is important in terms of applications \cite{Scudder22,Ando25-PRAppl}.

It should also be noted that the underestimation or overestimation of the transverse thermopower due to the thermal boundary conditions does not occur in thin films. The temperature of thin films is strongly thermalized to that of substrates, which have a large heat capacity. Thus, even if THE and structure-induced off-diagonal thermal conductivity are finite, the transverse thermopower is always obtained in the isothermal condition.

%%%%%%%%%%%%%%%%%%%%%%%%%%%%%%%%%%%%%%%%%
\section{Application of transverse thermoelectric conversion} \label{Sec:Application}
%%%%%%%%%%%%%%%%%%%%%%%%%%%%%%%%%%%%%%%%%

Although the transverse thermoelectric conversion is not yet in practical use, experimental verification is being actively conducted to realize next-generation heat flux sensing and energy harvesting technologies by utilizing the various characteristics introduced in Sec. \ref{Sec:Classification}. In this section, we show examples of experiments aimed towards applications of the transverse thermoelectric conversion, followed by presenting new concepts and future prospects for further performance improvements. 

%%%%%%%%%%%%%%%%%%%%%%%%%%%%%%%%%%%%% 
\subsection{Heat flux sensing} \label{Sec:heat-flux}
%%%%%%%%%%%%%%%%%%%%%%%%%%%%%%%%%%%%% 

A heat flux sensor is a device enabling simultaneous detection of the magnitude and direction of a heat flow, which can be an essential component for smart thermal management systems. However, practical applications of heat flux sensors have been limited due to several problems. The commercially available sensors are based on the Seebeck effect and consist of a serially connected three-dimensional array of two different thermoelectric materials. Therefore, the heat flux sensors based on the Seebeck effect require a durable substrate or thick plate to provide mechanical stability. Hence, the conventional sensors are mainly applicable to flat surfaces and their mechanical flexibility is limited. Due to the presence of thick substrates, these sensors have large thermal resistance that changes the heat flow distribution to be detected. Although the sensibility of the heat flux sensors based on the Seebeck effect is proportional to the number of the thermoelectric material junctions and to the sensor size, the complex structure and low mechanical durability make it difficult to construct highly sensitive sensors. As an alternative to the sensors having these drawbacks, thin-film-based heat flux sensors driven by the transverse thermoelectric effects, {\it e.g.} SSE-driven ISHE \cite{Kirihara16} and ANE \cite{Uchida-Zhou-Sakuraba21,Zhou-Sakuraba20,Modak22,Tanaka23}, were proposed and demonstrated. Owing to its simpler structure and improved transverse thermopower, ANE-based heat flux sensors have recently become mainstream. In the ANE-based sensor, a lateral thermopile structure consisting of alternately arranged and serially connected two different wires with different $S_{\rm ANE}$ values is used. The simple thermopile structure and the symmetry of ANE make the sensors thin, reducing their thermal resistance. If magnetic materials showing large ANE can be formed on thin flexible sheets, flexible heat flux sensors can be constructed \cite{Zhou-Sakuraba20,Modak22}. These features of the ANE-based heat flux sensors are advantageous over the conventional heat flux sensors based on the Seebeck effect. 

%%%%%%%%%%%%%%%%%%%%%%%%%%%%%%%%%%%%% 
%\begin{figure}[tb]
\begin{figure} 
  \begin{center}
    \includegraphics[width=5.9cm]{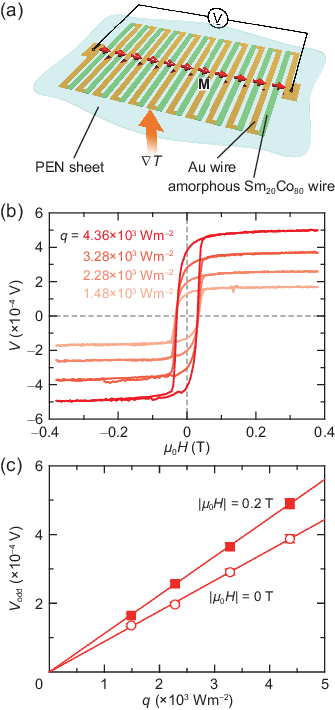}
  \end{center}
  \caption{(a) Schematic of the ANE-based heat flux sensor. The sensor comprises the amorphous Sm$_{20}$Co$_{80}$ wires and Au wires arranged alternately and connected in series formed on a flexible polyethylene naphthalate (PEN) sheet. ${\bf q}$ denotes the heat flux density with the magnitude $q$. (b) $H$ dependence of $V$ for the Sm$_{20}$Co$_{80}$/Au sensor for various values of $q$. (c) $q$ dependence of $V_{\rm odd}$ at $|\mu_0 H| = 0.2$ and 0.0 T. The data in (b) and (c) are reconstructed from Fig. 6 of Ref. \cite{Modak22}. }
  \label{fig:heat-flux}
\end{figure}
%%%%%%%%%%%%%%%%%%%%%%%%%%%%%%%%%%%%

The sensitivity of the heat flux sensor is determined by the output thermoelectric voltage over a heat flux density $q$. Thus, the open-circuit voltage is more important than the output power. To increase the open-circuit voltage due to ANE $V_{\rm ANE}$, one often uses an ANE-based heat flux sensor consisting of a thermopile structure in which one end of a magnetic wire is electrically connected to the opposite end of an adjacent magnetic wire [Fig. \ref{fig:heat-flux}(a)]. In this zigzag-shaped thermopile structure, when the magnetization (heat flux) is along the width (thickness) direction of the wires, {\it i.e.} in the IM configuration, the ANE voltage is generated along the length of the wires $l$ and the total output voltage is proportional to the number of the connected magnetic wires $n$. This structure is thus more advantageous for increasing the open-circuit voltage than plain blocks or sheets. The sensitivity of the ANE-based heat flux sensor $V_{\rm ANE}/q$ is thus proportional to $(S_{\rm ANE}ln)/\kappa$ with $\kappa$ being the thermal conductivity of the magnetic material in the out-of-plane direction. 

Here, we show a proof-of-concept demonstration of a flexible heat flux sensor based on ANE. We deposited the 100-nm-thick amorphous Sm$_{20}$Co$_{80}$ film on a 50-$\mu$m-thick flexible polyethylene naphthalate (PEN) sheet, where amorphous Sm$_{20}$Co$_{80}$ exhibits substantial $S_{\rm ANE}$, coercivity, and remanent magnetization and can be formed on any substrate \cite{Modak22}. We formed 50 Sm$_{20}$Co$_{80}$ wires with a width of 50 $\mu$m and a length of 10 mm arranged in a parallel configuration at an interval of 150 $\mu$m. The wires were then connected in series using Au wires to make the total device area of $10 \times 10~{\rm mm}^2$, where $q$ was estimated through the normalization by this area. The total length of the Sm$_{20}$Co$_{80}$ wires for this configuration reaches 500 mm. We measured the voltage $V$ between the ends of the thermopile structure with applying an in-plane external magnetic field along the width direction of the wires and an out-of-plane heat current. The details of the experimental procedures are shown in Ref. \cite{Modak22}. 

As shown in Fig. \ref{fig:heat-flux}(b), clear ANE signals with hysteresis behaviors were observed in the $H$ dependence of $V$ and their magnitude increases with increasing $q$ across the device. The saturated $V$ values for $\mu_0 H > 0.1~{\rm T}$ indicate the negligibly small ONE contribution, enabling the simple extraction of the ANE contribution. Figure \ref{fig:heat-flux}(c) shows the ANE voltage, represented by $V_{\rm odd}$, as a function of $q$ \cite{Modak22}, where $V_{\rm odd}$ at 0.2 T was extracted by the field-odd component of $V$ at $|\mu_0 H| = 0.2~{\rm T}$ and $V_{\rm odd}$ at zero field was extracted from the zero-field $V$ values for the positive-to-negative and negative-to-positive sweep of $H$. The sensitivity of the ANE-based heat flux sensor was estimated to be $V_{\rm ANE}/q \sim V_{\rm odd}/q = 1.12 \pm 0.02 \times 10^{-7}~{\rm VW}^{-1}{\rm m}^2$ and $0.89 \pm 0.02 \times 10^{-7}~{\rm VW}^{-1}{\rm m}^2$ in the saturation magnetization (0.2 T) and remanent magnetization (0.0 T) states, respectively. Importantly, the use of magnetic thin films with finite coercivity and remanent magnetization makes it possible to sense a heat flux using ANE even in the absence of an external magnetic field. Therefore, the design and control of magnetic anisotropy are essential for the application of ANE. Although the sensitivity of the sensor used in this demonstration is lower than that of commercial sensors based on the Seebeck effect, ANE has strong advantages in flexibility and low thermal resistance, extending applications of heat flux sensors. To further improve the performance of the ANE-based heat flux sensors, extensive efforts have been devoted to developing new materials with larger $S_{\rm ANE}$ and lower $\kappa$ as well as to optimizing the device structure. Since amorphous materials can be fabricated on any surface and have lower $\kappa$ than crystalline materials, they are promising candidates for the development of low-cost, high-performance heat flux sensors with large areas \cite{Gautam24,Park25,Park-ArXiv1,Park-ArXiv2,Modak22}. 

In addition to ANE, the off-diagonal Seebeck effect in anisotropic single-crystal metals \cite{McAfee23} and conductive layered oxides formed on off-cut substrates with tilted crystal planes \cite{Renk94,Roediger08,Zhang08,Takahashi09,Kanno14} is also expected to be used as a thermal sensor. Such artificially tilted conductive oxides are also being developed for other applications including light sensing and THz emission \cite{Yordanov23} via the transverse thermoelectric conversion. 

%%%%%%%%%%%%%%%%%%%%%%%%%%%%%%%%%%%%% 
\subsection{Thermal energy harvesting}
%%%%%%%%%%%%%%%%%%%%%%%%%%%%%%%%%%%%% 

Transverse thermoelectric devices are useful for harvesting and utilizing widely distributed thermal energy. By installing simple block- or sheet-shaped transverse thermoelectric devices on a heat source surface, one can extract a charge current and voltage along the surface direction from a temperature gradient perpendicular to the surface. Kirihara et al. pointed out that SSE-driven ISHE is suitable for the thermal energy harvesting from curved or complex-shaped heat source surfaces and demonstrated that SSE devices can be fabricated using simple coating \cite{Kirihara12} or spray \cite{Kirihara16} processes. Similar functionality can be obtained using ONE (ANE) in the in-plane magnetic field (magnetized) configuration, enabling the fabrication of flexible transverse thermoelectric devices based on thin films deposited on flexible substrates \cite{Modak22} or amorphous metal ribbons \cite{Gautam24,Park25,Park-ArXiv1,Park-ArXiv2}. However, challenges include the difficulty of extracting a large charge current and electrical power from thin films and the fact that the thermopower generated by ONE, ANE, and SSE-driven ISHE has not yet reached sufficient levels for practical applications. 

%%%%%%%%%%%%%%%%%%%%%%%%%%%%%%%%%%%%% 
%\begin{figure}[tb]
\begin{figure} 
  \begin{center}
    \includegraphics[width=7.5cm]{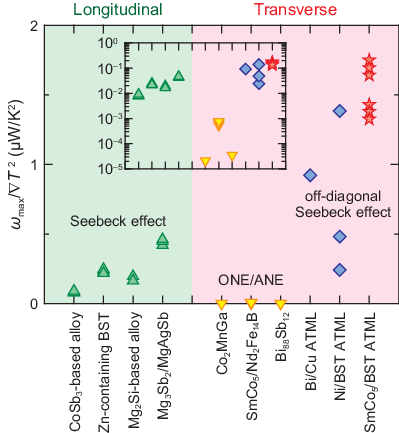}
  \end{center}
  \caption{Comparison of the maximum power density per heat transfer area $\omega_{\rm max}$ normalized by the square of the applied temperature gradient $\nabla T^2$ for various longitudinal and transverse thermoelectric modules. The modules corresponding to the green triangle, yellow triangle, and blue diamond data points are driven by the Seebeck effect \cite{Deng18,Skomedal16,Liu21,Salvador14}, ONE/ANE \cite{Murata21,Ando24,Chen24}, and off-diagonal Seebeck effect \cite{Kanno12,Takahashi13,Sakai14,Kanno14,Ando25}. The red stars show the data for the off-diagonal Seebeck effect in SmCo$_5$/BST ATML, which exhibits the current record-high $\omega_{\rm max} / \nabla T^2$ value \cite{Ando25}. The inset shows the same data on a logarithmic scale. }
  \label{fig:power-density}
\end{figure}
%%%%%%%%%%%%%%%%%%%%%%%%%%%%%%%%%%%%

Experimental demonstrations of thermal energy harvesting using the off-diagonal Seebeck effect, which exhibits high transverse thermoelectric conversion performance, have also been reported. Figure \ref {fig:power-density} compares the maximum power density normalized by the square of the temperature gradient ($\propto$ the power factor) for thermoelectric conversion modules driven by various principles \cite{Deng18,Skomedal16,Liu21,Salvador14,Murata21,Ando24,Chen24,Kanno12,Takahashi13,Sakai14,Kanno14,Ando25}. The maximum power density of the ATML-based modules driven by the off-diagonal Seebeck effect stands out, being 2-3 orders of magnitude larger than that of the modules driven by ONE and ANE. Recently developed SmCo$_5$/BST ATML exhibits excellent transverse thermoelectric conversion performance, with the maximum power density of the modules composed of 14 ATML elements reaching 56.7 mW/cm$^2$ at a temperature difference of 152 K \cite{Ando25}. SmCo$_5$/BST ATML has higher thermal conductivity than longitudinal thermoelectric materials, making it difficult to maintain a large temperature difference in steady-state conditions; the efficiency of the ATML-based module falls short of that of longitudinal modules driven by the Seebeck effect. However, the maximum power density normalized by the square of the temperature gradient in ATMLs exceeds that of the longitudinal modules based on the Seebeck effect (Fig. \ref{fig:power-density}). Therefore, ATML-based transverse thermoelectric modules are suitable for thermal energy harvesting because they can generate large power even in environments where only small temperature differences can be generated. Additionally, in transverse thermoelectric modules with thermopile structures, by alternately changing the height of each element, a built-in heat sink can be incorporated to promote heat release to the atmosphere \cite{Ando25}. Such a built-in heat sink structure cannot be achieved in longitudinal modules where rigid plates are present on the top and bottom surfaces. Even without sandwiching the thermoelectric module between hot and cold heat baths, the built-in heat sink structure enables substantially large power generation simply by placing it on a heat source owing to the improved heat release to the atmosphere. In particular, SmCo$_5$/BST ATML has remanent magnetization and functions as a permanent magnet; when it can be attached to a heat source through a magnetic attractive force, the contact thermal resistance between the module and heat source decreases even when thermal interface materials are inserted between them, enabling the effective generation of a temperature difference from the heat source. As shown in Ref. \cite{Ando25}, when the thermopile module comprising SmCo$_5$/BST ATML elements is placed on a hot plate, the electrical power generated solely by heat release to the air reaches a value sufficient to drive small sensors. 

%%%%%%%%%%%%%%%%%%%%%%%%%%%%%%%%%%%%% 
\subsection{Hybrid transverse magneto-thermoelectric conversion}
%%%%%%%%%%%%%%%%%%%%%%%%%%%%%%%%%%%%% 

In this subsection, we introduce one of the promising strategies for improving transverse thermoelectric conversion performance: hybrid transverse magneto-thermoelectric conversion \cite{Uchida24,Hirai24,Lee25}. As classified in Fig. \ref {fig:classification}, there are various driving principles for the transverse thermoelectric conversion, but the phenomena with time-reversal symmetry breaking and those with structural symmetry breaking have been studied independently in different fields. The interdisciplinary fusion of these phenomena not only leads to the creation of new research fields but also significantly contributes to improving the transverse thermoelectric conversion performance. Here, imagine a situation where multiple transverse thermoelectric conversion principles manifest in a single material, {\it i.e.} either a homogeneous or composite/hybrid material. As an example, if the transverse thermopower due to the off-diagonal Seebeck effect $S_{\rm OD}$ and that due to ONE $S_{\rm ONE}$ are simultaneously manifested, the total transverse thermopower is simply the sum of these two contribution: $S_{\rm OD} + S_{\rm ONE}$. Since $|S_{\rm OD}| > |S_{\rm ONE}|$ in many cases, the contribution of ONE to the total transverse thermopower is small. In this situation, the figure of merit is proportional to the square of the total transverse thermopower, and described as
%%%
\begin{eqnarray}
  z_{zy}T &\propto& (S_{\rm OD} + S_{\rm ONE})^2  \nonumber \\
  &=& S_{\rm OD}^2 + S_{\rm ONE}^2 + 2 S_{\rm OD} S_{\rm ONE}.
  \label{eq:hybrid}
\end{eqnarray}
%%%
The third term on the right-hand side of Eq. (\ref{eq:hybrid}) is important. Even if $S_{\rm ONE}$ is small, the modulation of the figure of merit due to ONE increases significantly owing to the synergistic effect with $S_{\rm OD}$. In this example, by increasing the magnetic-field/magnetization-independent $S_{\rm OD}$ contribution, the influence of magnetic fields or magnetization on the figure of merit can be enhanced. If three or more transverse thermoelectric conversion phenomena simultaneously occur in a single material, more diverse synergistic effects can be obtained. 

%%%%%%%%%%%%%%%%%%%%%%%%%%%%%%%%%%%%% 
%\begin{figure*}[tb]
\begin{figure*} 
  \begin{center}
    \includegraphics[width=17.4cm]{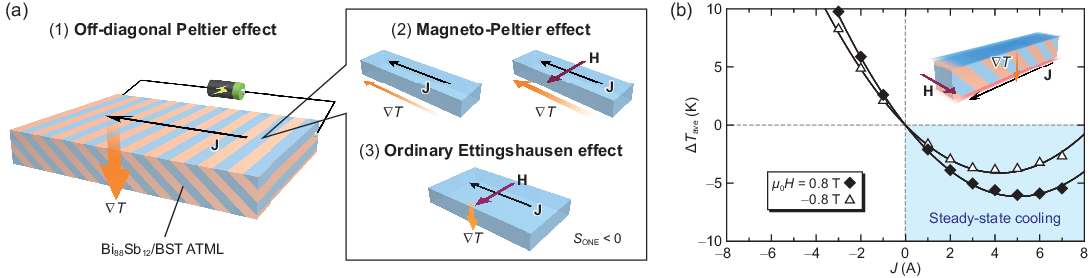}
  \end{center}
  \caption{(a) Schematic of the hybrid transverse magneto-thermoelectric cooling based on the off-diagonal Peltier, magneto-Peltier, and ordinary Ettingshausen effects in ATML. (b) Charge current $J$ dependence of $\Delta T_{\rm ave}$ at $\mu_0 H = \pm0.8~{\rm T}$ for Bi$_{88}$Sb$_{12}$/BST ATML. $\Delta T_{\rm ave}$ denotes the steady-state temperature change from room temperature averaged over one Bi$_{88}$Sb$_{12}$/BST pair. }
  \label{fig:hybrid}
\end{figure*}
%%%%%%%%%%%%%%%%%%%%%%%%%%%%%%%%%%%%

The effectiveness of the hybrid transverse magneto-thermoelectric conversion has been experimentally demonstrated in several systems. In Ref. \cite{Uchida24}, ATML composed of Bi$_{88}$Sb$_{12}$, which exhibits the large magneto-thermoelectric effects, and BST, which exhibits the large Seebeck and Peltier effects, was used to enhance the transverse thermoelectric conversion performance by applying a magnetic field. The observed performance enhancement in this system originates from the hybrid thermoelectric conversion involving three effects: the off-diagonal Seebeck (Peltier), magneto-Seebeck (Peltier), and ordinary Nernst (Ettingshausen) effects [Fig. \ref{fig:hybrid}(a)]. Figure \ref{fig:hybrid}(b) shows the charge current $J$ dependence of the surface temperature averaged over the Bi$_{88}$Sb$_{12}$/BST junction in ATML under the steady state, where the contribution of Joule heating results in a parabolic behavior and that of the thermoelectric effects results in the shift of the parabolic curves. The systematic experimental results reported in Ref. \cite{Uchida24} indicate that a temperature gradient is generated in a direction perpendicular to the applied charge current and the steady-state cooling performance due to the transverse thermoelectric conversion is dramatically modulated by the reversal of a magnetic field. In the experimental setup depicted in Fig. \ref{fig:hybrid}(b), when a positive magnetic field is applied, the direction of the net heat current generated by the off-diagonal Peltier effect is the same as that by the ordinary Ettingshausen effect, improving the transverse thermoelectric conversion performance. The figure of merit at room temperature for Bi$_{88}$Sb$_{12}$/BST ATML is 0.20 (0.14) when a positive (negative) magnetic field is applied, and this field-dependent modulation is much larger than the figure of merit for ONE alone in the Bi$_{88}$Sb$_{12}$ alloy. This is attributed to the contribution of the cross term in Eq. (\ref{eq:hybrid}). Furthermore, in this study, it has been demonstrated that by replacing BST in ATML with the permanent magnet Nd$_2$Fe$_{14}$B, the magneto-thermoelectric effects can be superimposed without applying an external magnetic field due to the stray magnetic field from the remanent magnetization of Nd$_2$Fe$_{14}$B. In Ref. \cite{Hirai24}, the candidate phenomena for the hybrid transverse magneto-thermoelectric conversion have been expanded by inducing ANE in ATML comprising the topological ferromagnet Co$_2$MnGa. In Co$_2$MnGa-based ATMLs combined with $p$-type BST and with $n$-type Bi$_2$Te$_3$, a magnetization-direction-dependent modulation of the figure of merit significantly larger than that of ANE alone was observed owing to the hybridization of the off-diagonal Seebeck effect and ANE. In Ref. \cite{Lee25}, the hybridization of the off-diagonal Seebeck effect and ANE has been realized in SmCo$_5$/BST ATML in the absence of an external magnetic field by utilizing the large in-plane remanent magnetization of SmCo$_5$. Note that in these ATMLs, the transverse thermopower generated by ONE (ANE), $S_{\rm ONE}$ ($S_{\rm ANE}$), is effectively reduced because of the shunting effect in adjacent conductors. Nevertheless, the gain from the cross term in Eq. (\ref{eq:hybrid}) is greater than the reduction in $S_{\rm ONE}$ ($S_{\rm ANE}$). 

Examples of the hybrid transverse magneto-thermoelectric conversion are still limited. In principle, the Seebeck-effect-driven ordinary/anomalous Hall effects and SSE/SdSE-driven ISHEs, not used in the above examples, are also integrable into the hybrid transverse magneto-thermoelectric conversion. In fact, although the thermoelectric output is small, the hybrid transverse thermoelectric generation based on the combination of ANE and SSE-driven ISHE has been demonstrated \cite{Uchida16,Boona16,Kikkawa13-PRB}. Originally, spin-caloritronic phenomena such as ANE and SSE have been extensively studied from the viewpoint of fundamental physics, but their low thermoelectric conversion performance has limited their impact on applications. The concept of the hybrid transverse magneto-thermoelectric conversion provides one of the solutions for practical applications of such spin-caloritronic phenomena. 

%%%%%%%%%%%%%%%%%%%%%%%%%%%%%%%%%%%%% 
\subsection{Thermoelectric permanent magnet}
%%%%%%%%%%%%%%%%%%%%%%%%%%%%%%%%%%%%% 

\begin{figure}[h] 
  \begin{center}
    \includegraphics[width=7.7cm]{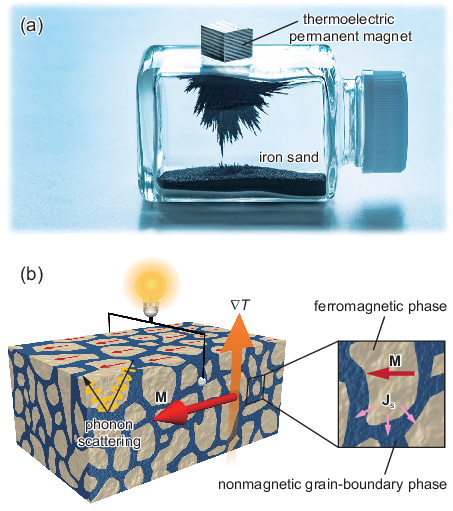}
  \end{center}
  \caption{(a) Photograph of a prototype thermoelectric permanent magnet, used in Ref. \cite{Uchida24}. (b) Schematic of the microstructure of a thermoelectric permanent magnet. }
  \label{fig:thermoelectric-permanent-magnet}
\end{figure}

The transverse thermoelectric conversion in permanent magnets is suitable for power generation and cooling applications as they enable the construction of large-area devices and their mass production. Their roles discussed so far are summarized as follows:
\begin{itemize}
  \item Manifestation of the giant anomalous Nernst and Ettingshausen effects \cite{Miura19,Miura20}, 
  \item Zero-magnetic-field operation of the anomalous Nernst and Ettingshausen effects \cite{Miura19,Miura20,Ando24,Lee25}, 
  \item Zero-magnetic-field operation of the magnetic-field-induced magneto-thermoelectric effects, {\it i.e.} the ordinary Nernst and Ettingshausen effects and the magneto-Seebeck and Peltier effects, utilizing stray magnetic fields from remanent magnetization \cite{Murata24,Uchida24}, 
  \item Reduction of contact thermal resistance between thermoelectric devices and heat sources through a magnetic attractive force \cite{Ando25}. 
\end{itemize}
Therefore, the creation of ``thermoelectric permanent magnets’’ that possess high transverse thermoelectric conversion performance with permanent magnet features is of importance. However, the above roles are based on the macroscopic properties of permanent magnets [Fig. \ref{fig:thermoelectric-permanent-magnet}(a)]. By reexamining permanent magnets from a microscopic perspective and engineering not only their magnetic properties but also their thermoelectric conversion characteristics, new prospects for the transverse thermoelectric conversion become apparent. 

Permanent magnets are not homogeneous materials but rather composite materials or engineered ensembles of interfaces with micrometer- to sub-micrometer-scale structures [Fig. \ref{fig:thermoelectric-permanent-magnet}(b)]. The goal of conventional permanent magnet development has been to achieve large maximum energy product $(BH)_{\rm max}$, and the optimization of microstructures has significantly improved the coercive force, remanent magnetization, and their temperature-dependent characteristics \cite{Hono12}. The design guidelines for the thermoelectric permanent magnets are completely different from those for conventional magnets; it is necessary to design and control microstructures to improve the transverse thermoelectric conversion characteristics. For example, by incorporating thermoelectric semiconductors into the microstructure of permanent magnets, the appearance of the Seebeck-effect-driven ordinary/anomalous Hall effects in the magnets is expected. Introducing spin-orbit interaction near the interface between ferromagnetic phases and nonmagnetic grain-boundary phases may result in the superposition of the transverse thermopower due to SSE/SdSE-driven ISHEs [see the right magnified illustration in Fig. \ref{fig:thermoelectric-permanent-magnet}(b)]. Furthermore, by appropriately designing the grain size and distribution, it may be possible to simultaneously reduce the phonon thermal conductivity \cite{Oyanagi25,Nomura18} [see the left illustration in Fig. \ref{fig:thermoelectric-permanent-magnet}(b)]. In other words, by optimally designing the microstructures of the thermoelectric permanent magnets, it is possible to independently engineer the transverse thermopower, electrical conductivity, and thermal conductivity, thereby significantly improving the figure of merit and energy conversion efficiency. To achieve this, it will be necessary to go beyond mere material exploration, and instead focus on the design of hierarchical structures and the control of phase boundaries within permanent magnets, with the aid of microstructure analysis techniques and precise transport property measurements. Microstructure engineering of magnetic materials aimed at realizing high-performance transverse thermoelectric conversion has already begun \cite{Oyanagi25,Gautam24,Park-ArXiv1,Park-ArXiv2,Kautsar25}.

%%%%%%%%%%%%%%%%%%%%%%%%%%%%%%%%%%%%% 
%\begin{figure}[tb]
%%%%%%%%%%%%%%%%%%%%%%%%%%%%%%%%%%%%

%%%%%%%%%%%%%%%%%%%%%%%%%%%%%%%%%%%%%%%%%
\section{Conclusions and prospects} \label{Sec:Conclusion}
%%%%%%%%%%%%%%%%%%%%%%%%%%%%%%%%%%%%%%%%%

In this review article, we have introduced various phenomena and principles that realize the transverse thermoelectric conversion, reclassified them into 8 categories based on modern knowledge (Fig. \ref{fig:classification}), and summarized their characteristics. We have formulated the figure of merit and efficiency for the transverse thermoelectric generator, the maximum attainable temperature difference for the thermoelectric refrigerator, and the effective thermal conductivity for the transverse active cooler. Importantly, the thermal boundary conditions in the electric field direction play an important role to precisely discuss and evaluate the transverse thermoelectric conversion performance. Each phenomenon exhibits different thermoelectric conversion characteristics and the research stage is also completely different. At present, excellent $z_{xy}T$ values have been obtained through ONE in some semimetals at low temperatures and the off-diagonal Seebeck effect in ATMLs and goniopolar materials. However, there are also issues such as limitations on material selection and the need to apply a magnetic field for transverse thermoelectric conversion operations. While many new physical phenomena and principles have been discovered in the field of spin caloritronics, the efficiency of the transverse thermoelectric generation using magnetic materials and the spin degree of freedom is currently very low, and it remains at the basic research stage. Nevertheless, the benefits of the transverse thermoelectric conversion, which enables the construction of thermoelectric conversion devices without junctions, are immeasurably significant, and it is expected that basic and applied studies will accelerate further. Various transverse thermoelectric phenomena that have been investigated independently in different fields have been integrated in recent years, and methods for further improving the performance have been proposed and demonstrated. If the transverse thermoelectric conversion that has high versatility and efficiency is achieved through material exploration, material development, microstructure engineering, optimal design of hybrid materials, and modularization technologies, one can overcome the technical issues faced by existing thermoelectric conversion modules.

%%%%%%%%%%%%%%%%%%%%%%%%%%%%%%%%%%%%%%%%%
\section*{Acknowledgments}
%%%%%%%%%%%%%%%%%%%%%%%%%%%%%%%%%%%%%%%%%

The authors thank Y. Goto, J. P. Heremans, K. Hirata, M. Hirschberger, Y. Lee, Y. Mizuguchi, S. Mori, M. Murata, H. Nagano, Y. Nakanishi, K. Oyanagi, S. J. Park, Y. Sakuraba, Y. Sato, H. Sepehri-Amin, A. Takahagi, T. Yagi, and W. Zhou for valuable discussions and many collaborators and group members for preliminary reviews of the manuscript. The previously reported experimental results introduced in this article were obtained from the collaboration with H. Masuda, T. Ohkubo, Y. Sakuraba, T. Seki, H. Sepehri-Amin, K. Takanashi, T. Yagi, and W. Zhou. This work was mainly supported by ERATO ``Magnetic Thermal Management Materials'' (grant no. JPMJER2201) from Japan Science and Technology Agency.

%%%%%%%%%%%%%%%%%%%%%%%%%%%%%%%%%%%%%%%%%
%\section*{References}
%%%%%%%%%%%%%%%%%%%%%%%%%%%%%%%%%%%%%%%%%


\begin{thebibliography}{999}

%%%Introduction%%%

\bibitem{Yan22}
  Q. Yan and M. G. Kanatzidis, 
  Nat. Mater. {\bf 21}, 503 (2022). 

\bibitem{Snyder17}
  G. J. Snyder and A. H. Snyder, 
  Energy Environ. Sci. {\bf 10}, 2280 (2017). 

\bibitem{Hao16}
  F. Hao, P. Qiu, Y. Tang, S. Bai, T. Xing, H.-S. Chu, Q. Zhang, P. Lu, T. Zhang, D. Ren, J. Chen, X. Shi, and L. Chen, 
  Energy Environ. Sci., {\bf 9}, 3120 (2016).

\bibitem{Deng18}
  R. Deng, X. Su, S. Hao, Z. Zheng, M. Zhang, H. Xie, W. Liu, Y. Yan, C. Wolverton, C. Uher, M. G. Kanatzidis, and X. Tang, 
  Energy Environ. Sci., {\bf 11}, 1520 (2018). 

\bibitem{KELK}
  https://www.kelk.co.jp/english/generation/index.html

\bibitem{Jia22}
  B. Jia, Y. Huang, Y. Wang, Y. Zhou, X. Zhao, S. Ning, X. Xu, P. Lin, Z. Chen, B. Jiang, and J. He, 
  Energy Environ. Sci. {\bf 15}, 1920 (2022).

\bibitem{Jiang22}
B. Jiang, W. Wang, S. Liu, Y. Wang, C. Wang, Y. Chen, L. Xie, M. Huang, and J. He,  
Science {\bf 377}, 208 (2022). 

\bibitem{Ando23}
F. Ando, H. Tamaki, Y. Matsumura, T. Urata, T. Kawabe, R. Yamamura, Y. Kaneko, R. Funahashi, and T. Kanno, 
Mater. Today Phys. {\bf 36}, 101156 (2023). 

\bibitem{Skomedal16}
  G. Skomedal, L. Holmgren, H. Middleton, I. S. Eremin, G. N. Isachenko, M. Jaegle, K. Tarantik, N. Vlachos, M. Manoli, T. Kyratsi, D. Berthebaud, N. Y. Dao Truong, and F. Gascoin, 
  Energy Convers. Manag. {\bf 110}, 13 (2016).

\bibitem{Liu21}
  Z. Liu, N. Sato, W. Gao, K. Yubuta, N. Kawamoto, M. Mitome, K. Kurashima, Y. Owada, K. Nagase, C.-H. Lee, J. Yi, K. Tsuchiya, and T. Mori,
  Joule {\bf 5}, 1196 (2021).

\bibitem{Wang24}
  L. Wang, W. Zhang, S. Y. Back, N. Kawamoto, D. Hieu Nguyen, and T. Mori, 
  Nat. Commun. {\bf 15}, 6800 (2024). 

\bibitem{Zong17}
  P. Zong, R. Hanus, M. Dylla, Y. Tang, J. Liao, Q. Zhang, G. J. Snyder, and L. Chen, 
  Energy Environ. Sci. {\bf 10}, 183 (2017). 

\bibitem{Yu20}
  J. Yu, Y. Xing, C. Hu, Z. Huang, Q. Qiu, C. Wang, K. Xia, Z. Wang, S. Bai, X. Zhao, L. Chen, and T. Zhu, 
  Adv. Energy Mater. {\bf 10}, 2000888 (2020).

\bibitem{Xing20}
  Y. Xing, R. Liu, J. Liao, C. Wang, Q. Zhang, Q. Song, X. Xia, T. Zhu, S. Bai, and L. Chen, 
  Joule {\bf 4}, 2475 (2020).

\bibitem{Rowe98}
  D. M. Rowe and G. Min, 
  J. Power Sources {\bf 73}, 193 (1998). 

\bibitem{Shittu20}
  S. Shittu, G. Li, X. Zhao, and X. Ma, 
  Appl. Energy {\bf 268}, 115075 (2020).  

\bibitem{Uchida-Heremans22}
  K. Uchida and J. P. Heremans,
  Joule {\bf 6}, 2240 (2022).

\bibitem{Mizuguchi-Nakatsuji19}
  M. Mizuguchi and S. Nakatsuji, 
  Sci. Technol. Adv. Mater. {\bf 20}, 262 (2019). 

%%%Spin caloritronics%%%

\bibitem{Bauer12}
  G. E. W. Bauer, E. Saitoh, and B. J. van Wees, 
  Nat. Mater. {\bf 11}, 391 (2012).

\bibitem{Boona14}
  S. R. Boona, R. C. Myers, and J. P. Heremans, 
  Energy Environ. Sci. {\bf 7}, 885 (2014).

\bibitem{Uchida21}
  K. Uchida,
  Proc. Jpn. Acad. Ser. B {\bf 97}, 69 (2021).

%%%ODSE%%%

\bibitem{Babin74}
  V. P. Babin, T. S. Gudkin, Z. M. Dashevskii, L. D. Dudkin, E. K. Iordanishvili, V. I. Kaidanov, N. V. Kolomoets, O. M. Narva, and L. S. Stil’bans, 
  Sov. Phys. Semicond. {\bf 8}, 478 (1974).

\bibitem{Goldsmid11}
  H. J. Goldsmid, 
  J. Electron. Mater. {\bf 40}, 1254 (2011). 

\bibitem{He19}
  B. He, Y. Wang, M. Q. Arguilla, N. D. Cultrara, M. R. Scudder, J. E. Goldberger, W. Windl, and J. P. Heremans, 
  Nat. Mater. {\bf 18}, 568 (2019).

\bibitem{Scudder21}
  M. R. Scudder, B. He, Y. Wang, A. Rai, D. G. Cahill, W. Windl, J. P. Heremans, and J. E. Goldberger,
  Energy Environ. Sci. {\bf 14}, 4009 (2021).

\bibitem{Scudder22}
  M. R. Scudder, K. G. Koster, J. P. Heremans, and J. E. Goldberger,
  Appl. Phys. Rev. {\bf 9}, 021420 (2022).  

\bibitem{Zhou13}
  C. Zhou, S. Birner, Y. Tang, K. Heinselman, and M. Grayson, 
  Phys. Rev. Lett. {\bf 110}, 227701 (2013). 

\bibitem{Tang15}
  Y. Tang, B. Cui, C. Zhou, and M. Grayson, 
  J. Electron. Mater. {\bf 44}, 2095 (2015).

%%%Transverse Thomson%%%

\bibitem{Takahagi25}
  A. Takahagi, T. Hirai, A. Alasli, S. J. Park, H. Nagano, and K. Uchida,
  Nat. Phys. {\bf 21}, 1283 (2025). 

%%%ONE%%%

\bibitem{Ettingshausen-Nernst86}
  A. V. Ettingshausen and W. Nernst, 
  Ann. Phys. {\bf 265}, 343 (1886).

\bibitem{Jandl-Birkholz94}
  P. Jandl and U. Birkholz, 
  J. Appl. Phys. {\bf 76}, 7351 (1994).

\bibitem{Pan22-ONE}
  Y. Pan, B. He, T. Helm, D. Chen, W. Schnelle, and C. Felser, 
  Nat. Commun. {\bf 13}, 3909 (2022). 

\bibitem{Li22}
  P. Li, P. Qiu, Q. Xu, J. Luo, Y. Xiong, J. Xiao, N. Aryal, Q. Li, L. Chen, and X. Shi, 
  Nat. Commun. {\bf 13}, 7612 (2022). 

\bibitem{Yang23}
  Y. Yang, Q. Tao, Y. Fang, G. Tang, C. Yao, X. Yan, C. Jiang, X. Xu, F. Huang, W. Ding, Y. Wang, Z. Mao, H. Xing, and Z.-A. Xu, 
  Nat. Phys. {\bf 19}, 379 (2023). 

\bibitem{Pasquale24}
  G. Pasquale, Z. Sun, G. Migliato Marega, K. Watanabe, T. Taniguchi, and A. Kis, 
  Nat. Nanotechnol. {\bf 19}, 941 (2024). 

\bibitem{Murata24}
  M. Murata, T. Hirai, T. Seki, and K. Uchida,
  Appl. Phys. Lett. {\bf 124}, 193901 (2024).

%%%ANE%%%

\bibitem{Uchida-Zhou-Sakuraba21}
  K. Uchida, W. Zhou, and Y. Sakuraba,
  Appl. Phys. Lett. {\bf 118}, 140504 (2021). 

\bibitem{Pu08}
  Y. Pu, D. Chiba, F. Matsukura, H. Ohno, and J. Shi, 
  Phys. Rev. Lett. {\bf 101}, 117208 (2008). 

\bibitem{He19-Joule}
  B. He, C. Şahin, S. R. Boona, B. C. Sales, Y. Pan, C. Felser, M. E. Flatt\'e, and J. P. Heremans, 
  Joule {\bf 5}, 3057 (2021). 

\bibitem{Toyama2024}
  R. Toyama, W. Zhou, and Y. Sakuraba, 
  Phys. Rev. B {\bf 109}, 054415 (2024). 

\bibitem{Sakai18}
  A. Sakai, Y. P. Mizuta, A. A. Nugroho, R. Sihombing, T. Koretsune, M.-T. Suzuki, N. Takemori, R. Ishii, D. Nishio-Hamane, R. Arita, P. Goswami, and S. Nakatsuji, 
  Nat. Phys. {\bf 14}, 1119 (2018). 

\bibitem{Reichlova18}
  H. Reichlova, R. Schlitz, S. Beckert, P. Swekis, A. Markou, Y.-C. Chen, D. Kriegner, S. Fabretti, G. H. Park, A. Niemann, S. Sudheendra, A. Thomas, K. Nielsch, C. Felser, and S. T. B. Goennenwein, 
  Appl. Phys. Lett. {\bf 113}, 212405 (2018).

\bibitem{Guin19}
  S. N. Guin, K. Manna, J. Noky, S. J. Watzman, C. Fu, N. Kumar, W. Schnelle, C. Shekhar, Y. Sun, J. Gooth, and C. Felser, 
  NPG Asia Mater. {\bf 11}, 16 (2019).

\bibitem{Sakuraba20}
  Y. Sakuraba, K. Hyodo, A. Sakuma, and S. Mitani, 
  Phys. Rev. B {\bf 101}, 134407 (2020). 

\bibitem{Sumida20}
  K. Sumida, Y. Sakuraba, K. Masuda, T. Kono, M. Kakoki, K. Goto, W. Zhou, K. Miyamoto, Y. Miura, T. Okuda, and A. Kimura, 
  Commun. Mater. {\bf 1}, 89 (2020). 

\bibitem{Oyanagi25}
  K. Oyanagi, H. Sepehri-Amin, K. Takamori, T. Tadano, T. Imamura, R. Nagasawa, K. Mahalingam, T. Hirai, F. Ando, Y. Sakuraba, S. Kobayashi, and K. Uchida, 
  Acta Mater. {\bf 296}, 121239 (2025). 

\bibitem{Nakayama19}
  H. Nakayama, K. Masuda, J. Wang, A. Miura, K. Uchida, M. Murata, and Y. Sakuraba, 
  Phys. Rev. Mater. {\bf 3}, 114412 (2019).

\bibitem{Zhou-Sakuraba20}
  W. Zhou and Y. Sakuraba, 
  Appl. Phys. Express {\bf 13}, 043001 (2020).

\bibitem{Sakai20}
  A. Sakai, S. Minami, T. Koretsune, T. Chen, T. Higo, Y. Wang, T. Nomoto, M. Hirayama, S. Miwa, D. Nishio-Hamane, F. Ishii, R. Arita, and S. Nakatsuji, 
  Nature {\bf 581}, 53 (2020).

\bibitem{Chen22}
  T. Chen, S. Minami, A. Sakai, Y. Wang, Z. Feng, T. Nomoto, M. Hirayama, R. Ishii, T. Koretsune, R. Arita, and S. Nakatsuji, 
  Sci. Adv. {\bf 8}, eabk1480 (2022).

\bibitem{Fujiwara23}
  K. Fujiwara, Y. Kato, H. Abe, S. Noguchi, J. Shiogai, Y. Niwa, H. Kumigashira, Y. Motome, and A. Tsukazaki, 
  Nat. Commun. {\bf 14}, 3399 (2023). 

\bibitem{Pan22-ANE}
  Y. Pan, C. Le, B. He, S. J. Watzman, M. Yao, J. Gooth, J. P. Heremans, Y. Sun, and C. Felser, 
  Nat. Mater. {\bf 21}, 203 (2022). 

\bibitem{Guin19-AM}
  S. N. Guin, P. Vir, Y. Zhang, N. Kumar, S. J. Watzman, C. Fu, E. Liu, K. Manna, W. Schnelle, J. Gooth, C. Shekhar, Y. Sun, and C. Felser, 
  Adv. Mater. {\bf 31}, 1806622 (2019). 

\bibitem{Ding19}
  L. Ding, J. Koo, L. Xu, X. Li, X. Lu, L. Zhao, Q. Wang, Q. Yin, H. Lei, B. Yan, Z. Zhu, and K. Behnia, 
  Phys. Rev. X {\bf 9}, 041061 (2019). 

\bibitem{Noguchi24}
  S. Noguchi, K. Fujiwara, Y. Yanagi, M.-T. Suzuki, T. Hirai, T. Seki, K. Uchida, and A. Tsukazaki,
  Nat. Phys. {\bf 20}, 254 (2024). 

\bibitem{Asaba21}
  T. Asaba, V. Ivanov, S. M. Thomas, S.Y. Savrasov, J. D. Thompson, E. D. Bauer, and F. Ronning, 
  Sci. Adv. {\bf 7}, eabf1467 (2021).

\bibitem{Uchida15}
  K. Uchida, T. Kikkawa, T. Seki, T. Oyake, J. Shiomi, Z. Qiu, K. Takanashi, and E. Saitoh, 
  Phys. Rev. B {\bf 92}, 094414 (2015).

\bibitem{Seki21}
  T. Seki, Y. Sakuraba, K. Masuda, A. Miura, M. Tsujikawa, K. Uchida, T. Kubota, Y. Miura, M. Shirai, and K. Takanashi, 
  Phys. Rev. B {\bf 103}, L020402 (2021). 

\bibitem{Gautam24}
  R. Gautam, T. Hirai, A. Alasli, H. Nagano, T. Ohkubo, K. Uchida, and H. Sepehri-Amin,
  Nat. Commun. {\bf 15}, 2184 (2024). 

\bibitem{Park25}
  S. J. Park, K. M. Bang, J. Park, and H. Jin, 
  Appl. Therm. Eng. {\bf 265}, 125555 (2025). 

\bibitem{Park-ArXiv1}
  S. J. Park, H. Lee, J. M. Lee, J. Ha, H.-W. Lee, and H. Jin,
  arXiv:2411.04433

\bibitem{Park-ArXiv2}
  S. J. Park, R. Modak, R. Gautam, A. Alasli, T. Hirai, F. Ando, H. Nagano, H. Sepehri-Amin, and K. Uchida, 
  Acta Mater. {\bf 301}, 121422 (2025). 

\bibitem{Miura19}
  A. Miura, H. Sepehri-Amin, K. Masuda, H. Tsuchiura, Y. Miura, R. Iguchi, Y. Sakuraba, J. Shiomi, K. Hono, and K. Uchida,
  Appl. Phys. Lett. {\bf 115}, 222403 (2019). 

\bibitem{Miura20}
  A. Miura, K. Masuda, T. Hirai, R. Iguchi, T. Seki, Y. Miura, H. Tsuchiura, K. Takanashi, and K. Uchida,
  Appl. Phys. Lett. {\bf 117}, 082408 (2020). 

\bibitem{Lee04}
  W.-L. Lee, S. Watauchi, V. L. Miller, R. J. Cava, and N. P. Ong, 
  Phys. Rev. Lett. {\bf 93}, 226601 (2004). 

%%%AHE%%%

\bibitem{Nagaosa10}
  N. Nagaosa, J. Sinova, S. Onoda, A. H. MacDonald, and N. P. Ong, 
  Rev. Mod. Phys. {\bf 82}, 1539 (2010). 

%%%Topological NE%%%

\bibitem{Smejkal20}
  L. \v{S}mejkal, R. Gonz\'{a}lez-Hern\'{a}ndez, T. Jungwirth, and J. Sinova, 
  Sci. Adv. {\bf 6}, eaaz8809 (2020). 

\bibitem{Hirschberger20}
  M. Hirschberger, L. Spitz, T. Nomoto, T. Kurumaji, S. Gao, J. Masell, T. Nakajima, A. Kikkawa, Y. Yamasaki, H. Sagayama, H. Nakao, Y. Taguchi, R. Arita, T. Arima, and Y. Tokura, 
  Phys. Rev. Lett. {\bf 125}, 076602 (2020). 

\bibitem{Kolincio21}
  K. K. Kolincio, M. Hirschberger, J. Masell, S. Gao, A. Kikkawa, Y. Taguchi, T. Arima, N. Nagaosa, and Y. Tokura, 
  Proc. Natl. Acad. Sci. U.S.A. {\bf 118}, e2023588118 (2021). 

\bibitem{Khanh25}
  N. D. Khanh, S. Minami, M. M. Hirschmann, T. Nomoto, M.-C. Jiang, R. Yamada, N. Heinsdorf, D. Yamaguchi, Y. Hayashi, Y. Okamura, H. Watanabe, G.-Y. Guo, Y. Takahashi, S. Seki, Y. Taguchi, Y. Tokura, R. Arita, and M. Hirschberger, 
  Nat. Commun. {\bf 16}, 2654 (2025). 

%%%SAHE%%%

\bibitem{Zhou21}
  W. Zhou, K. Yamamoto, A. Miura, R. Iguchi, Y. Miura, K. Uchida, and Y. Sakuraba,
  Nat. Mater. {\bf 20}, 463 (2021).

\bibitem{Zhou23}
  W. Zhou, A. Miura, T. Hirai, Y. Sakuraba, and K. Uchida,
  Appl. Phys. Lett. {\bf 122}, 062402 (2023). 

\bibitem{Zhou24}
  W. Zhou, T. Sasaki, K. Uchida, and Y. Sakuraba, 
  Adv. Sci. {\bf 11}, 2308543 (2024). 

\bibitem{Yamamoto21}
  K. Yamamoto, R. Iguchi, A. Miura, W. Zhou, Y. Sakuraba, Y. Miura, and K. Uchida,
  J. Appl. Phys. {\bf 129}, 223908 (2021). 

%%%SSE and SdSE%%%

\bibitem{Uchida08}
  K. Uchida, S. Takahashi, K. Harii, J. Ieda, W. Koshibae, K. Ando, S. Maekawa, and E. Saitoh,
  Nature {\bf 455}, 778 (2008).

\bibitem{Uchida10-Nmat}
  K. Uchida, J. Xiao, H. Adachi, J. Ohe, S. Takahashi, J. Ieda, T. Ota, Y. Kajiwara, H. Umezawa, H. Kawai, G. E. W. Bauer, S. Maekawa, and E. Saitoh,
  Nat. Mater. {\bf 9}, 894 (2010). 

\bibitem{Jaworski10}
  C. M. Jaworski, J. Yang, S. Mack, D. D. Awschalom, J. P. Heremans, and R. C. Myers, 
  Nat. Mater. {\bf 9}, 898 (2010).

\bibitem{Uchida10-APL}
  K. Uchida, H. Adachi, T. Ota, H. Nakayama, S. Maekawa, and E. Saitoh,
  Appl. Phys. Lett. {\bf 97}, 172505 (2010). 

\bibitem{Azevedo05}
  A. Azevedo, L. H. Vilela Le$\tilde{a}$o, R. L. Rodriguez-Suarez, A. B. Oliveira, and S. M. Rezende, 
  J. Appl. Phys. {\bf 97}, 10C715 (2005). 

\bibitem{Saitoh06}
  E. Saitoh, M. Ueda, H. Miyajima, and G. Tatara, 
  Appl. Phys. Lett. {\bf 88}, 182509 (2006). 

\bibitem{Valenzuela06}
  S. O. Valenzuela and M. Tinkham, 
  Nature {\bf 442}, 176 (2006). 

\bibitem{Rojas-Sanchez13}
  J. R. Rojas S\'anchez, L. Vila, G. Desfonds, S. Gambarelli, J. Attan\'e, J. M. De Teresa, C. Magen, and A. Fert, 
  Nat. Commun. {\bf 4}, 2944 (2013). 

\bibitem{Yagmur16}
  A. Yagmur, S. Karube, K. Uchida, K. Kondou, R. Iguchi, T. Kikkawa, Y. Otani, and E. Saitoh,
  Appl. Phys. Lett. {\bf 108}, 242409 (2016). 

\bibitem{Xiao10}
  J. Xiao, G. E. W. Bauer, K. Uchida, E. Saitoh, and S. Maekawa,
  Phys. Rev. B {\bf 81}, 214418 (2010). 

\bibitem{Adachi11}
  H. Adachi, J. Ohe, S. Takahashi, and S. Maekawa, 
  Phys. Rev. B {\bf 83}, 094410 (2011). 

\bibitem{Adachi13}
  H. Adachi, K. Uchida, E. Saitoh, and S. Maekawa,
  Rep. Prog. Phys. {\bf 76}, 036501 (2013). 

\bibitem{Rezende14}
  S. M. Rezende, R. L. Rodr\'iguez-Su\'arez, R. O. Cunha, A. R. Rodrigues, F. L. A. Machado, G. A. Fonseca Guerra, J. C. Lopez Ortiz, and A. Azevedo, 
  Phys. Rev. B {\bf 89}, 014416 (2014). 

\bibitem{Kikkawa15}
  T. Kikkawa, K. Uchida, S. Daimon, Z. Qiu, Y. Shiomi, and E. Saitoh,
  Phys. Rev. B {\bf 92}, 064413 (2015). 

\bibitem{Jin15}
  H. Jin, S. R. Boona, Z. Yang, R. C. Myers, and J. P. Heremans, 
  Phys. Rev. B {\bf 92}, 054436 (2015). 

\bibitem{Kirihara12}
  A. Kirihara, K. Uchida, Y. Kajiwara, M. Ishida, Y. Nakamura, T. Manako, E. Saitoh, and S. Yorozu,
  Nat. Mater. {\bf 11}, 686 (2012).

\bibitem{Uchida16}
  K. Uchida, H. Adachi, T. Kikkawa, A. Kirihara, M. Ishida, S. Yorozu, S. Maekawa, and E. Saitoh,
  Proc. IEEE {\bf 104}, 1946 (2016).

\bibitem{Ramos15}
  R. Ramos, T. Kikkawa, M. H. Aguirre, I. Lucas, A. Anad\'on, T. Oyake, K. Uchida, H. Adachi, J. Shiomi, P. A. Algarabel, L. Morell\'on, S. Maekawa, E. Saitoh, and M. R. Ibarra,
  Phys. Rev. B {\bf 92}, 220407(R) (2015). 

\bibitem{Boona16}
  S. R. Boona, K. Vandaele, I. N. Boona, D. W. McComb, and J. P. Heremans, 
  Nat. Commun. {\bf 7}, 13714 (2016). 

\bibitem{Slachter10}
  A. Slachter, F. L. Bakker, J.-P. Adam, and B. J. van Wees, 
  Nat. Phys. {\bf 6}, 879 (2010). 

\bibitem{Iguchi18}
  R. Iguchi, A. Yagmur, Y.-C. Lau, S. Daimon, E. Saitoh, M. Hayashi, and K. Uchida,
  Phys. Rev. B {\bf 98}, 014402 (2018). 

%%%ATML%%%
  
\bibitem{Zahner98}
  Th. Zahner, R. F\"org, and H. Lengfellner, 
  Appl. Phys. Lett. {\bf 73}, 1364 (1998). 

\bibitem{Kyarad06}
  A. Kyarad and H. Lengfellner, 
  Appl. Phys. Lett. {\bf 89}, 192103 (2006). 

\bibitem{Kanno09}
  T. Kanno, S. Yotsuhashi, A. Sakai, K. Takahashi, and H. Adachi, 
  Appl. Phys. Lett. {\bf 94}, 061917 (2009). 

\bibitem{Kanno12}
  T. Kanno, A. Sakai, K. Takahashi, A. Omote, H. Adachi, and Y. Yamada, 
  Appl. Phys. Lett. {\bf 101}, 011906 (2012). 

\bibitem{Takahashi13}
  K. Takahashi, T. Kanno, A. Sakai, H. Tamaki, H. Kusada, and Y. Yamada, 
  Sci. Rep. {\bf 3}, 1501 (2013). 

\bibitem{Sakai14}
  A. Sakai, T. Kanno, K. Takahashi, H. Tamaki, H. Kusada, Y. Yamada, and H. Abe, 
  Sci. Rep. {\bf 4}, 6089 (2014). 

\bibitem{Mu19}
  X. Mu, H. Zhou, W. Zhao, D. He, W. Zhu, X. Nie, Z. Sun, and Q. Zhang, 
  J. Power Sources {\bf 430}, 193 (2019). 

\bibitem{Li20}
  Y. Li, P. Wei, H. Zhou, X. Mu, W. Zhu, X. Nie, X. Sang, and W. Zhao, 
  J. Electron. Mater. {\bf 49}, 5980 (2020). 

\bibitem{Zhou22}
  H. Zhou, H. Liu, G. Qian, H. Yu, X. Gong, X. Li, and J. Zheng, Micromachines {\bf 13}, 233 (2022). 

\bibitem{Yue22}
  K. Yue, W. Zhu, Q. He, X. Nie, X. Qi, C. Sun, W. Zhao, and Q. Zhang, 
  ACS Appl. Mater. Inter. {\bf 14}, 39053 (2022). 

\bibitem{Uchida24}
  K. Uchida, T. Hirai, F. Ando, and H. Sepehri-Amin,
  Adv. Energy Mater. {\bf 14}, 2302375 (2024). 

\bibitem{Hirai24}
  T. Hirai, F. Ando, H. Sepehri-Amin, and K. Uchida,
  Nat. Commun. {\bf 15}, 9643 (2024). 

\bibitem{Ando25}
  F. Ando, T. Hirai, A. Alasli, H. Sepehri-Amin, Y. Iwasaki, H. Nagano, and K. Uchida,
  Energy Environ. Sci. {\bf 18}, 4068 (2025). 

\bibitem{Lee25}
  Y. Lee, F. Ando, T. Hirai, R. Modak, H. Sepehri-Amin, and K. Uchida, 
  Ann. Phys. {\bf 537}, e00127 (2025).

%%%Unipolar anisotropy%%%

\bibitem{Chandrasekhar59}
  B. S. Chandrasekhar, 
  J. Phys. Chem. Solids, {\bf 11}, 268 (1959). 

\bibitem{Sill65}
  L. R. Sill and S. Legvold, 
  Phys. Rev. {\bf 137}, A1139 (1965). 

\bibitem{Saunders65}
  G. A. Saunders, C. Miziumski, G. S. Cooper, and A. Lawson, 
  J. Phys. Chem. Solids, {\bf 26}, 1299 (1965). 

\bibitem{Shao19}
  Q. Shao, A. M. Kanakkithodi, Y. Xia, M. K. Y. Chan, and M. Grayson, 
  MRS Adv. {\bf 4}, 491 (2019). 

\bibitem{McAfee23}
  K. McAfee, P. B. Sunderland, and O. Rabin, 
  Sens. Actuators A: Phys. {\bf 363}, 114729 (2023). 

%%%%ADCP%%%

\bibitem{Chung03}
  D.-Y. Chung, S. D. Mahanti, W. Chen, C. Uher, and M. G. Kanatzidis, 
  Mater. Res. Soc. Symp. Proc. {\bf 793}, 206 (2003). 

\bibitem{Gu05}
  J.-J. Gu, M.-W. Oh, H. Inui, and D. Zhan, 
  Phus. Rev. B {\bf 71}, 113201 (2005). 

\bibitem{Ong10}
  K. P. Ong, D. J. Singh, and P. Wu, 
  Phys. Rev. Lett. {\bf 104}, 176601 (2010). 

\bibitem{Cohn12}
  J. L. Cohn, B. D. White, C. A. M. dos Santos, and J. J. Neumeier,
  Phys. Rev. Lett. {\bf 108}, 056604 (2012). 

\bibitem{Ochs24}
  A. M. Ochs, G. H. Fecher, B. He, W. Schnelle, C. Felser, J. P. Heremans, and J. E. Goldberger, 
  Adv. Mater. {\bf 36}, 2308151 (2024). 

\bibitem{Goto24}
  Y. Goto, H. Usui, M. Murata, J. E. Goldberger, J. P. Heremans, and C.-H. Lee, 
  Chem. Mater. {\bf 36}, 2018 (2024). 

\bibitem{Manako24}
  H. Manako, S. Ohsumi, Y. J. Sato, R. Okazaki, and D. Aoki, 
  Nat. Commun. {\bf 15}, 3907 (2024). 

\bibitem{Ohsumi24}
  S. Ohsumi, Y. J. Sato, and R. Okazaki, 
  PRX Energy {\bf 3}, 043007 (2024). 

%%%Calculation part%%%

\bibitem{Ioffe57}
  A. F. Ioffe, {\it Semiconductor Thermoelements and Thermoelectrics Cooling} (Infosearch, London, 1957).

\bibitem{Heikes-Ure61}
  R. R. Heikes and R. W. Ure, Jr.,
  {\it Thermoelectricity: Science and Engineering} (Interscience, New York, 1961).

\bibitem{Harman-Honig67} 
  T. C. Harman and J. M. Honig,
  {\it Thermoelectric and Thermomagnetic Effects and Applications}
  (McGraw-Hill, New York, 1967).

\bibitem{Goldsmid10}
  H. J. Goldsmid, {\it Introduction to Thermoelectricity}
  (Springer Berlin, Heidelberg, 2010). 
  
\bibitem{Harman58}
  T. Harman, J. Appl. Phys. {\bf 29}, 1471 (1958).

\bibitem{Sherman60}
  B. Sherman, R. R. Heikes, and R. W. Ure, Jr.,
  J. Appl. Phys. {\bf 31}, 1 (1960).

\bibitem{Wright62}
  D. A. Wright, 
  Brit. J. Appl. Phys. {\bf 13}, 583 (1962).

\bibitem{Mahan98}
  G. D. Mahan,
  {\it Solid State Physics} (Academic Press, New York, 1998),
  {\bf 51}, pp. 81–157.
  
\bibitem{Horst63}
  R. B. Horst,
  J. Appl. Phys. {\bf 34}, 3246 (1963). 

\bibitem{Delves64}
  R. T. Delves,
  Brit. J. Appl. Phys. {\bf 15}, 105 (1964). 

\bibitem{Onose08}
  Y. Onose, Y. Shiomi, and Y. Tokura,
  Phys. Rev. Lett. {\bf 100}, 016601 (2008). 

\bibitem{Osterle-Angrist63}
  J. F. Osterle and S. W. Angrist,
  J. Appl. Mech. Trans ASME {\bf 11}, Ser. E, 426 (1963).

\bibitem{Adams19}
  M. J. Adams, M. Verosky, M. Zebarjadi, and J. P. Heremans, 
  Phys. Rev. Appl. {\bf 11}, 054008 (2019).

\bibitem{Tang16}
  Y. Tang, M. Ma, and M. Grayson, 
  Proc. SPIE {\bf 9821}, 98210K (2016). 

\bibitem{Landau-electrodyn}
  L. D. Landau, E. M. Lifshitz, and L. P. Pitaevskii, 
  {\it Electrodynamics of Continuous Media}
  (Pergamon Press, Oxford, 1984).

\bibitem{Callen-textbook}
  H. B. Callen, 
  {\it Thermodynamics} 
  (John Wiley \& Sons, Inc., New York, 1960).

\bibitem{Franco18}
  V. Franco, J. S. Bl\'{a}zquez, J. J. Ipus, J. Y. Law, L. M. Moreno-Ram\'{i}rez, and A. Conde, 
  Prog. Mater. Sci. {\bf 93}, 112 (2018). 

\bibitem{Domenicali54}
  C. A. Domenicali,
  J. Appl. Phys. {\bf 25}, 1310 (1954).

\bibitem{Harman-Honig62b}
  T. C. Harman and J. M. Honig,
  J. Appl. Phys. {\bf 33}, 3188 (1962).   

\bibitem{Kooi63}
  C. F. Kooi, R. B. Horst, K. F. Cuff, and S. R. Hawkins,
  J. Appl. Phys. {\bf 34}, 1735 (1963).

\bibitem{Goldsmid17}
  H. J. Goldsmid, 
  {\it The Physics of Thermoelectric Energy Conversion} 
  (Morgan \& Claypool Publishers, San Rafael, 2017). 

\bibitem{Harman63}
  T. C. Harman, 
  Adv. Energy Convers. {\bf 3}, 667 (1963).

\bibitem{Polash21}
  M. M. H. Polash and D. Vashaee, 
  Phys. Rev. Appl. {\bf 15}, 014011 (2021). 

\bibitem{Grayson18}
  M. Grayson, Q. Shao, B. Cui, Y. Tang, X. Yan, and C. Zhou, 
  Introduction to (p $\times$ n)-Type Transverse Thermoelectrics in {\it Bringing Thermoelectricity into Reality} (IntechOpen, Rijeka, 2018). 

%%%Configuration%%%

\bibitem{Uchida-Iguchi21}
  K. Uchida and R. Iguchi, 
  J. Phys. Soc. Jpn. {\bf 90}, 122001 (2021).

\bibitem{Itoh17}
  R. Itoh, R. Iguchi, S. Daimon, K. Oyanagi, K. Uchida, and E. Saitoh, 
  Phys. Rev. B {\bf 96}, 184422 (2017).

\bibitem{Flipse14}
  J. Flipse, F. K. Dejene, D. Wagenaar, G. E. W. Bauer, J. Ben Youssef, and B. J. van Wees, 
  Phys. Rev. Lett. {\bf 113}, 027601 (2014).

\bibitem{Breitenstein10}
  O. Breitenstein, W. Warta, and M. Langenkamp, 
  {\it Lock-in Thermography: Basics and Use for Evaluating Electronic Devices and Materials Introduction} 
  (Springer, Berlin/Heidelberg, 2010). 

\bibitem{Wid16}
  O. Wid, J. Bauer, A. M\"uller, O. Breitenstein, S. S. P. Parkin, and G. Schmidt, 
  Sci. Rep. {\bf 6}, 28233 (2016).

\bibitem{Daimon16}
  S. Daimon, R. Iguchi, T. Hioki, E. Saitoh, and K. Uchida, 
  Nat. Commun. {\bf 7}, 13754 (2016).

\bibitem{Seki18}
  T. Seki, R. Iguchi, K. Takanashi, and K. Uchida, 
  Appl. Phys. Lett. {\bf 112}, 152403 (2018).

\bibitem{Uchida18}
  K. Uchida, S. Daimon, R. Iguchi, and E. Saitoh, 
  Nature {\bf 558}, 95 (2018). 

\bibitem{Yamazaki20}
  T. Yamazaki, R. Iguchi, T. Ohkubo, H. Nagano, and K. Uchida, 
  Phys. Rev. B {\bf 101}, 020415(R) (2020).

\bibitem{Yamazaki24}
  T. Yamazaki, T. Hirai, T. Yagi, Y. Yamashita, K. Uchida, T. Seki, and K. Takanashi,
  Phys. Rev. Appl. {\bf 21}, 024039 (2024). 

\bibitem{Kikkawa13}
  T. Kikkawa, K. Uchida, Y. Shiomi, Z. Qiu, D. Hou, D. Tian, H. Nakayama, X.-F. Jin, and E. Saitoh. 
  Phys. Rev. Lett. {\bf 110}, 067207 (2013). 

\bibitem{Kikkawa13-PRB}
  T. Kikkawa, K. Uchida, S. Daimon, Y. Shiomi, H. Adachi, Z. Qiu, D. Hou, X.-F. Jin, S. Maekawa, and E. Saitoh, 
  Phys. Rev. B {\bf 88}, 214403 (2013). 

\bibitem{Ando25-PRAppl}
  F. Ando, T. Hirai, H. Adachi, and K. Uchida, 
  Phys. Rev. Appl. {\bf 23}, 064061 (2025). 

%%%Heat flux sensing%%%

\bibitem{Kirihara16}
  A. Kirihara, K. Kondo, M. Ishida, K. Ihara, Y. Iwasaki, H. Someya, A. Matsuba, K. Uchida, E. Saitoh, N. Yamamoto, S. Kohmoto, and T. Murakami,
  Sci. Rep. {\bf 6}, 23114 (2016). 

\bibitem{Modak22}
  R. Modak, Y. Sakuraba, T. Hirai, T. Yagi, H. Sepehri-Amin, W. Zhou, H. Masuda, T. Seki, K. Takanashi, T. Ohkubo, and K. Uchida,
  Sci. Technol. Adv. Mater. {\bf 23}, 767 (2022).

\bibitem{Tanaka23}
  H. Tanaka, T. Higo, R. Uesugi, K. Yamagata, Y. Nakanishi, H. Machinaga, and S. Nakatsuji, 
  Adv. Mater. {\bf 35}, 2303416 (2023). 

%%%Tilted oxide%%%

\bibitem{Renk94}
  K. F. Renk, J. Betz, S. Zeuner, H. Lengfellner, and W. Prettl, 
  Phys. C: Supercond. {\bf 235–240}, 37 (1994).

\bibitem{Roediger08}
  T. Roediger, H. Knauss, U. Gaisbauer, E. Kraemer, S. Jenkins, and J. von Wolfersdorf, 
  J. Turbomach. {\bf 130}, 011018 (2008). 

\bibitem{Zhang08}
  P. X. Zhang and H.-U. Habermeier, 
  J. Nanomater. {\bf 2008}, e329601 (2008). 

\bibitem{Takahashi09}
  K. Takahashi, A. Sakai, T. Kanno, and H. Adachi, 
  Appl. Phys. Lett. {\bf 95}, 051913 (2009). 

\bibitem{Kanno14}
T. Kanno, K. Takahashi, A. Sakai, H. Tamaki, H. Kusada, and Y. Yamada, 
J. Electron. Mater. {\bf 43}, 2072 (2014).   

\bibitem{Yordanov23} %THz%
  P. Yordanov, T. Priessnitz, M.-J. Kim, G. Cristiani, G. Logvenov, B. Keimer, and S. Kaiser, 
  Adv. Mater. {\bf 35}, 2305622 (2023). 

%%%Performance comparison%%%

\bibitem{Salvador14}
  J. R. Salvador, J. Y. Cho, Z. Ye, J. E. Moczygemba, A. J. Thompson, J. W. Sharp, J. D. Koenig, R. Maloney, T. Thompson, J. Sakamoto, H. Wang, and A. A. Wereszczak, 
  Phys. Chem. Chem. Phys. {\bf 16}, 12510 (2014). 

\bibitem{Murata21}
  M. Murata, K. Nagase, K. Aoyama, A. Yamamoto, and Y. Sakuraba, 
  iScience {\bf 24}, 101967 (2021). 
  
\bibitem{Ando24}
  F. Ando, T. Hirai, and K. Uchida,
  APL Energy {\bf 2}, 016103 (2024). 

\bibitem{Chen24}
  M. Chen, J. Wang, K. Liu, W. Fan, Y. Sun, C. Felser, T. Zhu, and C. Fu, 
  Adv. Energy Mater. {\bf 14}, 2400411 (2024). 

%%%Permanent magnet%%%

\bibitem{Hono12}
  K. Hono and H. Sepehri-Amin, 
  Scr. Mater. {\bf 67}, 530 (2012). 

\bibitem{Nomura18}
  M. Nomura, J. Shiomi, T. Shiga, and R. Anufriev, 
  Jpn. J. Appl. Phys. {\bf 57}, 080101 (2018). 

\bibitem{Kautsar25}
  Z. H. Kautsar, B. Madavali, T. Hirai, K. Uchida, and H. Sepehri-Amin,
  Energy Mater. {\bf 5}, 500129 (2025). 

\end{thebibliography}
\end{document}